\documentclass[11pt]{article}

\usepackage{graphicx}
\usepackage{xcolor}
\usepackage{amsmath}
\usepackage{amssymb}
\usepackage{amsthm}
\usepackage{color}			
\usepackage{booktabs}
\usepackage{multirow}
\usepackage{bmpsize}			
\usepackage{amssymb}
\usepackage{algorithmicx}
\usepackage[ruled]{algorithm}
\usepackage[title,titletoc]{appendix}
\usepackage{algpseudocode}
\usepackage{rotating}
\usepackage{hyperref,cite,cleveref}
\usepackage{caption}
\usepackage{subfig}
\usepackage{accents}

\numberwithin{equation}{section}
\newtheorem{theorem}{Theorem}

\newtheorem{problem}{Approach}
\newtheorem{definition}{Definition}
\newtheorem{proposition}{Proposition}

\usepackage{algorithmicx}
\usepackage[ruled]{algorithm}
\usepackage[title,titletoc]{appendix}
\usepackage{algpseudocode}
\usepackage{comment}

\newcommand{\E}{\mathbb{E}}

\newcommand{\R}{\mathbb{R}}
\newcommand{\s}{\mathcal{S}}
\newcommand{\mc}[1]{\mathcal#1}
\newcommand{\ttg}{\hat T_n}
\newcommand{\wh}[1]{\hat{#1}}

\title{On Efficient and Scalable Computation of the Nonparametric  \\ Maximum Likelihood Estimator in Mixture Models}							
\author{Yangjing Zhang\thanks{Institute of Applied Mathematics, Academy of Mathematics and Systems Science, Chinese Academy of Sciences, Beijing, People's Republic of China (yangjing.zhang@amss.ac.cn). Part of this work was done while Y. Zhang was with Department of Mathematics, National University of Singapore.}
\and
Ying Cui\thanks{
Department of Industrial and
Systems Engineering, University of Minnesota, Minneapolis, MN 55414 (yingcui@umn.edu); supported by NSF grant CCF-2153352}
\and
 Bodhisattva Sen\thanks{Department of Statistics, Columbia University, New York, NY 10027
(bodhi@stat.columbia.edu); supported by NSF grant DMS-2015376}
\and
Kim-Chuan Toh\thanks{Department of Mathematics, and Institute of Operations Research and Analytics, National University of Singapore,  Singapore 119076 (mattohkc@nus.edu.sg)}
}

\begin{document}
\maketitle

\begin{abstract}
In this paper we study the computation of the  nonparametric maximum likelihood estimator (NPMLE) in multivariate mixture models. Our first approach discretizes this infinite dimensional convex optimization problem by fixing the support points of the NPMLE and optimizing over the mixture proportions. In this context we propose, leveraging the sparsity of the solution, an efficient and scalable semismooth Newton based augmented Lagrangian method (ALM). Our algorithm beats the state-of-the-art methods~\cite{koenker2017rebayes, kim2020fast} and can handle $n \approx 10^6$ data points with $m \approx 10^4$ support points. Our second procedure, which combines the expectation-maximization (EM) algorithm with the ALM approach above, allows for joint optimization of both the support points and the probability weights. For both our algorithms we provide formal results on their (superlinear) convergence properties.

The computed NPMLE can be immediately used for denoising the observations in the framework of empirical Bayes. We propose new denoising estimands in this context along with their consistent estimates. Extensive numerical experiments are conducted to illustrate the effectiveness of our methods. In particular, we employ our procedures to analyze two astronomy data sets: (i) Gaia-TGAS Catalog~\cite{anderson2018improving} containing $n \approx 1.4 \times 10^6$ data points in two dimensions, and (ii) the $d=19$ dimensional data set from the APOGEE survey~\cite{majewski2017apache} with $n \approx 2.7 \times 10^4$.
\end{abstract}


\section{Introduction}\label{sec:intro}
\begin{figure}
  \centering
  \includegraphics[width=0.3\textwidth]{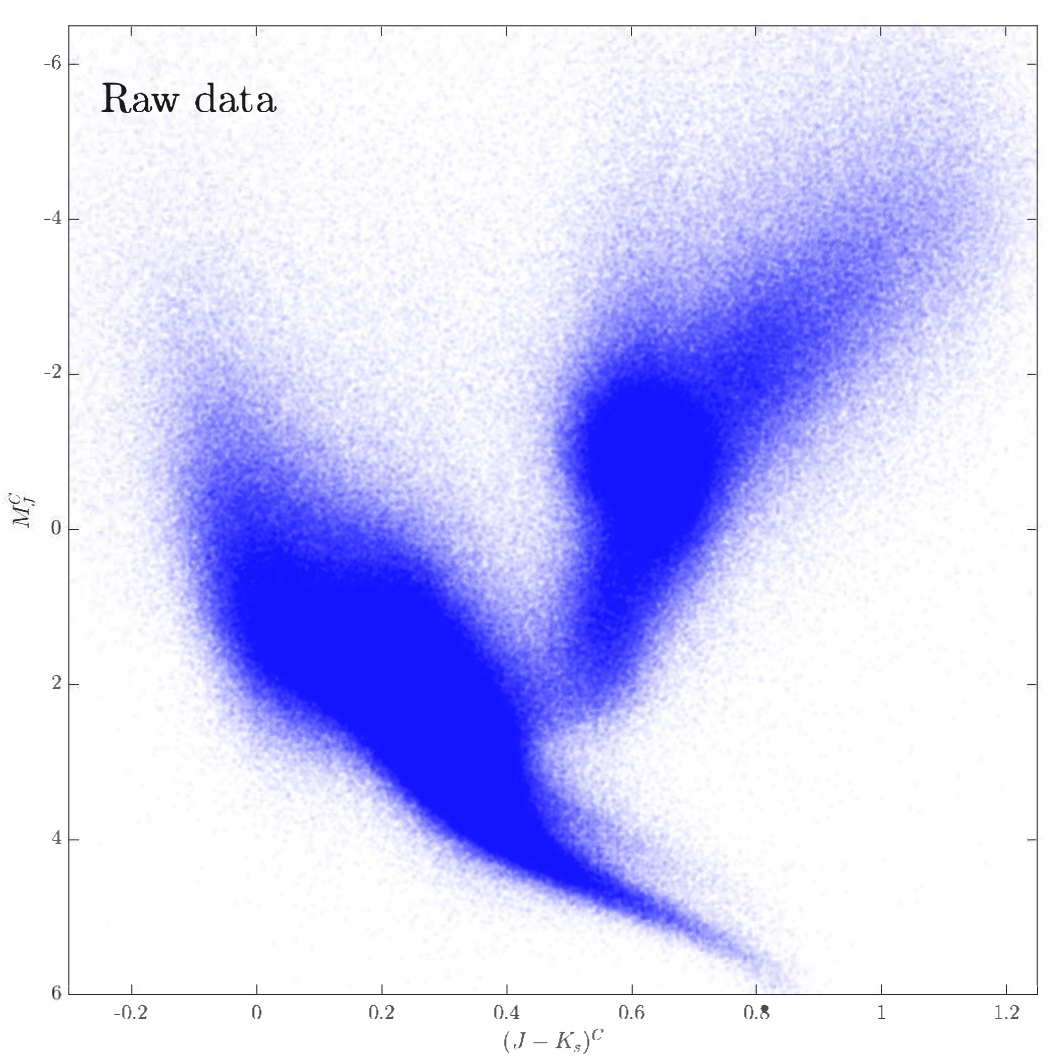}
  \includegraphics[width=0.3\textwidth]{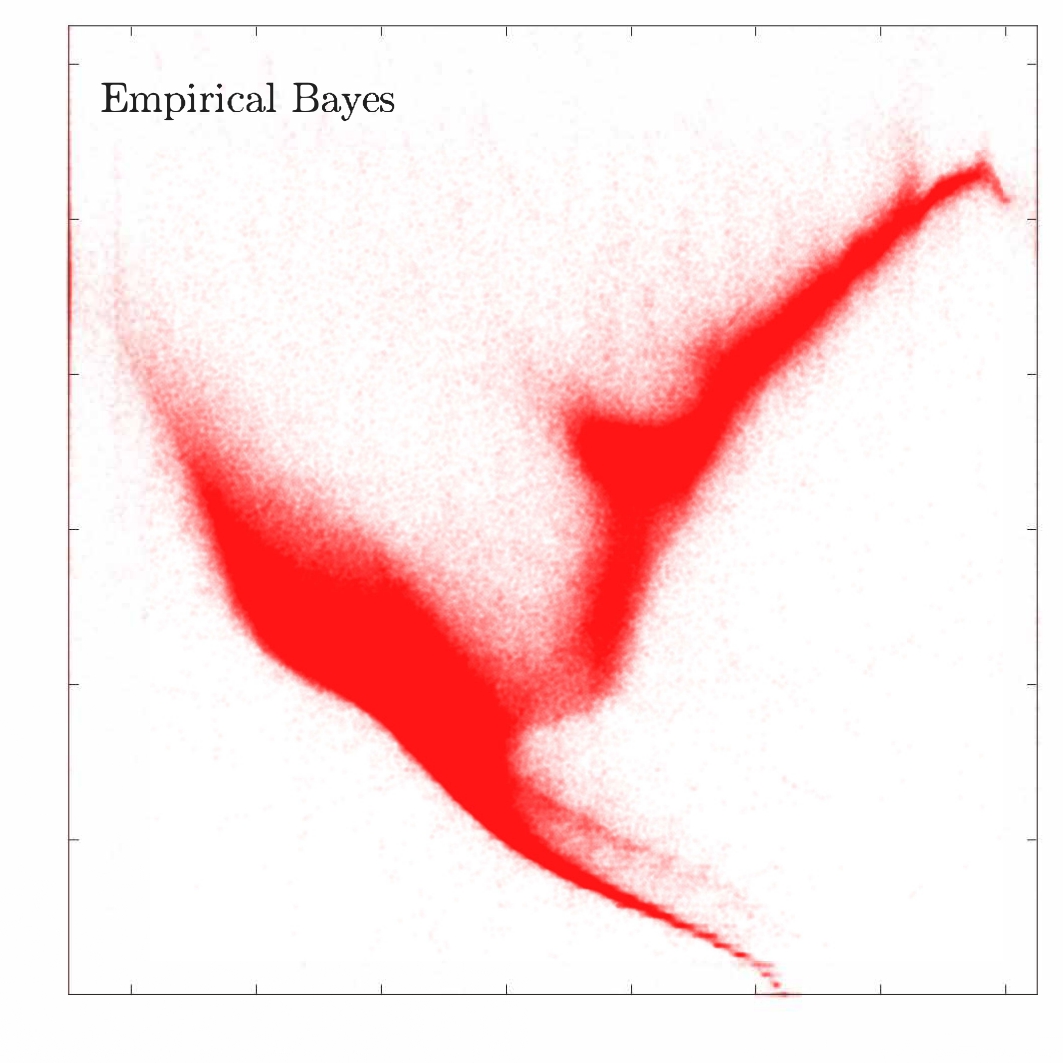}
  \includegraphics[width=0.3\textwidth]{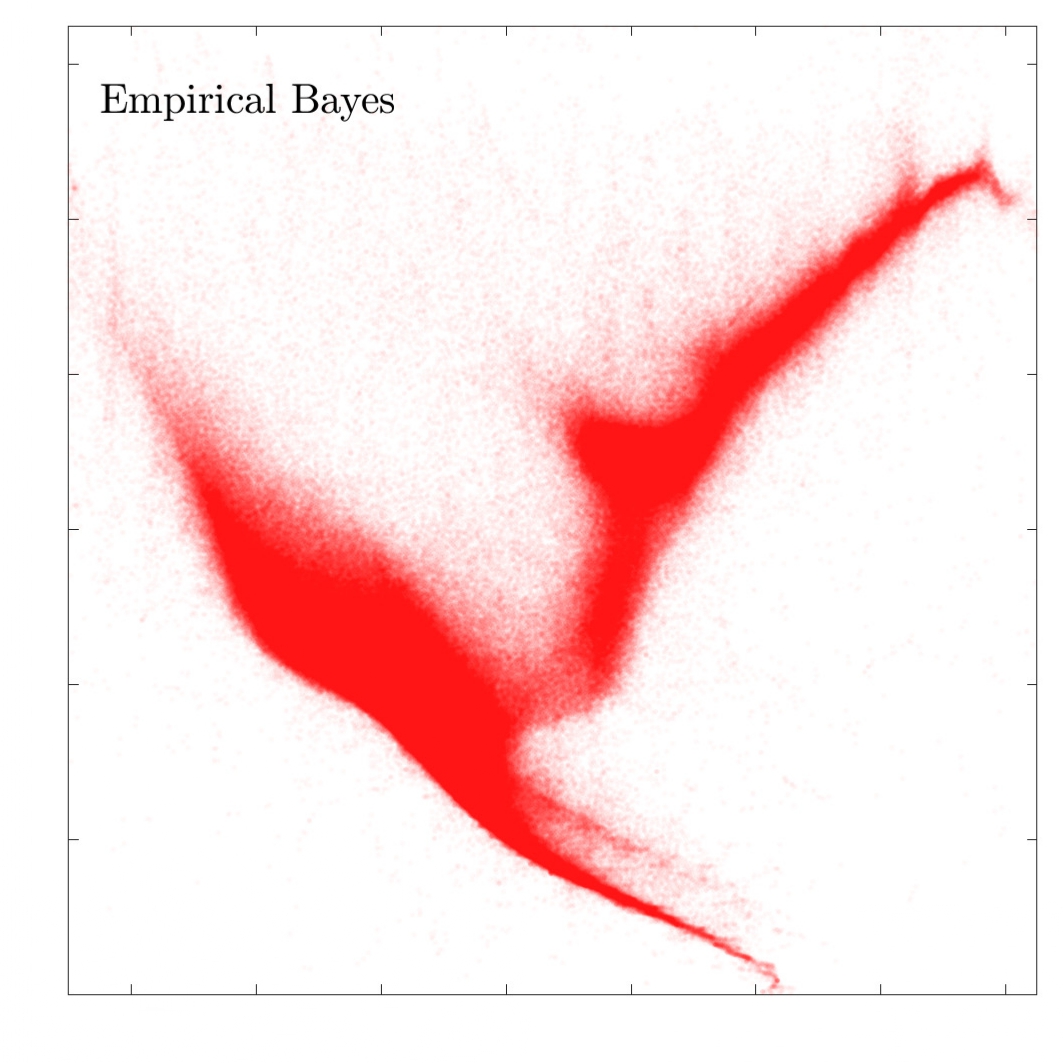}
  \caption{\small Left: The noisy CMD corresponding to data  $\{Y_i\}_{i=1}^n \subset \R^2$ for $n \approx 1.4 \times 10^6$ stars obtained from the Gaia-TGAS Catalog.
  Middle: The denoised CMD using empirical Bayes estimates $\{\widehat{\theta}_i\}_{i=1}^n$ based on the NPMLE computed via the {\it augmented Lagrangian} method (see Section~\ref{sec:ALM0}). Right: The denoised CMD computed via the {\it partial expectation maximization} algorithm (see Section~\ref{sec:adaptive supports}). Both the denoised CMDs have rather sharp tails in the bottom of the plots (i.e., the main sequence) and the top right (i.e., the tip of the red-giant branch) as well as a definitive cluster in the center-right (i.e., the red clump).
}\label{fig-CMD}
\end{figure}

We observe data $Y_1, \ldots, Y_n$ in $\mathbb{R}^d$ (for $d \ge 1$) from the heteroscedastic Gaussian location  mixture model
\begin{equation}\label{denoising data}
Y_i  = \theta_i + Z_i, \quad \mbox{with $\;\;\theta_i \overset{iid}{\sim} G^*\;\;$ and $\;\;Z_i \overset{ind}{\sim} {\cal N}(0, \Sigma_i)$}
\end{equation}
where the underlying (unknown) latent parameters $\{\theta_i\}^n_{i=1}$ are assumed to be drawn i.i.d.~from a common unknown distribution $G^*$ on $\mathbb{R}^d$, and $\{\Sigma_i\}_{i=1}^n$ is a collection of known $d \times d$ positive definite heteroscedastic covariance matrices; assume further that $\theta_i$ and $Z_i$ are independent for each $i= 1, \ldots, n$. It is of importance to nonparametrically estimate $G^*$ and the latent variables $\{\theta_i\}^n_{i=1}$ that are observed with errors. Such mixture models arise naturally in various applications~\cite{Carlin-Louis-96, Efron-10, Efron-Hastie-21}, including in the analysis of astronomy data~\cite{kelly2012measurement, AB96, HDB10}; see the left panel of  Figure~\ref{fig-CMD} which shows the noisy color-magnitude diagram (CMD) corresponding to observations $\{Y_i\}_{i=1}^n$ for $n \approx 1.4 \times 10^6$ stars from the Gaia-TGAS Catalog~\cite{anderson2018improving}.

Observe that the marginal density of $Y_i$ in~\eqref{denoising data} is given by
\begin{equation}\label{mixture-dis}
f_{G^*, \Sigma_i}(y) := \int \phi_{\Sigma_i}(y - \theta) dG^*(\theta),
\end{equation}
where $\phi_{\Sigma_i}(y) := [\det(2\pi\Sigma_i)]^{-1/2} \exp(-y^\top \Sigma_i^{-1}y/2)$ is the density function of ${\cal N}(0, \Sigma_i)$; further the observed $Y_i$'s are independent. A classical approach to estimating the unknown probability distribution $G^*$ in~\eqref{denoising data}, which goes back to the works of Robbins~\cite{robbins1950generalization} and Kiefer and Wolfowitz~\cite{kiefer1956consistency}, is via the following {\it nonparametric maximum likelihood estimator} (NPMLE) which maximizes the marginal likelihood of the observations $Y_i$'s \cite{robbins1950generalization,kiefer1956consistency,lindsay1983geometry,lindsay1995mixture,jiang2009general}:
\begin{equation}\label{NPMLE1}
  \widehat{G}_n \in \underset{G\in\mathcal{G}}{\arg\max} \;\,  \frac{1}{n} \sum_{i=1}^{n} \log\,f_{G, \Sigma_i}(Y_i),
\end{equation}
where the set $\mathcal{G}$ consists of all probability distributions on $\mathbb{R}^d$. Based on the solution $\widehat{G}_n$ of~\eqref{NPMLE1},  the marginal density $f_{G^*,\Sigma_i}$ of $Y_i$ can be estimated by $f_{\widehat{G}_n,\Sigma_i}$, and each observation $Y_i$ can be {\it denoised} via the empirical Bayes estimator (see e.g.,~\cite{soloff2021multivariate}):
\begin{equation}\label{EB}
\widehat{\theta}_i:=\mathbb{E}_{\widehat{G}_n}[\theta_i\mid Y_i], \qquad \mbox{ where }  \theta_i \sim \widehat{G}_n\;\mbox{ and } \;Y_i\mid \theta_i \sim {\cal N}(\theta_i,\Sigma_i),
\end{equation}
to obtain an `estimate' of the underlying latent parameter $\theta_i$; see e.g.,~\cite{jiang2009general, efron19, soloff2021multivariate}. For the noisy CMD from~\cite{anderson2018improving}, the middle and right panels of Figure~\ref{fig-CMD} (barely distinguishable) show the  denoised empirical Bayes estimates $\{\widehat{\theta}_i\}^n_{i=1}$ based on the NPMLE and solved via two different algorithms --- the  {\it augmented Lagrangian} method and the {\it partial expectation maximization} algorithm --- proposed and studied in this paper.

The main goal of this paper is to develop efficient and scalable optimization algorithms for the computation of the NPMLE $\widehat{G}_n$ of $G^*$ based on the observed data $\{Y_i\}^n_{i=1}$, especially when $d > 1$ and $n$ is large (e.g., $n\approx 10^6$). Note that~\eqref{NPMLE1} is an {\it infinite dimensional convex optimization} problem (in the sense that the decision variable $G$ in~\eqref{NPMLE1} ranges over all probability measures on $\R^d$) that is challenging to solve computationally, especially when $n$ is large and $d >1$. Many numerical methods for approximately computing the NPMLE have been considered --- including the expectation maximization (EM) algorithm~\cite{laird1978nonparametric}, vertex direction and exchange methods \cite{bohning1985numerical}, semi-infinite methods \cite{lesperance1992algorithm}, constrained-Newton methods \cite{wang2007fast}, 
and hybrid methods \cite{liu2007partially, bohning2003algorithm} --- typically described for the special case when $d=1$ and homoscedastic errors. Broadly speaking, there are two main strategies to solve~\eqref{NPMLE1} --- for each of which we develop a computationally efficient algorithm with provable convergence guarantees --- as described below.

\begin{problem}[Finite mixture model with known atoms]

A natural way to alleviate the computational difficulty of~\eqref{NPMLE1}
is to discretize (a compact region of) the whole space $\mathbb{R}^d$ and restrict ${\cal G}$ to the class of all distributions with a finite fixed support, say
$\{ \mu_1,\dots,\mu_m\} \subseteq \mathbb{R}^d$; see e.g.,~\cite{koenker2014convex, kim2020fast}. Namely, we assume that every $G\in {\cal G}$ takes the form
\begin{equation}\label{eq:discretized prob}
  G = \sum_{j=1}^{m} x_j \delta_{\mu_j},\quad \mbox{where }\;\; x_j \geq 0 \; \forall j, \quad \mbox{and }\;\;\sum_{j=1}^{m} x_j = 1
\end{equation}
for unknown mixture proportion $x = (x_1, \dots,x_m)^\top$ and fixed $\{ \mu_1,\dots,\mu_m\}$ with $m$ large; here by $\delta_{a}$ we mean the Dirac delta measure at $a$. Under the above reduction,~\eqref{NPMLE1} reduces to the following finite dimensional convex optimization problem:
\begin{equation}\label{primal-0}
\begin{array}{cl}
\displaystyle\operatornamewithlimits{maximize}_{x=(x_1,\dots,x_m)^\top\in\mathbb{R}^m} & \displaystyle \frac{1}{n} \sum_{i=1}^{n}\log\,\left(\sum_{j=1}^{m}L_{ij} x_j\right)\\[0.25in]
\mbox{subject to} & {\bf 1}_m^\top \, x = 1,\quad  x_j \geq 0, \;\forall j = 1,\ldots, m,
\end{array}
\end{equation}
where $L := (L_{ij} ) \in \mathbb{R}^{n\times m}$ is a fixed matrix with nonnegative entries such that $L_{ij} :=  \phi_{\Sigma_i}(Y_i - \mu_j)$ 
and ${\bf 1}_m$ denotes the vector of all ones in $\mathbb{R}^m$.

Observe that the optimization problem~\eqref{primal-0} can also arise in other contexts, e.g., it encompasses the maximum likelihood estimation of mixture
proportions in a finite mixture model where the component densities are known. That is, suppose that we observe $n$ i.i.d.~data $Y_1, \ldots, Y_n$ following the mixture density $\sum_{j=1}^m x_j f_j(\cdot)$ with unknown mixture proportion $x = (x_1,\ldots, x_m)^\top$ and known density functions $f_1,\ldots,f_m$. Taking $L_{ij} = f_j(Y_i)$, the MLE of the vector of mixture
proportions $x$ reduces to problem~\eqref{primal-0}.
\end{problem}

The most classical approach to solving~\eqref{primal-0} is to use the EM algorithm~\cite{dempster1977maximum}. However, the EM may converge very slowly; see e.g.,~\cite{redner1984mixture, varadhanEM2008, koenker2014convex}.  Compared to the EM, modern convex optimization methods would be more efficient and stable. Among them, first order methods are natural choices for solving \eqref{primal-0}, although the convergence of first order methods for solving this problem may slow down considerably as they approach the solution as shown in \cite[Section~4.3.5]{kim2020fast}, especially when $m$ and $n$ are large.
In principle,  the convex problem \eqref{primal-0} can also be solved by off-the-shelf interior point based solvers. In fact, the routine \texttt{KWDual} in the  \texttt{R} package \texttt{REBayes} \cite{koenker2017rebayes} adopts the interior point method implemented by the commercial interior point solver Mosek \cite{andersen2000mosek,mosek} to solve the dual formulation of \eqref{primal-0}. Although very stable and efficient for small to medium sized problems,  the interior point method has  inherently ill-conditioned normal equations that are extremely costly to solve by an iterative method when both $m$ and $n$ are large\footnote{In particular, we found that the \texttt{REBayes} solver~\cite{koenker2017rebayes} reports failure for a synthetic data set with $m=10^4$ and $n=7 \times 10^4$; see Figure~\ref{fig-sim2} in Section~\ref{sec:syn} for details.}. In the recent paper \cite{wang2021nonparametric}, the authors have proposed a cubic regularized Newton method to solve \eqref{primal-0} for $d=1$ under additional shape constraints.

\begin{figure}[!ht]
\centering
\includegraphics[width=0.43\textwidth]{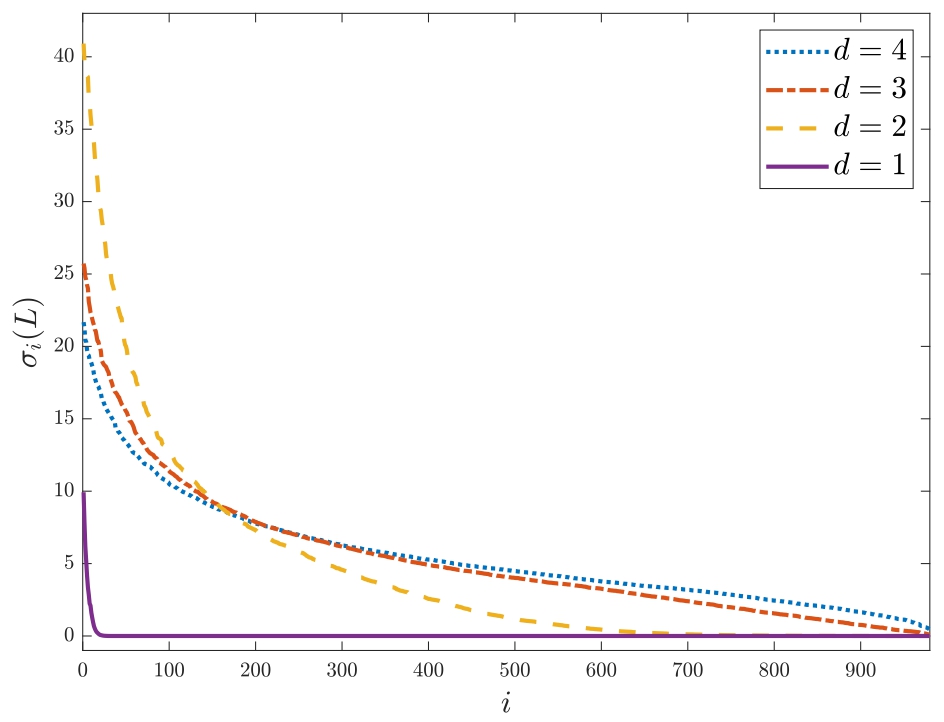}\,
\includegraphics[width=0.43\textwidth]{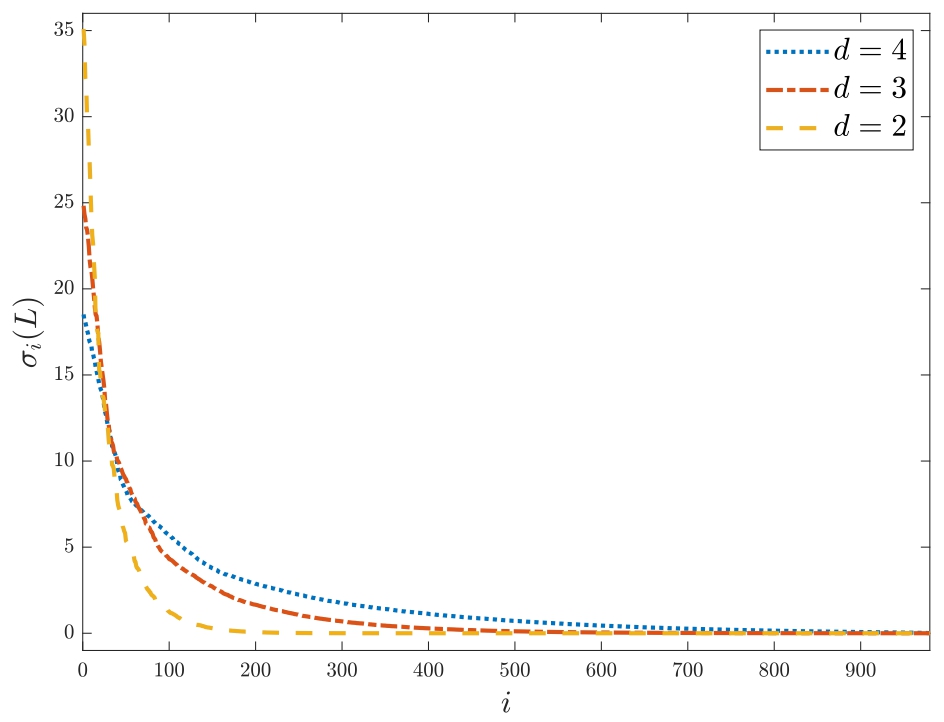}
  \caption{\small The distribution of the singular values $\sigma_i(L)$ of $L$ computed from (left) the APOGEE data (here $n=27,135$, and $m=1,000$) and (right) the synthetic Example 3(a)  (here $n=5,000$, $m=1,000$) as $d$ varies. The top $20$ singular values are excluded so that the others are not overshadowed. {Observe the slow decay of the singular values when $d >1$.}}
  \label{lowrankL}
\end{figure}

In order to solve  \eqref{primal-0} more efficiently, especially when $n$ is large, an active set based sequential quadratic programming (SQP) method was recently proposed by Kim et al.~\cite{kim2020fast}. The proposed algorithm \texttt{mix-SQP} is able to solve~\eqref{primal-0} with large $n$ (up to $10^6$) and small to medium $m$ (up to several hundreds) very efficiently, by leveraging a low rank approximation of the matrix $L$ (in~\eqref{primal-0}) for univariate probability distributions (i.e., $d=1$). However,
the low rank approximation in the SQP method \cite{kim2020fast} may not work well for estimating multivariate (i.e., when $d\geq 2$) probability distributions $G^*$. Figure~\ref{lowrankL} shows the distributions of the singular values of the matrix $L$ computed from the APOGEE survey~\cite{majewski2017apache} (see Section \ref{sec:real} for details) and the synthetic Example 3(a) (see Section \ref{sec:d3} for details). We can see that most singular values of $L$ are close to $0$ when $d=1$, but $L$ has a significant proportion of nonzero singular values when $d\geq2$, and the situation is more pronounced when $d$ is larger. This observation suggests that a low rank approximation of $L$ may lose crucial information in data fitting when $d\geq 2$. In addition,  for problem~\eqref{primal-0},  the number of grid points $m$ needed to obtain a good approximation to the infinite dimensional problem~\eqref{NPMLE1} may be large for $n$ large, especially when $d >1$; see Figure~\ref{fig-Gaia} where we show plots for the noisy CMD data from~\cite{anderson2018improving} with different values of $m$.

The first primary goal of the present paper is to provide a highly efficient, stable, and scalable numerical algorithm for solving problem \eqref{primal-0} that can handle large $n$ and $m$ (e.g., $n \approx 10^6$ and $m \approx 10^4$).
Our proposal is to apply the {\it augmented Lagrangian method} (ALM) for solving the dual problem of \eqref{primal-0}. Briefly, the ALM is an iterative method that solves a sequence of unconstrained subproblems to approximate the targeted constrained problem; see~\cite{hestenes1969multiplier, powell1969method, rockafellar1976augmented}. The ALM subproblem for the dual of  problem~\eqref{primal-0} can be converted into the minimization of a tractable continuously differentiable function, with the aid of the powerful tool of Moreau-Yosida regularization
(see, e.g., \cite[Chapter 1.G]{rockafellar2009variational}).  Further, to solve this ALM subproblem 
we employ a {\it semismooth Newton method} (see e.g.,~\cite[Chapters 7 and 8]{facchinei2007finite}) that can
exploit the sparsity in the corresponding generalized Hessian matrix (a nonsmooth counterpart of the Hessian matrix that arises in a second order optimization problem; see \eqref{eq:hessian} for details) leveraging  the sparsity of the solution $x$ in~\eqref{primal-0} (see e.g.,~\cite{koenker2017rebayes,polyanskiy2020self}).

Although the sparsity of the solution $x$ is also exploited in the paper \cite{kim2020fast} in computing the search direction for each SQP subproblem, their Hessian matrix itself is dense and the computational cost in evaluating each Hessian is $O((n+m)k^2)$ with $k$ being the (approximate) rank of the matrix $L$. In contrast, 
the generalized Hessian matrix arising from the proposed semismooth Newton method (in each ALM subproblem) is inherently sparse and  the computational cost can be substantially reduced to $O(ns\min(n,s))$, where $s$ is the number of nonzero elements in a certain vector in $\mathbb{R}^m$ closely related to $x$; see~\eqref{jac-plus-fct},~\eqref{eq:hessian} and~\eqref{eq:s}. When  $s < n$ (in fact, $s \ll n$ in most situations; see~\cite{polyanskiy2020self}), the computational cost is $O(ns^2)$; see Section~\ref{subsec: computational cost} for details.

Theoretically, we show that both the ALM for the outer loop and the semismooth Newton method for the inner loop have global convergence (see Proposition~\ref{prop:convergence}) and superlinear convergence rate (see Propositions~\ref{prop:rate} and~\ref{prop:convergence ssn}).
We illustrate the scalability and efficiency of our proposed method via extensive numerical experiments.
In particular, as far as we are aware, for the $2$-dimensional noisy CMD data in the left panel of Figure~\ref{fig-CMD}, our ALM is the only known convex optimization method that can handle $m \approx 10^4$ grid points (with $n \approx 1.4 \times 10^6$). For such a large $m$, even if we randomly subsample $n=10^5$ data points  to make the package \texttt{REBayes}~\cite{koenker2017rebayes} applicable (the \texttt{mix-SQP} solver~\cite{kim2020fast} still does not work), the latter package is about $15$ times slower than our ALM; see Section~\ref{sec:real} for a detailed comparison of these methods based on their run time.

\begin{problem}[Finite mixture model with unknown atoms] Although {Approach 1} works well when $d=1$ or $2$ where one can cover the support of $\widehat{G}_n$ (in~\eqref{NPMLE1}) using finely chosen grid points\footnote{See~\cite{soloff2021multivariate} where rigorous results are provided on the choice of these grid points.}, it may no longer work when $d$ is large (e.g., for the APOGEE data~\cite{majewski2017apache} in Section~\ref{sec:real}, where the abundances are in $\mathbb{R}^{19}$). However, it is known that every solution $\widehat{G}_n$ of~\eqref{NPMLE1} is discrete with $\hat k$ atoms (say); in fact, there always exists a $\widehat{G}_n$ with $\hat k \le n$ atoms (see~\cite{soloff2021multivariate}). This implies that $\widehat{G}_n$ may be taken to be the maximum likelihood solution to a $\hat k$-component, heteroscedastic Gaussian mixture model where $\hat k \le n$. As a result, we are naturally interested in the following {\it nonconvex} optimization problem:
\begin{equation}\label{adaptive supports 0}
\begin{array}{cl}
\displaystyle\operatornamewithlimits{maximize}_{x\in \mathbb{R}^m, \{\mu_j\}_{j=1}^m\in \mathbb{R}^d} &  \displaystyle \frac{1}{n} \sum_{i=1}^{n}\log\,\left(\sum_{j=1}^{m} x_j  \phi_{\Sigma_i}(Y_i-\mu_j) \right)\\[0.25in]
\mbox{subject to} & {\bf 1}_m^\top \, x = 1,\qquad  x_j \geq 0, \;\forall j = 1,\ldots, m,
\end{array}
\end{equation}
where $m$ can always be taken to be $n$, but in practice $m \ll n$ would also work well.
Problem~\eqref{adaptive supports 0} can be viewed as simultaneously estimating the  Gaussian centers and the mixture proportions.
\end{problem}

One classical way to tackle the nonconvexity of \eqref{adaptive supports 0} is via the EM algorithm~\cite{dempster1977maximum,redner1984mixture},  where the objective function is successively minorized by the expectation of the `complete data' log-likelihood (with respect to the posterior distribution of the latent variables based on the current parameter estimates). Since the minorization may not tightly approximate the objective function  in \eqref{adaptive supports 0}, the  EM algorithm may suffer from slow convergence rate and  poor solution quality\footnote{The EM solution for $x$ in~\eqref{adaptive supports 0} is not usually sparse, although theory guarantees sparsity; see e.g.,~\cite{polyanskiy2020self}.}; see Table~\ref{table-EM1}.

To improve the performance of the classical EM algorithm, we propose in Section~\ref{sec:adaptive supports} a new algorithm --- the {\it partial EM} (PEM) --- that adopts the same update rule for $\{\mu_j\}_{j=1}^m$ as in the EM while directly computing the vector of mixture proportions $x$ at the currently computed $\{\mu_j\}_{j=1}^m$ via our fast ALM solver developed for {Approach 1}.  Since this PEM strategy generates tighter bounds on the objective function at each step of the $x$-update, it produces solutions with larger objective values and exhibits faster convergence compared to the EM. We further show that the sequence generated by the PEM algorithm is bounded with any of its accumulation point being a stationary point of the nonconvex problem~\eqref{adaptive supports 0} (see Proposition~\ref{prop:PEM}).  The numerical performance of the PEM on the synthetic data and the APOGEE survey~\cite{majewski2017apache} for $d\geq 3$ 
can be found in Section~\ref{subsec: numerical PEM}.

\vskip 0.1in
The third main contribution of this paper is related to the denoising of the observations $\{Y_i\}_{i=1}^n$. So far we have focussed on denoising the $Y_i$'s by the empirical Bayes estimates defined in~\eqref{EB}. In Section~\ref{sec:Denoising} we argue that these empirical Bayes estimates are not necessarily guaranteed to lie `close' to the support of $G^*$, which may be undesirable in certain applications. We  propose new denoising estimands defined via the theory of optimal transport (see e.g.,~\cite{Villani2003, Villani2009}) that can mitigate this shortcoming of the empirical Bayes estimates. Moreover, in Section~\ref{sec:Denoising-NPMLE} we propose and study sample estimates of these new denoising estimands and prove, via a finite sample high probability bound (see Theorem~\ref{thm:Rate-OT-Map}), that the sample estimates are close to their population counterparts.

\vskip 0.1in
In the following we briefly summarize the main contributions of the paper:
\begin{itemize}
  \item We propose a highly efficient and scalable semismooth Newton based ALM for problem \eqref{primal-0} in Section~\ref{sec:ALM0}.
  The key technique to guarantee the efficiency and scalability of our ALM is the full exploitation of the second order sparsity in the generalized Hessian matrix arising in the ALM subproblem. Due to this sparsity, the semismooth Newton direction can be computed cheaply and the subproblem can be solved efficiently (see Table~\ref{table:complexity} for a comparison of the computational costs among~\texttt{REBayes}~\cite{koenker2017rebayes}, the SQP method~\cite{kim2020fast}, and our ALM).

 \item We propose a PEM method to solve \eqref{adaptive supports 0} in Section~\ref{sec:adaptive supports}, which allows for joint optimization of both the support points $\mu_j$'s and the probability weights $x_j$'s, by combining the advantages of the convex program~\eqref{primal-0} and  the classical EM algorithm.  In particular, like the EM algorithm, the objective value in our PEM method is guaranteed to increase with iteration; but unlike the EM, our PEM exhibits fast convergence and can produce sparse solutions for $x$. In fact, we find that the PEM algorithm almost always yields the highest objective (i.e., log-likelihood) value when compared to the other competing methods. 

\item In Section~\ref{sec:Denoising} we propose new denoising estimands for model~\eqref{denoising data} that are guaranteed to lie close to the support of $G^*$. We also propose consistent estimates for these denoising estimands.

  \item We conduct extensive numerical experiments on synthetic and real astronomy data sets in Section~\ref{sec:numerical}. We illustrate that, for problem~\eqref{primal-0}, our ALM is much faster and scalable when compared to other existing solvers. For solving~\eqref{adaptive supports 0}, we show that the proposed PEM algorithm usually provides solutions that: (i) have higher objective value, (ii) are sparser, and (iii) require less run time, when compared to the classical EM algorithm. Our proposed methods work very well when $d$ is moderately large (e.g., when $d \le 10$).
Further, our methods are implemented in Matlab and the relevant codes, including simulation experiments, are available in the first author's \texttt{GitHub}
page at \url{https://github.com/YangjingZhang/Dual-ALM-for-NPMLE}.

\end{itemize}
The proofs of the main results and further numerical experiments are relegated to the Appendix.


\section{An augmented Lagrangian method for solving~\eqref{primal-0}}\label{sec:ALM0}
\subsection{Preliminaries}\label{sec:prelim}
We first introduce some basic notions from convex analysis, including the concept of Moreau-Yosida regularization of a proper closed convex function that will be useful in the sequel. A convex function $f:\R^n \to [-\infty, \infty]$ is said to be {\it proper} if $f(x) < +\infty$ for at least one $x$ and $f(x) > -\infty$ for every $x$. The convex function $f$ is said to be {\it closed} if $\{x\,|\,f(x)\leq \alpha\}$ is closed for every $\alpha\in\mathbb{R}$.
Let  $f:\, \mathbb{R}^n\rightarrow(-\infty,+\infty]$ be a proper closed convex function. Parametrized by a scalar $\sigma>0$, the {\it Moreau-Yosida regularization} of $f$ (also called the {\it Moreau envelope} of $f$) is defined as
\begin{equation}\label{def-MY}
\mathcal{M}_{f}^{\sigma}(x):=\min\limits_{z\in \mathbb{R}^n}\left\{f(z)+\frac{\sigma}{2}\|z-x\|_2^2\right\},\quad  x\in \mathbb{R}^n;
\end{equation}
here $\|\cdot\|_2$ denotes the usual Euclidean norm. The unique optimal solution of \eqref{def-MY} for any given $x$, denoted as
\[
{\rm Prox}_{f}^{\sigma}(x):=\displaystyle\operatornamewithlimits{argmin}_{z\in \mathbb{R}^n}\,\left\{f(z)+\frac{\sigma}{2}\|z-x\|_2^2\right\},
\]
 is called the {\it proximal point} of $x$ associated with $f$. The corresponding function
${\rm Prox}_{f}^{\sigma}$ is called the {\it proximal mapping} of $f$.
This regularization is a powerful tool to smooth a possibly nonsmooth convex function such that its gradient can be computed easily based on the proximal mapping of the original function. In fact, one important property is that the Moreau envelope $\mathcal{M}_f^{\sigma}$ is always continuously differentiable (and convex),
regardless of whether the original function $f$ is smooth or not, and the function $\mathcal{M}_f^{\sigma}$ has a Lipschitz gradient given by
\begin{equation}\label{def-grad-MY}
\nabla \mathcal{M}_{f}^{\sigma} (x) = \sigma \left[ \, x - {\rm Prox}_{f}^{\sigma}(x) \, \right], \quad x\in\mathbb{R}^n.
\end{equation}
Interested readers may consult \cite[Chapter 1.G]{rockafellar2009variational} for more properties of the Moreau envelope and the proximal mapping.

Next we introduce the concept of {\it semismoothness} starting from some basic variational analysis. Let $F:\mathbb{R}^n \to \mathbb{R}^m$ be a  vector-valued locally Lipschitz continuous function. It follows from
Rademacher's theorem that $F$ is differentiable almost everywhere.
We can thus define the {\it Clarke generalized Jacobian}  of $F$ at any $x\in \mathbb{R}^n$ as
\[
\partial F(x) :=\mbox{conv} \left\{\, \lim_{k\to \infty} JF(x^k) \, \mid \, \mbox{$\{x^k\}_{k\geq 1}$ is a sequence of  differentiable points of $F$ converging to $x$} \,\right\}.
\]
where $JF(x)$ denotes the Jacobian matrix of $F$; here by conv$(S)$ we mean the convex hull of a given set $S$.
We say $F$ is {\sl semismooth} at $x\in \mathbb{R}^n$ if $F$ is directionally differentiable at $x$ and for any $V_h\in \partial F(x+h)$,
\[
F(x+h) - F(x) - V_h h = o(\|h\|_2) \quad \mbox{as $h\to 0$}.
\]
Detailed properties of semismooth functions can be found in the monograph \cite{facchinei2007finite}.

\subsection{An augmented Lagrangian method for the dual of~\eqref{primal-0}}\label{sec:ALM}
In this subsection, we describe our proposed algorithm to solve the optimization problem~\eqref{primal-0}. As mentioned in the Introduction, our framework involves  applying the augmented Lagrangian method (ALM) to solve the dual problem (see \eqref{dual} below) of the convex optimization problem~\eqref{primal-0}.

Since the   $-\log(\cdot)$ function in the objective is convex nonincreasing and scale invariant (i.e., for any $t, \alpha >0$, it holds that $-\log( \alpha t) = -\log t - c$ for $c := \log \alpha$), it has been observed in \cite{kim2020fast} that  problem \eqref{primal-0} is equivalent to the following convex optimization problem with nonnegative constraints only (see \cite[Proposition 3.2]{kim2020fast}):
\begin{equation*}
\begin{array}{cl}
\displaystyle\operatornamewithlimits{maximize}_{x=(x_1,\dots,x_m)^\top\in \mathbb{R}^m} & \displaystyle \frac{1}{n} \sum_{i=1}^{n}\log\,\left(\sum_{j=1}^{m}L_{ij} x_j\right) - {\bf 1}_m^\top x + 1 \\[0.2in]
\mbox{subject to} & x_j \geq 0 , \;\forall j = 1,\ldots, m.
\end{array}
\end{equation*}
We introduce an auxiliary variable $y$ to separate the components in the objective function and obtain the following primal problem:
\begin{equation}\label{primal}
\begin{array}{cl}
\displaystyle\operatornamewithlimits{maximize}_{x\in \mathbb{R}^m,\, y\in \mathbb{R}^n} & \displaystyle \frac{1}{n} \sum_{i=1}^{n}\log\, y_i  - {\bf 1}_m^\top x + 1\\[0.2in]
\mbox{subject to} &  \displaystyle\frac{1}{n}(Lx - y) = 0,\quad x \geq 0.
\end{array}\tag{P}
\end{equation}
Here by $x\geq 0$ we mean that every coordinate of $x$ is nonnegative.  One can obtain the dual problem of \eqref{primal} by minimizing the  Lagrangian function associated with \eqref{primal}, i.e.,
$$
\operatornamewithlimits{minimize}_{x\geq 0 \,  \in \mathbb{R}^m,\, y\in \mathbb{R}^n}\quad
\frac{1}{n}\sum_{i=1}^{n}\log\, y_i  - {\bf 1}_m^\top x + 1 +  \frac{1}{n} u^\top(Lx-y),
$$
where $u\in \mathbb{R}^n$ is the Lagrange multiplier. The dual problem admits the following formulation:
\begin{equation}\label{dual}
\begin{array}{cl}
\displaystyle\operatornamewithlimits{minimize}_{u, v\in \mathbb{R}^n} & h(u):=\displaystyle -\frac{1}{n} \sum_{i=1}^{n}\log\, u_i  \\[0.2in]
\mbox{subject to} & \displaystyle\frac{1}{n} L^\top v \leq {\bf 1}_m,\quad u-v = 0.
\end{array}\tag{D}
\end{equation}
Here the auxiliary variable $v\in \mathbb{R}^n$ is introduced to separate the difficulties in dealing with the $-\log(\cdot)$ objective function and the inequality constraint simultaneously.

Equipped with the above notions, we are now able to introduce the ALM applied to the dual problem \eqref{dual}. The  ALM was first proposed by Hestenes \cite{hestenes1969multiplier} and Powell \cite{powell1969method} for equality-constrained nonlinear programs.
The augmented Lagrangian function involves quadratic penalties on the violation of equality constraints, and  the ALM converts the minimization of an equality constrained problem into the minimization of a sequence of unconstrained problems. For a general convex nonlinear program (having both equality and inequality constraints), we can follow \cite{rockafellar1976augmented} for the derivation of the ALM. For problem \eqref{dual} having an inequality constraint $\displaystyle\frac{1}{n}\, L^\top v \leq {\bf 1}_m$, the augmented Lagrangian function \cite[(1.4)]{rockafellar1976augmented} is
\begin{align}
\displaystyle L_{\sigma}(u,v;x,y) \displaystyle :=  & \;\; h(u) + y^\top (u-v) + \displaystyle\frac{\sigma}{2}\|u-v\|_2^2\nonumber \\
&  \quad +
\left\{
\begin{array}{ll}
 x^\top \left(\, \displaystyle\frac{1}{n} L^\top v - {\bf 1}_m\,\right) + \displaystyle\frac{\sigma}{2} \left\|\,\frac{1}{n} L^\top v - {\bf 1}_m \,\right\|_2^2 & \quad \mbox{if $\;\;\displaystyle\frac{1}{n} L^\top v - {\bf 1}_m \geq - \frac{x}{\sigma}$} \\
-\displaystyle\frac{1}{2\sigma}\|x\|_2^2 & \quad \mbox{if $\;\;\displaystyle\frac{1}{n} L^\top v - {\bf 1}_m \leq - \frac{x}{\sigma}$}
\end{array}\right.
\nonumber \\[0.1in]
 & \hspace{-0.87in} = \;\; {\small h(u) + \displaystyle\frac{\sigma}{2} \left\| \, \max\left(\, \frac{1}{n} L^\top v - {\bf 1}_m + \frac{1}{\sigma} x,\, 0 \, \right) \, \right\|_2^2 -\frac{1}{2\sigma}\left( \|x\|_2^2 + \|y\|_2^2\right) 
+ \frac{\sigma}{2}\Big\| \, u-v + \frac{1}{\sigma}\, y \, \Big\|_2^2}, \label{AL-fct}
\end{align}
for a scalar $\sigma >0$ and a primal variable $(x,y)\in \mathbb{R}^m \times \mathbb{R}^n$; here $\max$ is a componentwise notation, and \eqref{AL-fct} is obtained by the completion of squares. Given an initial point $(x^0, y^0)$ and a positive scalar sequence $\{\sigma_k\}_{k\geq 0}$, the iterative framework of the ALM consists of the following steps:
\[
\left\{
\begin{array}{l}
(u^{k+1},v^{k+1}) \approx \displaystyle\operatornamewithlimits{argmin}_{u,v\in \mathbb{R}^n} \, L_{\sigma_k}(u,v;x^k,y^k) \quad \mbox{with a proper stopping criterion} \\[0.2in]
\displaystyle x^{k+1} := \sigma_k \max\left(\,\displaystyle \frac{1}{n} L^\top v^{k+1} - {\bf 1}_m + \frac{1}{\sigma_k} x^k, \,0\,\right) = \max\left(\,\displaystyle \frac{\sigma_k}{n} L^\top v^{k+1} - \sigma_k{\bf 1}_m + x^k, \,0\,\right) \\[0.2in]
\displaystyle y^{k+1} := y^k + \sigma_k (u^{k+1} - v^{k+1}).
\end{array}
\right.
\]
In Section~\ref{subsec:ALM}, we shall specify the stopping criterion for the update of $(u^{k+1},v^{k+1})$ and show that the above algorithm converges globally and it enjoys asymptotically superlinear convergence rate. The major computational cost in the above iterative framework is to find $(u^{k+1}, v^{k+1})$ at given $(x^k, y^k)$, which does not have a closed form expression. In the next subsection, we present a semismooth Newton method to solve this subproblem that exploits the special structure of the generalized Hessian of~\eqref{AL-fct}.

\subsection{Semismooth Newton method for the ALM subproblem}
\label{sec:semiNewton}
To design the semismooth Newton method for minimizing the augmented Lagrangian function $L_\sigma(u,v;x^k,y^k)$ in the ALM subproblem,
we first
eliminate the variable $u$  and transform the subproblem into a tractable continuously differentiable problem as follows.
Adopting the notation of the Moreau-Yosida regularization \eqref{def-MY} and the formulation of the augmented Lagrangian function \eqref{AL-fct}, 
we observe that
\begin{align}
& \displaystyle\operatornamewithlimits{min}_{u,v\in \mathbb{R}^n} \, L_{\sigma_k}(u,v; x^k,y^k) \label{AL0} \\[0.1in]
\Leftrightarrow & \displaystyle\operatornamewithlimits{min}_{u,v\in \mathbb{R}^n} \left\{ \, h(u) + \displaystyle\frac{\sigma_k}{2} \left\| \, \max\left(\, \frac{1}{n} L^\top v - {\bf 1}_m + \frac{1}{\sigma_k} x^k,\, 0 \, \right) \, \right\|_2^2+ \frac{\sigma_k}{2}\left\| \, u-v + \frac{1}{\sigma_k}\, y^k \, \right\|_2^2 \,\right\} \label{AL1} \\[0.2in]
\Leftrightarrow & \displaystyle\operatornamewithlimits{min}_{v\in \mathbb{R}^n} \left\{ \,  \displaystyle\frac{\sigma_k}{2} \left\| \, \max\left(\, \frac{1}{n} L^\top v - {\bf 1}_m  +\frac{1}{\sigma_k} x^k,\, 0 \, \right) \, \right\|_2^2 + \mathcal{M}_h^{\sigma_k}(v - \sigma_k^{-1} y^k) \,\right\}. \label{AL2}
\end{align}
Note that the minimization in \eqref{AL1} with respect to $u$ is achieved at $u^*:={\rm Prox}_h^{\sigma_k}(v - \sigma_k^{-1} y^k)$ for any given $v$, and substituting $u^*$ back into \eqref{AL1} yields \eqref{AL2}.
The above observation indicates that the joint minimization of the augmented Lagrangian function over $(u,v)$ can be achieved by a sequential update of $v$ and $u$ in the following way:
\[
\left\{\begin{array}{l}
\displaystyle v^{k+1} \approx \displaystyle\operatornamewithlimits{argmin}_{v\in \mathbb{R}^n}  \left\{ \, \phi_k(v):= \displaystyle\frac{\sigma_k}{2} \left\| \, \max\left(\, \frac{1}{n} L^\top v - {\bf 1}_m  +\frac{1}{\sigma_k} x^k,\, 0 \, \right) \, \right\|_2^2 +  \mathcal{M}_h^{\sigma_k}(v - \sigma_k^{-1} y^k) \,\right\},\\[0.3in]

\displaystyle u^{k+1} =  {\rm Prox}_h^{\sigma_k}(v^{k+1} - \sigma_k^{-1} y^k).
\end{array}\right.
\]
Therefore, the ALM subproblem \eqref{AL0} is transformed into a tractable continuously differentiable problem $\min\limits_v\,\phi_k(v)$
since $\phi_k$ is convex and continuously differentiable, as both the squared max function $\|\max(\bullet,0)\|^2$ and the Moreau envelope $\mathcal{M}_h^{\sigma_k}$ are continuously differentiable. As we know,  to minimize a convex and continuously differentiable function, it suffices to set its gradient to zero. Therefore, we solve the
problem $\min\limits_v\,\phi_k(v)$ via finding the solution of the following equation:
{\begin{equation}\label{eq:nonsmooth eq}
\nabla \phi_k(v) \, =  \displaystyle\frac{\sigma_k}{n} L \max\left(\, \frac{1}{n} L^\top v - {\bf 1}_m + \frac{1}{\sigma_k} x^k, \, 0 \,\right)   + \sigma_k \left( v - \displaystyle\frac{1}{\sigma_k} y^k - {\rm Prox}_h^{\sigma_k}\left(v - \frac{1}{\sigma_k} y^k\right) \right) = 0,
\end{equation}}
where the gradient of the Moreau envelope $ \mathcal{M}_h^{\sigma_k}$ is obtained via the general formula \eqref{def-grad-MY}.

Following the ALM discussed in the last subsection, it is clear that the cornerstone of the overall algorithm is fast and scalable computation of~\eqref{eq:nonsmooth eq}. Due to the nonsmoothness of the componentwise max operation on the left side of \eqref{eq:nonsmooth eq}, the classical Newton method for solving a smooth nonlinear equation may not be applicable here. Fortunately, the gradient $\nabla \phi_k$ is a so-called semismooth function (defined formally in Section~\ref{sec:prelim}), in fact piecewise smooth, so that one may apply the {\it semismooth Newton} (see e.g.,~\cite{facchinei2007finite}) method to solve~\eqref{eq:nonsmooth eq}. It turns out that the nonsmoothness of this gradient equation is the key reason that our ALM is scalable.

      The semismooth Newton method is a generalization of the classical Newton method for solving semismooth equations \cite{kojima1986extension,kummer1988newton, kummer1992newton,qi1993nonsmooth}. The basic idea of the semismooth Newton method is that for a semismooth function $F:\mathbb{R}^n \to \mathbb{R}^m$, one can still approximate the function value $F(x)$ locally at any given point $\bar{x}\in \mathbb{R}^n$ by a linear mapping $F(\bar{x}) + V(x-\bar{x})$ with residual $o(\|x-\bar{x}\|)$, where instead of taking $V=JF(\bar{x})$ as in the smooth case, we set $V$ to be an arbitrary Clarke generalized Jacobian in the set $\partial F(x)$; see Section~\ref{sec:prelim} for a review of these concepts from variational analysis.

Now, coming back to solving problem \eqref{eq:nonsmooth eq},
the Clarke generalized Jacobian  of the piecewise linear function $ F_{\max}(x) = F_{\max}(x_1,\dots,x_m):=(\max(x_1,0),\dots,\max(x_m,0))$ 
for $x\in \mathbb{R}^m$ is given by
\begin{equation}\label{jac-plus-fct}
\partial F_{\max}(x)
= \left\{
{\rm Diag}(d)\,:\,
d_i =  \left\{\begin{array}{ll}
1     & \mbox{ if } x_i>0;\\[0.02in]
[0,1] & \mbox{ if } x_i=0; \\[0.02in]
0     & \mbox{ if } x_i<0;
\end{array}\right.
i = 1,\ldots,m
\right\}.
\end{equation}
Equipped with this Clarke generalized Jacobian, we can consider the following set-valued mapping as the collection of generalized Hessians of the function $\phi_k$:
\begin{equation}\label{eq:hessian}
{\small \partial^2 \phi_k (v) =
\Bigg\{ \sigma_k \Big[\,\displaystyle\frac{1}{n^2} LSL^\top + \underbrace{I_n - \nabla {\rm Prox}_h^{\sigma_k}\left(v - \frac{y^k}{\sigma_k}  \right)}_{\mbox{denoted $D^k$}}
\,\Big]:  S\in\partial F_{\max}\left(\displaystyle\frac{1}{n} L^\top v - {\bf 1}_m + \frac{x^k}{\sigma_k}\right)
\Bigg\}.}
\end{equation}
Two critical remarks regarding the above set of generalized Hessians are in order.
One, by noticing that
\[
\begin{array}{cl}
\displaystyle \nabla {\rm Prox}_{h}^{\sigma}(y)  = \displaystyle \frac{1}{2} I_n + \frac{1}{2} {\rm Diag}\left(\frac{y_1}{\sqrt{y_1^2 + 4/(\sigma n)}},\dots,\frac{y_n}{\sqrt{y_n^2 + 4/(\sigma n)}} \right),
\end{array}
\]
we get that $D^k$ in \eqref{eq:hessian} is an $n\times n$ positive definite diagonal matrix.
Two,  one may derive from
 \eqref{jac-plus-fct}
  that each $S\in\partial F_{\max}\left(\displaystyle\frac{1}{n} L^\top v - {\bf 1}_m + \frac{x^k}{\sigma_k}\right)$ is an $m\times m$ diagonal matrix with either 0 or 1 in the diagonal entries. Let
  \begin{equation}\label{eq:s}
  s:=|\{i\,:\,S_{ii}=1\}|,
  \end{equation} which represents the number of nonzero entries in $S$.  Notice that if the dual variable $v$ is feasible, then $\frac{1}{n}L^\top v\leq {\bf 1}_m$ and   so the number of positive entries in $\max ({n}^{-1} L^\top v - {\bf 1}_m + {\sigma_k}^{-1}{x^k},0 \, )$ cannot exceed the number of positive entries in the vector $x^k$. Since the primal solution is usually sparse,  one may expect that the matrix $S$ has only a few nonzero entries during the
semismooth Newton iterations, especially near the optimal solution.


The above two facts together indicate that the elements in $\partial^2 \phi_k (v)$ are always positive definite and potentially very sparse --- this is referred to as {\sl second order sparsity}. 
When $\partial^2 \phi_k (v)$ consists of more than one matrix, we always take the sparsest one in our implementation.

The entire algorithm of the semismooth Newton method for solving the semismooth equation $\nabla \phi_k(v) = 0$ in \eqref{eq:nonsmooth eq} is presented below; see Algorithm~\ref{alg:ssnal}. Similar to the Newton method for solving smooth nonlinear equations, the semismooth Newton method with the unit step length only works locally near the optimal solution. In order to make sure that the overall algorithm converges, we adopt the standard line search strategy as the semismooth Newton direction computed from Step 1 below is always a descent direction of the objective function $\phi_k$; for details, see \cite[Section 8.3.3]{facchinei2007finite}. We shall prove the convergence and the superlinear convergence rate of the generated sequence $\{v^t\}_{t\geq 1}$ in the next subsection.

\begin{algorithm}[H]
\caption{A semismooth Newton method for solving $\displaystyle\min_{v} \phi_k(v)$}
\label{alg:ssnal}
Choose $\bar{\eta} \in (0,1)$ and $\tau \in (0,1]$ (parameters for Step~1); $\mu \in (0,1/2)$ and $\beta \in (0,1)$ (parameters for Step~2), and $v^0\in\mathbb{R}^n$. Execute the following steps for $t=0, 1, \ldots$:
\begin{description}
\item[Step 1: (finding the direction).] Choose  $H_t \in \partial^2 \phi_k (v^t)$ and solve the following linear system
\begin{equation}\label{newton-sys}
H_t \, d = -\nabla\phi_k(v^{\,t})
\end{equation} to find the semismooth Newton direction $d^{\,t}$ such that $\| H_t \, d^{\,t}+\nabla\phi_k(v^t)\|_2 \leq \min(\bar{\eta},\|\nabla\phi_k(v^t)\|_2^{1+\tau})$.
\item[Step 2: (line search).] Set $ \alpha_t = \beta^{m_t}$, where $m_t$ is the smallest nonnegative integer $m$ such that
\begin{equation*}
\phi_k(v^t + \beta^m \, d^{\,t}) \leq \phi_k(v^t) + \mu\beta^m \langle \nabla\phi_k(v^t),d^{\,t} \rangle.
\end{equation*}
\item[Step 3.] Set $ v^{t+1} = v^t + \alpha_t \,d^{\,t} $.		
\end{description}
\end{algorithm}


\subsection{Convergence results for the ALM and the semismooth Newton method}\label{subsec:ALM}
In this subsection, we provide convergence guarantees and the rates for both the  ALM algorithm and the semismooth Newton method discussed in the previous two subsections. 
Let $(\bar{u}, \bar{v})$ be an optimal solution of \eqref{dual}, i.e., there exists $(\bar{x}, \bar{y})\in \mathbb{R}^{n+m}$ such that the following Karush-Kuhn-Tucker
(KKT) optimality conditions hold:
\begin{equation}\label{kkt}
\left\{\begin{array}{l}
L\bar{x}=\bar{y},\qquad \bar{x}\geq 0,\qquad \Big(\displaystyle\frac{1}{n} L^\top \bar{v}- {\bf 1}_m\Big)^\top \bar{x} = 0,\\[6pt]
\displaystyle\frac{1}{n} L^\top \bar{v} \leq {\bf 1}_m,\qquad \bar{u}-\bar{v}=0,\\[6pt]
\bar{u}_i>0, \qquad \bar{u}_i\bar{y}_i=1,\,i=1,\dots,n.
\end{array}\right.
\end{equation}

In the seminal work of Rockafellar \cite{rockafellar1976augmented,rockafellar1976monotone}, the global convergence and the asymptotically superlinear convergence rate of the ALM for solving convex problems were derived under the following two stopping criteria:
\[
\left\{\begin{array}{l}
\mbox{(S1)} \;  L_{\sigma_k}(u^{k+1},v^{k+1};x^k,y^k) - \displaystyle\inf_{u,v\in \mathbb{R}^n} L_{\sigma_k}(u,v;x^k,y^k) \leq \varepsilon_k^2/(2\sigma_k), \\[0.2in]
\mbox{(S2)} \; L_{\sigma_k}(u^{k+1},v^{k+1};x^k,y^k) - \displaystyle\inf_{u,v\in \mathbb{R}^n} L_{\sigma_k}(u,v;x^k,y^k) \leq {\eta_k^2}\|(x^{k+1}, y^{k+1}) - (x^k, y^k)\|_2^2/(2\sigma_k),
\end{array}
\right.
\]
where $\{\varepsilon_k\}_{k\geq 0}$ and $\{\eta_k\}_{k\geq 0}$ are two prescribed positive summable sequences satisfying
\begin{equation}\label{eq:condition for stop}
\max\left(\,\sum_{k=0}^\infty \varepsilon_k, \sum_{k=0}^\infty \eta_k \,\right) <+\infty.
\end{equation}
The positiveness of $\varepsilon_k$ and $\eta_k$ allows for inexact computation of the ALM subproblems. In practice, one may choose $\varepsilon_k = \eta_k = \beta^{-k}$ for some $\beta >1$. Under (S1) we will show (in Proposition \ref{prop:convergence} below) that the sequence $\{(x^k, y^k)\}_{k\geq 1}$ is convergent. This further implies that $\lim_{k\to \infty} \|(x^{k+1}, y^{k+1}) - (x^k, y^k)\|_2 = 0$ so that the stopping criterion (S2) is in fact stronger than (S1). This stronger criterion yields a convergence rate for $\{(x^k, y^k)\}_{k\geq 1}$  (see  Proposition \ref{prop:rate} below).

Notice that the Slater condition trivially holds for problem \eqref{primal} by taking $x_j = 1/m$ for all $j = 1, \ldots, m$ so that a  KKT solution $(\bar{u}, \bar{v}, \bar{x}, \bar{y})$ (satisfying the KKT conditions \eqref{kkt}) always exists; see \cite[Proposition 4.3.9]{bertsekas16}. The following proposition regarding the global convergence of the sequence generated by the ALM is a consequence of \cite[Theorem 4]{rockafellar1976augmented}.

\begin{proposition}\label{prop:convergence}
Let $\{\sigma_k\}_{k\geq 0}$ be a nondecreasing positive sequence converging to $\sigma_\infty \leq \infty$. Let $\{(u^k, v^k, x^k, y^k)\}_{k\geq 1}$ be the sequence generated by the ALM with each subproblem satisfying the stopping criterion {\rm (S1)}. Then the primal sequence $\{(x^k, y^k)\}_{k\geq 1}$ converges to a solution $(\bar{x}, \bar{y})$ that solves problem \eqref{primal}.
\end{proposition}

Next we discuss the convergence rate of the ALM. Recall that a  sequence $\{w^k\}_{k\ge 1}$ in $\mathbb{R}^n$ is said to converge to $\bar{w}$ (with $w^k\neq \bar{w}$ for all $k$) {\it superlinearly} if $$\lim_{k\to \infty} \frac{\|w^{k+1} - \bar{w}\|_2}{\|w^k - \bar{w}\|_2} = 0.$$
The superlinear convergence rate of the ALM has been extensively studied in the existing literature since the pioneering work of Powell~\cite{powell1969method}.
For convex nonlinear programming, the convergence rate of $\{(x^k, y^k)\}_{k\geq 1}$ can be derived under  the so-called quadratic growth condition  of problem~\eqref{dual} \cite{rockafellar1976monotone, rockafellar1976augmented,cui2017quadratic}.
Recall that $h(\cdot)$ is the objective function of the dual problem defined in \eqref{dual}.
Let $(\bar{u},\bar{v})$ be the  optimal solution of \eqref{dual}, which must be unique since the dual objective function $h$ is strictly convex and $\bar{u} = \bar{v}$ due to the constraints.
The quadratic growth condition  of problem~\eqref{dual} pertains to the existence of a positive scalar $\kappa$ and a neighborhood ${\cal N}$ of $(\bar{u}, \bar{v})$ such that
for any dual feasible solution $(u,v)\in {\cal N}$  satisfying $\displaystyle\frac{1}{n} L^\top v \leq {\bf 1}_m$ and $u=v$, the following inequality holds:
\begin{equation}\label{eq:quadratic growth}
h(u) \geq h(\bar{u}) + \kappa \left(\, \|u-\bar{u}\|_2^2 + \|v-\bar{v}\|_2^2 \,\right).
\end{equation}
We show in the next result (see Section~\ref{app:rate ALM} for a proof) that problem~\eqref{dual} satisfies this requirement, and therefore the ALM for solving \eqref{dual} has asymptotically superlinear convergence rate.
\begin{proposition}\label{prop:rate}
Let $\{\sigma_k\}_{k\geq 0}$ be a nondecreasing positive sequence converging to $\sigma_\infty \leq \infty$. Let $\{(u^k, v^k, x^k, y^k)\}_{k\geq 1}$ be the sequence generated by the ALM with each subproblem satisfying the stopping criterion {\rm(S2)}, and $(\bar{x}, \bar{y})$ be the optimal solution of \eqref{primal}. Then either the algorithm converges in finite steps, or
\[
\frac{\|(x^{k+1}, y^{k+1}) - (\bar{x}, \bar{y})\|_2}{\|(x^{k}, y^{k}) - (\bar{x}, \bar{y})\|_2} \leq
\left(\frac{\kappa}{\sqrt{\kappa^2 + \sigma_k^2}} + \eta_k \right)(1-\eta_k)^{-1}.\]
\end{proposition}
The above proposition states that when the ALM subproblem is solved approximately under criterion (S2), the sequence $(x^k,y^k)$ converges to an optimal pair $(\bar{x},\bar{y})$ at a linear rate $(\kappa/\sqrt{\kappa^2 + \sigma_k^2} + \eta_k )(1-\eta_k)^{-1}$. Since $\eta_k\to 0$ due to \eqref{eq:condition for stop} and $\sigma_k >0$, we know that the rate $\left( \frac{\kappa}{\sqrt{\kappa^2 + \sigma_k^2} } + \eta_k \right)(1-\eta_k)^{-1}$ is smaller than $1$ when $k$ is sufficiently large. In addition, as $\sigma_k \to \sigma_\infty$ as $k\to \infty$, the rate eventually
converges to $\kappa/\sqrt{\kappa^2 + \sigma_{\infty}^2}$. It is, roughly speaking, inversely proportional to $\sigma_{\infty}$ if $\sigma_{\infty}$ is large. If $\sigma_{\infty} = \infty$, the convergence is superlinear. This is the reason that we say the ALM has asymptotically superlinear convergence rate.


Finally, we provide the global convergence and the local convergence rate of the semismooth Newton method (Algorithm~\ref{alg:ssnal}) discussed in Section~\ref{sec:semiNewton}. These are standard results and one may consult the monograph~\cite[Chapters 7 and 8]{facchinei2007finite} for the detailed proofs.
\begin{proposition}\label{prop:convergence ssn}
Let $\{v^t\}_{t \ge 1}$ be the sequence generated by Algorithm~\ref{alg:ssnal}. Then $\{v^t\}_{t \ge 1}$ converges globally to the unique solution $v^*$ of \eqref{eq:nonsmooth eq}. Furthermore, the convergence rate is superlinear, i.e., $\displaystyle\lim_{t\to \infty} \frac{\|v^{t+1} - v^*\|_2}{\|v^t - v^*\|_2} = 0$.
\end{proposition}

\subsection{Comparison of the computational cost with other second order methods}
\label{subsec: computational cost}

In this subsection, we compare the computational cost per iteration for three second order methods for solving~\eqref{primal-0}:  our semismooth Newton based ALM, the interior point method (implemented in the \texttt{REBayes} package \cite{koenker2017rebayes}) and the SQP method (implemented in \texttt{mix-SQP}~\cite{kim2020fast}). One can see clearly from the comparison in Table~\ref{table:complexity} how the second order sparsity helps to reduce the computational burden in our approach.

\vskip 0.1in
\noindent
\underline{\bf Semismooth Newton based ALM}.
The most expensive step in our ALM is to find the semismooth Newton direction from the linear system \eqref{newton-sys}. It follows from the expression of $\partial^2 \phi_k(v)$ in \eqref{eq:hessian} that the linear equation in \eqref{newton-sys} takes the following abstract form:
\begin{equation}\label{eq:semiNewton abstract}
\left(\, D + LSL^\top \,\right)d = {\rm rhs},
\end{equation}
where $D$ is an $n\times n$ positive definite diagonal matrix, $S$ is an $m\times m$ diagonal matrix with  diagonal entries being either $0$ or $1$, and rhs is a given vector in $\mathbb{R}^n$.
Denote $J:=\{i\,:\,S_{ii} = 1\}$ and $s:=|J|$, and write $L_J\in\mathbb{R}^{n\times s}$ as the sub-matrix of $L$ with columns in $J$. Based on the special diagonal structure of $S$, one can observe that
\[
D + LSL^\top = D + L_J L_J^\top.
\]
Therefore,  the cost of evaluating the generalized Hessian matrix once via $D + L_J L_J^\top$ is $O(n^2 s)$. When $s < n$, one can also solve~\eqref{eq:semiNewton abstract}
via the following Sherman-Morrison-Woodbury formula:
\[
(D + L_J L_J^\top)^{-1} = D^{-1} - D^{-1} L_J(I_s + L_J^\top D^{-1} L_J)^{-1} L_J^\top D^{-1}.
\]
Therefore, it suffices to solve a reduced linear system with the coefficient matrix being $I_s + L_J^\top D^{-1} L_J \in\mathbb{R}^{s\times s}$. The cost of computing $I_s + L_J^\top D^{-1} L_J$  is $O(n s^2)$, which is smaller than the direct evaluation of the Hessian matrix when $s<n$. Notice that for both cases, the computational cost for solving the linear equation \eqref{eq:semiNewton abstract} is independent of $m$.

Each gradient evaluation $\nabla \phi_k(\cdot)$ needs $O(nm)$ operations due to the multiplications of $L$ and $L^\top$ with  vectors; see~\eqref{eq:nonsmooth eq}. Since the number of gradient evaluations is the total number of semismooth Newton iterations for all ALM subproblems, one may expect that such evaluations do not need to be done many times.

In fact, we can also incorporate a low rank approximation of $L$, as in the \texttt{mix-SQP} solver (see~\eqref{lowrank-approx} below), if the rank of $L$ is indeed small to further reduce the computational cost of our gradient evaluations. With such techniques, the cost of each gradient evaluation is $O((n+m)k + \min(n,m)^2)$, where $k$ is the rank of the matrix $L$. 

\vskip 0.1in
\noindent
\underline{\bf Interior point method}.
We have found from the source code of~\texttt{REBayes} \cite{koenker2017rebayes} that it calls the exponential cone optimization\footnote{\url{https://docs.mosek.com/modeling-cookbook/expo.html}} in Mosek to solve \eqref{dual}. In fact, problem \eqref{dual} can be formulated equivalently as the following exponential cone optimization problem:
\begin{equation}\label{op:ipm}
\begin{array}{cl}
\displaystyle\operatornamewithlimits{minimize}_{t,u \in \mathbb{R}^n} & \displaystyle  -\frac{1}{n}\sum_{i=1}^n t_i \\[0.1in]
\mbox{subject to} & \displaystyle \frac{1}{n} L^\top u \leq {\bf 1}_m,\qquad (t_i,u_i,1)\in K_{\exp},\,i=1,\ldots,n,
\end{array}
\end{equation}
where $K_{\exp}:= {\rm closure}\{ (x,y,z)\in\mathbb{R}^3\,|\, z > 0, y \geq z \exp(x/z) \}$ is the nonsymmetric exponential cone studied by Charez \cite{chares2009cones}; here ${\rm closure}(\cdot)$ denotes the closure of a convex set. It is well known that the interior point method for \eqref{op:ipm} generally relies on a logarithmically homogeneous self-concordant barrier (LHSCB) (and its conjugate barrier)  of the exponential cone (and its dual cone).
Notice that the constraints of problem \eqref{op:ipm} involves $n$ numbers of exponential cones.
 From \cite[Proposition~1.2.4]{gao2017homogeneous}, we can see that the cost of computing the gradient and Hessian of LHSCBs for all these exponential cones is $O(n)$.
The most expensive step in the interior point method is to find a search direction of a linear system (e.g., \cite[(2.2)]{gao2017homogeneous}, \cite[(4)]{dahl2021primal}) for the central path. In particular, the Schur complement equation (e.g., \cite[(2.13)]{gao2017homogeneous}) of the linear system involves computing $L^\top g$ and $L^\top H L$, where $g\in\mathbb{R}^n$ and $H\in\mathbb{R}^{n\times n}$  are associated with the gradient and Hessian of the LHSCB. The cost of computing  $L^\top g$ and $L^\top H L$ is $O(nm)$ and $O(n^2m)$ respectively.

\vskip 0.1in
\noindent
\underline{\bf SQP}.
For the SQP method implemented in the \texttt{mix-SQP} solver \cite{kim2020fast}, the gradient $g$ and Hessian $H$ for each SQP subproblem are given by
\[
g = -\displaystyle\frac{1}{n} L^\top d + {\bf 1}_m \quad \mbox{and} \quad
H = \displaystyle\frac{1}{n}L^\top {\rm diag}(d)^2 L,
\]
where $d = (1/(Lx)_1, \cdots, 1/(Lx)_n)^\top \in \mathbb{R}^n$ for some given $x\in \mathbb{R}^m$. Recall that $L$ is an $n\times m$ matrix.
The cost of naively computing the gradient and Hessian is $O(nm)$ and $O(nm^2)$ respectively.
In the solver \texttt{mix-SQP}, when the matrix $L$ is numerically rank deficient, say $\mbox{rank}\approx k$, then the matrix $L$ can be approximated by the following truncated QR decomposition (if $m\leq n$):
\begin{equation}\label{lowrank-approx}
L \approx QRP^\top,\quad \mbox{with $Q\in\mathbb{R}^{n\times k},\,R\in\mathbb{R}^{k\times m},\,P\in\mathbb{R}^{m\times m}$}.
\end{equation}
The cost of computing the gradient and Hessian in \texttt{mix-SQP} then reduces to $O((n+m)k + \min(n,m)^2)$ and $O((n+m)k^2)$ respectively.

\begin{table}[h]
\centering
\begin{tabular}{r|c|cc|cc}
    \toprule
      &  IPM &  \multicolumn{2}{c|}{ALM} &  \multicolumn{2}{c}{\texttt{mix-SQP}}\\
     & & {\sl full $L$} & {\sl rank $k$ approx. of $L$} & {\sl full $L$} & {\sl rank $k$ approx.~of $L$} \\
    \midrule
    gradient  & $O(nm)$ & $O(nm)$ & $O((n+m)k + \min(n,m)^2)$ & $O(nm)$ & $O((n+m)k + \min(n,m)^2)$\\[6pt]
    Hessian  & $O(n^2m)$ & \multicolumn{2}{c|}{$O(ns \min(n,s))$} & $O(nm^2)$ & $O((n+m)k^2)$  \\
    \bottomrule
\end{tabular}
\caption{\small Computational cost of ALM, interior point method (IPM), and \texttt{mix-SQP} for evaluating the gradient and Hessian. The columns `{\sl full $L$}' represent the cost of plain evaluation of the gradient and Hessian, while the columns `{\sl rank $k$ approx.~of $L$}' list the reduced cost with the low rank approximation of $L$ as in~\eqref{lowrank-approx}.}
\label{table:complexity}
\end{table}


\section{The partial EM algorithm for solving~\eqref{adaptive supports 0}}
\label{sec:adaptive supports}
So far we have worked with the matrix $L$ (in~\eqref{primal-0}) being fixed. In computing the NPMLE of the Gaussian location mixture problem in~\eqref{eq:discretized prob} this corresponds to fixing the support points $\mu_1,\ldots, \mu_m$ in advance. Although this works well when $d=1$ or $2$ where one can cover the support of $\widehat{G}_n$ (in~\eqref{NPMLE1}) using finely chosen grid points (see~\cite{soloff2021multivariate} where rigorous results are provided on the choice of these grid points), 
it may no longer work when $d$ is large (e.g., for the APOGEE data set~\cite{majewski2017apache} in Section~\ref{sec:real}, where the abundances are in $\mathbb{R}^{19}$). In this section, we focus on the nonconvex problem~\eqref{adaptive supports 0} where $\{\mu_j\}_{j=1}^m$ are also unknown variables. For ease of reference, we repeat model \eqref{adaptive supports 0} below:
\begin{equation}\label{adaptive supports}
\begin{array}{cl}
\displaystyle\operatornamewithlimits{maximize}_{x\in \mathbb{R}^m, \, \mu = \{\mu_j \in \mathbb{R}^d\}_{j=1}^m} & \ell_n(x, \mu) := \displaystyle \frac{1}{n} \sum_{i=1}^{n}\log\,\left(\sum_{j=1}^{m} x_j  \phi_{\Sigma_i}(Y_i-\mu_j) \right)\\[0.25in]
\mbox{subject to} & {\bf 1}_m^\top \, x = 1,\quad  x \geq 0.
\end{array}
\end{equation}
In the classical EM algorithm~\cite{dempster1977maximum,redner1984mixture} for solving  the above problem, the data is viewed as generated from the latent model
\[
Y_i \sim \sum_{j=1}^m x_{j}\,  {\cal N}(\mu_j,\Sigma_i),
\]
and $\{\Delta_{i}\}_{i=1}^n$ are latent random variables taking values in $\{1,\ldots, m\}$, with $\Delta_i = j$ meaning $Y_i$ comes from the distribution ${\cal N}(\mu_j,\Sigma_i)$. It follows from Jensen's inequality that for any $(\bar{x}, \bar{\mu})$ and $(x,\mu)$,
\begin{equation}\label{eq:Jensen inequality}
\begin{array}{rl}
\ell_n(x,\mu)  = & \displaystyle \frac{1}{n} \sum_{i=1}^{n} \log\,\left( \sum_{j=1}^{m}   \mathbb{P}_{\bar{x}, \bar{\mu}}(\Delta_i=j \mid Y_i) \, \frac{ x_j \phi_{\Sigma_i}(Y_i-\mu_j)}{\mathbb{P}_{\bar{x}, \bar{\mu}}(\Delta_i=j \mid Y_i)} \,\right)\\[0.15in]
\;\; \geq   & \displaystyle \frac{1}{n} \sum_{i=1}^{n}\sum_{{j\in \{1, \ldots, m:\, x_j\neq 0\}}}   \mathbb{P}_{\bar{x}, \bar{\mu}}(\Delta_i=j \mid Y_i) \log\,\left(\,\frac{x_j \phi_{\Sigma_i}(Y_i-\mu_j)}{\mathbb{P}_{\bar{x}, \bar{\mu}}(\Delta_i=j \mid Y_i) }\,\right)\\[0.15in]
 =  &  \underbrace{\displaystyle \frac{1}{n} \sum_{i=1}^{n}\sum_{{j\in \{1, \ldots, m:\, x_j\neq 0\}}}   \mathbb{P}_{\bar{x}, \bar{\mu}}(\Delta_i=j \mid Y_i) \log\,\left(\,\frac{x_j}{\mathbb{P}_{\bar{x}, \bar{\mu}}(\Delta_i=j \mid Y_i) }\,\right)}_{\mbox{denoted as $\widehat{\ell}_{n;x}(x; \bar{x}, \bar\mu)$}} \\
 & +  \qquad \underbrace{\displaystyle \frac{1}{n} \sum_{i=1}^{n}\sum_{{j\in \{1, \ldots, m:\, x_j\neq 0\}}}   \mathbb{P}_{\bar{x}, \bar{\mu}}(\Delta_i=j \mid Y_i) \log\,\left(\,\frac{\phi_{\Sigma_i}(Y_i-\mu_j)}{\mathbb{P}_{\bar{x}, \bar{\mu}}(\Delta_i=j \mid Y_i) }\,\right)}_{\mbox{denoted as $\widehat{\ell}_{n;\mu}(\mu,{x}; \bar{x}, \bar\mu)$}}.
 \end{array}
\end{equation}
Note that the inequality in the second line of the above display becomes an equality if $(x,\mu) = (\bar{x}, \bar{\mu})$.
Given any initial point $(x^0, \mu^0)$, the EM algorithm can be thought of as a minorization-maximization method that iteratively maximizes the minorizer $\left[\,\widehat{\ell}_{n;x}(x; x^k, \mu^k) + \widehat{\ell}_{n;\mu}(\mu,x; x^k, \mu^k)\,\right]$ subject to the constraints in \eqref{adaptive supports}; i.e.,
\begin{equation}\label{EM}
\left\{\begin{array}{ll}
\mu^{k+1} = \displaystyle\operatornamewithlimits{argmax}_{\mu= \{\mu_j\}_{j=1}^m} \widehat{\ell}_{n; \mu}(\mu,x^{k}; x^k, \mu^k) = \left(\, \left(\displaystyle\sum_{i=1}^n \widehat{\gamma}_{ij}^{k+1} \Sigma_i^{-1}\right)^{-1} \left(\displaystyle \sum_{i=1}^n \widehat{\gamma}_{ij}^{k+1} \Sigma_i^{-1} Y_i\right)\,\right)_{j=1}^m,\\[0.2in]
x^{k+1}  = \displaystyle\operatornamewithlimits{argmax}_{{\bf 1}_m^\top x=1, \, x\geq 0} \widehat{\ell}_{n; x}(x; x^k, \mu^k)= \left(\displaystyle \displaystyle\frac{1}{n}\sum_{i=1}^n \widehat{\gamma}^{k+1}_{ij}\right)_{j=1}^m,
\end{array}\right.
\end{equation}
where $$\widehat{\gamma}_{ij}^{k+1}:=  \mathbb{P}_{x^k, \mu^k}(\Delta_i=j \mid Y_i) = \frac{x_j^k \, \phi_{\Sigma_i}(Y_i-\mu_j^k)}{\sum_{j^\prime=1}^m x_{j^\prime}^k \phi_{\Sigma_i}(Y_i-\mu_{j^\prime}^k)}, \qquad \mbox{ for } \; i= 1, \ldots, n, \;\; j=1, \ldots, m.$$ However, Jensen's inequality in \eqref{eq:Jensen inequality} may be loose and so the overall algorithm may not be efficient. To improve the performance of the EM, our proposal is to
 take advantage of the ALM solver discussed in Section \ref{sec:ALM0} to directly maximize $x$ in \eqref{adaptive supports} at the latest $\mu$, instead of maximizing the approximate function $\widehat{\ell}_{n;x}$ to update $x$. Specifically,
the iterative framework we consider is
\begin{equation}\label{eq:PEM}
\left\{\begin{array}{ll}
\left\{\mu_j^{k+1}\right\}_{j=1}^m = \left(\,\left\{
\begin{array}{ll}
\left(\,\displaystyle\sum_{i=1}^n \widehat{\gamma}_{ij}^{k+1} \Sigma_i^{-1}\right)^{-1} \left(\displaystyle \sum_{i=1}^n \widehat{\gamma}_{ij}^{k+1} \Sigma_i^{-1} Y_i\right) & \mbox{if $x_j^{k+1}  \neq 0$} , \\[0.2in]
\mu_j^k & \mbox{if $x_j^{k+1}  = 0$}
\end{array}\right. \right)_{j=1}^m \\[0.2in]
\qquad \quad \qquad  \in  \displaystyle\operatornamewithlimits{argmax}_{\mu= \{\mu_j\}_{j=1}^m} \widehat{\ell}_{n;\mu}(\mu, x^{k}; x^{k}, \mu^k), \\[0.2in]

x^{k+1}  \in \displaystyle\operatornamewithlimits{argmax}_{{\bf 1}_m^\top x=1, \, x\geq 0} \ell_n(x,\mu^{k+1}),
\end{array}\right.
\end{equation}
where $\widehat{\gamma}_{ij}^{k+1}$ is the same as in \eqref{EM}. For those components $j$'s such that $x_j^{k+1} = 0$, the value of $\widehat{\ell}_{n;\mu}(\mu, x^{k+1}; x^{k+1}, \mu^k)$ would not depend on the vector $\mu_j$ so that $\mu_j^{k+1}$ could be any vector in $\mathbb{R}^d$; we simply take $\mu_j^{k+1} = \mu_j^k$ for such $j$'s.

Compared to the iterative framework of the EM algorithm in \eqref{EM}, it is clear that our proposed algorithm is a {\sl partial expectation maximization} (PEM) method --- only the support points $\{\mu_j\}_{j=1}^m$ are estimated by the EM procedure; the vector of mixture proportions $x$ is computed to maximize the original log-likelihood in~\eqref{adaptive supports} based on the latest estimates of the support points.
This apparent simple tweak to \eqref{EM}  dramatically changes the behavior of the resulting algorithm. The resulting solution in $x$ is very sparse (as expected from the results in \cite{soloff2021multivariate}); see Section~\ref{subsec: numerical PEM} where we illustrate this phenomenon through various simulated and real data analyses.   Compare this with the usual EM algorithm \eqref{EM} which cannot produce sparse $x$ solutions\footnote{If the initial point $x^0$ has no zero entry, then $x^k_j \neq 0$ for all $j=1, \ldots, m$ and all $k\geq 1$, as can be easily seen for the explicit update of \eqref{EM}.}.  Also, \eqref{eq:PEM}  converges in very few steps (e.g.,$10\sim15$) compared to the EM which can take more than $1000$ iterations  to come close to the optima. Further, it is easy to derive that
\begin{equation*}\label{PEM:decrease}
\begin{array}{rl}
\ell_n(x^{k+1}, \mu^{k+1})
\geq & \widehat{\ell}_{n;x}(x^{k+1}; x^{k+1}, \mu^k)  + \widehat{\ell}_{n;\mu}(\mu^{k+1}, x^{k+1}; x^{k+1}, \mu^k) \\[0.1in]
\geq &  \widehat{\ell}_{n;x}(x^{k+1}; x^{k+1}, \mu^k)  + \widehat{\ell}_{n;\mu}(\mu^{k}, x^{k+1}; x^{k+1}, \mu^k)
= \ell_n(x^{k+1}, \mu^k) \geq  \ell_n(x^{k}, \mu^{k}),
\end{array}
\end{equation*}
where the first inequality is due to \eqref{eq:Jensen inequality},  the second and last inequalities are due to \eqref{eq:PEM}.
Hence, the PEM inherits the nice property of the classical EM algorithm --- the log-likelihood objective never decreases along iterations.
 In fact, we can obtain the following result on the limit point of the iterative sequence; see  Appendix~\ref{app:PEM} for its proof.

\begin{proposition}\label{prop:PEM}
The sequence $\{(x^k, \mu^k)\}_{k\geq 1}$ generated by the PEM algorithm \eqref{eq:PEM} is bounded and any of its accumulation point, say $(x^*, \mu^*)$, is a stationary point of the constrained  problem \eqref{adaptive supports} satisfying
\begin{equation}\label{eq: stationary of PEM}
\left\{\begin{array}{ll}
\nabla_x \, \ell_n(x^*, \mu^*)^\top (x-x^*) \leq 0 \quad  \mbox{for all $x$ feasible to  \eqref{adaptive supports}} \\[0.1in]
\nabla_\mu \, \ell_n(x^*, \mu^*) = 0.
\end{array}\right.
\end{equation}
\end{proposition}
It follows from the theory of nonlinear programming (see, e.g., \cite[section~1.3.1]{facchinei2007finite}) that the conditions in \eqref{eq: stationary of PEM} are necessary for $(x^*, \mu^*)$ to be a local maximizer of problem~\eqref{adaptive supports}.
Unfortunately, since  \eqref{adaptive supports} is nonconvex, condition \eqref{eq: stationary of PEM} cannot guarantee the global optimality of $(x^*, \mu^*)$. Nevertheless, when the ALM, the EM and the PEM start from the same initial point, the PEM leads to the maximum value of the log-likelihood among the three methods, signifying that its solutions are closest to the original NPMLE problem in \eqref{NPMLE1}.

\section{Denoising via optimal transport}\label{sec:Denoising}
In this section we consider the Gaussian location mixture model~\eqref{denoising data} and present new denoising {\it estimands} defined through a {\it matching} idea (via the theory of optimal transport~\cite{Villani2003, Villani2009}) along with their sample estimates. To motivate our proposal, let us first describe the rationale  behind~\eqref{EB}. The problem of denoising the observed $Y_i$'s can be formally described using the following Bayesian framework. It is known that if the goal is to minimize the expected squared error {\it Bayes risk}
\begin{equation}\label{eq:Bayes-risk}
\E \left[\|\mathfrak{d}(Y_i) - \theta_i \|^2_2 \right] \equiv \int \int \| \mathfrak{d}(y) - \theta \|^2_2 \, \phi_{\Sigma_i}(y - \theta)\, dG^*(\theta) \, dy
\end{equation}
over all measurable functions $\mathfrak{d}: \R^d \to \R^d$, where $\theta_i \sim G^*$ and $Y_i \mid \theta_i \sim {\cal N}(\theta_i, \Sigma_i)$, then the best estimator for $\theta_i$ is the {\it oracle posterior mean}
\begin{equation}\label{eq:Oracle-Post-Mean}
\mathfrak{d}^*(Y_i) := \E[\theta_i \mid Y_i].
\end{equation}
In empirical Bayes, given an estimate $\widehat G_n$ of the unknown prior $G^*$, one imitates the optimal Bayesian analysis and estimates of the oracle posterior means by the empirical Bayes estimates~\eqref{EB}; see e.g.,~\cite{jiang2009general, efron19}, etc. Although this is a natural strategy which has been studied extensively in  the literature, there are a few drawbacks of this plug-in approach:
\begin{figure}[!h]
  \centering
  \includegraphics[width=0.9\textwidth]{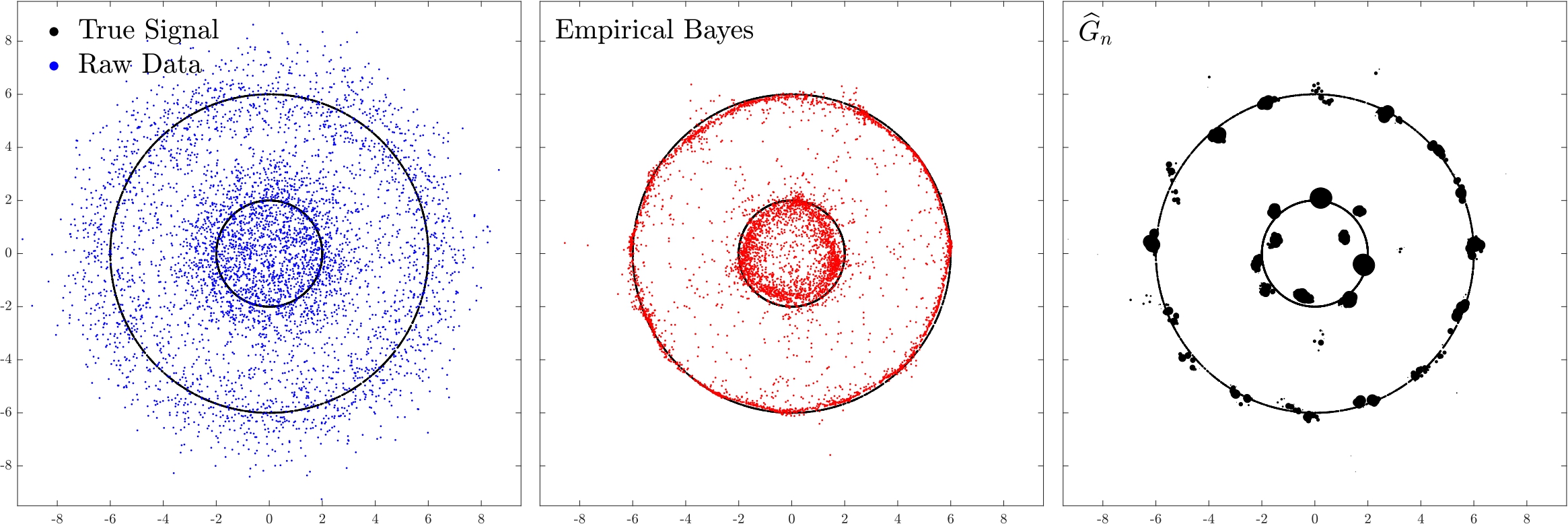}\\
  \caption{\small Plots of the raw data (in blue) with $n= 5,000$ in $d=2$, the corresponding empirical Bayes estimates (in red), the true $G^*$ (in black), and $\widehat{G}_n$ (in black dots) obtained from our ALM. Here half of the true signals $\theta_i \in \R^d$ are drawn uniformly at random from each of the two concentric circles of radii 2 and 6 respectively (centered at $(0,0) \in \R^2$), and $Y_i \mid \theta_i \sim {\cal N}(\theta_i, I_2)$ for $i=1,\dots,n$. {Observe that some of the empirical Bayes estimates $\widehat{\theta}_i$'s are far from the support of $G^*$ (and $\widehat{G}_n$).}}\label{fig-onecircle}
\end{figure}

\vskip 0.1in
\noindent
(1) The {\it oracle posterior mean} $\mathfrak{d}^*(Y_i)$ in~\eqref{eq:Oracle-Post-Mean} and the empirical Bayes estimates in~\eqref{EB} are not necessarily guaranteed to lie `close' to the support of $G^*$ (say $\s \subset \R^d$). In fact, if the goal is to estimate $\theta_i \sim G^*$, it is reasonable to restrict $\mathfrak{d}(\cdot)$ in~\eqref{eq:Bayes-risk} to all estimators such that $\mathfrak{d}(Y_i)$ is distributed (approximately) as $G^*$.
	
	To illustrate this phenomenon suppose that $G^*$ has {\it structure}\footnote{For example, suppose that $G^*$ is discrete with a few atoms, which corresponds to the {clustering problem} (see e.g., Figures~\ref{table-5c-d6} and \ref{table-5c-d9}).} (e.g., the $\theta_i$'s are supported on a lower dimensional manifold $\s$ in $\R^d$); see e.g., Figure~\ref{fig-onecircle}. The empirical Bayes estimator $\widehat{\theta}_i$, as defined in~\eqref{EB}, may {\it not} necessarily, in general, lie `close' to the set $\s$ (see the red points in the middle panel of Figure~\ref{fig-onecircle}). Thus, if the emphasis is on estimating $\theta_i$'s focussing on recovering the support $\s$, the estimates $\widehat{\theta}_i$'s are not necessarily ideal. 	
\vskip 0.1in
\noindent (2)  It is worth noting that although we call $\widehat{\theta}_i$'s as natural estimates of $\theta_i$, they are not {\it consistent} estimates\footnote{This is not at all surprising, as intuitively we have $n$ unknowns $\theta_i$, and only have access to $n$ data points $Y_i$'s to estimate them, and without further information/restriction, it is difficult to estimate the $n$ parameters accurately.}, in the sense that generally, $\widehat{\theta}_i$ does not converge (e.g., in probability) to $\theta_i$.  \newline

To motivate our alternative approach, first suppose that $\Sigma_i \equiv \Sigma$ for all $i=1,\ldots, n$ (such that the $Y_i$'s are marginally i.i.d.) and that the $\theta_i$'s are known up to a permutation, i.e., the empirical distribution
\begin{equation}\label{eq:G_n}
G_n := \frac{1}{n} \sum_{i=1}^n \delta_{\theta_i}
\end{equation}
of the $\theta_i$'s is known. Then, it seems natural to associate $Y_i$ to a $\theta_j$ by solving the {\it matching} (optimization) problem:
$\min_{\sigma: [n] \to [n]} \frac{1}{n} \sum_{i=1}^n \|Y_i - \theta_{\sigma(i)} \|^2,$
where $\sigma = (\sigma(1), \ldots, \sigma(n))$ is a permutation of $[n] := \{1,\ldots, n\}$. In other words, we match the data points $Y_i$'s to the $\theta_j$'s such that the average cost of the matching is smallest. Letting $\nu_n := \frac{1}{n} \sum_{i=1}^n \delta_{Y_i}$ denote the empirical distribution of the observed $Y_i$'s, this matching problem can be formulated as an {\it optimal transport} (OT) problem (see Appendix~\ref{app:optimaltransport} for a brief introduction to this topic):
\begin{equation}\label{eq:Oracle-Assign-2}
\min_{T: T \#\nu_n = G_n } \frac{1}{n} \sum_{i=1}^n \|Y_i - T(Y_i) \|^2_2  
\end{equation}
where the above minimization is over all maps $T$ such that $T \#\nu_n = G_n$ which means that $T$ transports the distribution of $\nu_n$ to $G_n$, i.e., $T: \{Y_1,\ldots, Y_n\} \to \{\theta_1, \ldots, \theta_n\}$  is a bijection. Note that~\eqref{eq:Oracle-Assign-2} can be viewed as an {\it assignment problem} for which algorithms with worst case complexity $O(n^3)$ are available in the literature (see e.g.,~\cite{munkres1957, bertsekas1988}). It is known from the theory of OT that the minimum value of the above objective matches the Wasserstein (squared) distance between $\nu_n$ and $G_n$.

The above approach can be easily cast in the population setting by considering the minimization problem (cf.~\eqref{eq:Bayes-risk})
\begin{equation}\label{eq:Opt-Trans}
\min_{\mathfrak{T} \# \nu = G^*}\E \left[\|Y-\mathfrak{T}(Y) \|^2_2 \right] \equiv \min_{\pi \in \Pi(\nu, G^*)} \int \|y - \theta \|^2 \, d \pi(y, \theta) =: W_2^2(\nu,G^*)
\end{equation}
over all measurable functions $\mathfrak{T}: \R^d \to \R^d$ such that $\mathfrak{T} \#\nu = G^*$, which means that $\mathfrak{T}$ transports $\nu$ --- the marginal distribution of $Y$ --- to $G^*$, i.e., $\mathfrak{T}(Y) \sim G^*$ where $Y \sim \nu$. The right side of~\eqref{eq:Opt-Trans} involves minimization over $\Pi(\nu, G^*)$ --- the class of all joint distributions $\pi$ with marginals $\nu$ and $G^*$, and gives the equivalence\footnote{In this case Monge's problem is equivalent to Kantorovich's relaxation as the reference distribution $\nu$ is absolutely continuous.} of Monge's problem to Kantorovich's relaxation (see e.g.,~\cite{Villani2003, Villani2009}).

Suppose that $\mathfrak{T}^*$ is the optimal solution to~\eqref{eq:Opt-Trans}; i.e., $\mathfrak{T}^*$ is the OT map such that $\mathfrak{T}^*\#\nu = G^*$. It is known from the theory of OT that such a $\mathfrak{T}^*$ exists, is unique a.e., and can be expressed as the gradient of a convex function (say $\mathfrak{T}^* =  \nabla \psi$ where $\psi:\R^d \to \R$ is a convex function); see e.g.,~\cite{Villani2003, Villani2009}. Then,
\begin{equation}\label{eq:OT-Estimand}
\tilde{\theta}_{i} := \mathfrak{T}^*(Y_i), \qquad \mbox{for }\;\; i =1,\ldots, n,
\end{equation}
could be considered as a natural {\it denoising} target for $Y_i$. Observe that, by~\eqref{eq:Opt-Trans}, we have $\tilde{\theta}_{1}, \ldots, \tilde{\theta}_{n}$ are i.i.d.~$G^*$ (compare this with the fact that ${\theta}_{1}, \ldots, {\theta}_{n}$ are also i.i.d.~$G^*$). Further, the new estimand $\tilde{\theta}_{i}$ is related to $Y_i$ directly through the map $\mathfrak{T}^*$ via~\eqref{eq:OT-Estimand}. One can think of $\tilde{\theta}_{i} \sim G^*$ as the `closest' (in the sense of distributions) to $Y_i \sim \nu$. In the following we will consider estimation of $\tilde{\theta}_{1}, \ldots, \tilde{\theta}_{n}$.

\subsection{NPMLE based estimation of the OT based estimands}\label{sec:Denoising-NPMLE}
In order to estimate our new denoising targets $\tilde{\theta}_{1}, \ldots, \tilde{\theta}_{n}$ we first need to estimate $\mathfrak{T}^*$ defined via~\eqref{eq:Opt-Trans}. A natural plug-in approach here would be to replace $\nu$ and $G^*$ in~\eqref{eq:Opt-Trans} with $\nu_n$ (the empirical distribution of $Y_1,\ldots, Y_n$) and $\widehat{G}_n$ (obtained by solving~\eqref{NPMLE1}) respectively.  Thus, we solve the linear program
\begin{equation}\label{eq:Assign-Est}
\min_{\pi \in \Pi(\nu_n, \widehat{G}_n)}  \int \|y - \theta \|^2 \, d \pi(y, \theta) \equiv W_2^2(\nu_n, \widehat{G}_n).
\end{equation}
As $\widehat{G}_n$, obtained via~\eqref{NPMLE1} has finite support (see~\cite{soloff2021multivariate}), and $\nu_n$ is a discrete distribution, the optimal coupling in~\eqref{eq:Assign-Est} can be represented by a matrix $\hat \pi = ((\hat \pi_{ij}))_{n \times \hat{k}}$ which has marginals $\nu_n$ and $\widehat{G}_n$; here  we suppose that $\widehat{G}_n = \sum_{j=1}^{\hat k} \hat \alpha_j \delta_{\hat{a}_j}$, where $\hat a_1,\ldots, \hat a_{\hat k} \in \R^d$ and $\hat \alpha_j$'s are nonnegative weights summing up to 1. To obtain a transport map from this joint coupling $\hat \pi$ we can use the idea of {\it barycentric projection}  (see~\cite{deb2021rates}) and define
\begin{equation}\label{eq:Cond-Mean}
\ttg(Y_i) := \E_{\hat \pi}[\theta \mid Y_i] = n \sum_{j = 1}^{\hat k} \hat \pi_{ij} \, \hat{a}_j,
\end{equation}
as an estimator of $\tilde{\theta}_i \equiv \mathfrak{T}^*(Y_i)$. As $\widehat{G}_n$ is a discrete distribution with much fewer atoms than $n$ (in practice), most of the $Y_i$'s will be essentially transported to one element in $\widehat{G}_n$; see the right panel of Figure~\ref{fig-onecircle}. Thus the estimates $\ttg(Y_i) $ will essentially lie in the support of $\widehat{G}_n$; this rectifies the drawbacks of the empirical Bayes approach outlined at the beginning of this section.

In the following result (proved in Appendix~\ref{pf:Denoising}) we show that our proposed estimand $\tilde{\theta}_{i} \equiv \mathfrak{T}^*(Y_i)$, in~\eqref{eq:OT-Estimand}, can be consistently estimated by the estimator $\ttg(Y_i)$ defined in~\eqref{eq:Cond-Mean}. In fact, the above result provides a finite sample bound on the rate of convergence of $\ttg$ in average (squared) Euclidean norm.
\begin{theorem}\label{thm:Rate-OT-Map}
Suppose that we have data from~\eqref{denoising data} where $\Sigma_i \equiv \Sigma$ for all $i=1,\ldots, n$, and $\Sigma$ is a $d \times d$ positive definite matrix with minimum eigenvalue $\sigma >0$. Suppose that the denoising estimands $\tilde{\theta}_i$'s are defined via~\eqref{eq:OT-Estimand} where $\mathfrak{T}^* =  \nabla \psi$ with $\psi:\R^d \to \R$ being a convex function. We assume that $\psi$ is $\lambda$-strongly convex\footnote{Strong convexity here refers to $\psi(z) \ge \psi(x) + \nabla \psi(x)^\top (z-x) + \frac{\lambda}{2} \|x-z\|^2, \quad \mbox{for all } \;\; x,z \in \R^d.$} and $L$-smooth\footnote{By $L$-smoothness we mean: $\psi(z) \le \psi(x) + \nabla \psi(x)^\top (z-x) + \frac{L}{2} \|x-z\|^2, \quad \mbox{for all } \;\; x,z \in \R^d.$}, for $\lambda, L >0$, and that $G^*$ is compactly supported, i.e., $G^*([-M,M]^d) = 1$, for some $M > 0$. Then there is a function $n(d, \sigma, M)$ and a constant $C_{d,\sigma} >0$ such that, for all sample sizes $n$ with $n \ge n(d,\sigma, M)$,
\begin{equation}\label{eq:Emp-Risk}
\frac{1}{n} \sum_{i=1}^n \|\hat T_n(Y_i) - \tilde \theta_i \|^2_2 \le C_{d, \sigma} \frac{L}{\lambda}\, \frac{1}{\log n}
\end{equation}
with probability at least $1- \frac{4 d}{n^8}$.
\end{theorem}
Our proof technique for Theorem~\ref{thm:Rate-OT-Map} first relates the left-hand side of~\eqref{eq:Emp-Risk} to $W_2^2(G_n,  \widehat{G}_n)$ --- the deconvolution error in the Wasserstein metric (see~\cite{deb2021rates, manole2021plugin}). It is well known that the smoothness of the Gaussian errors makes the deconvolution problem difficult; in fact, the logarithmic rate 
is minimax optimal for deconvolution with Gaussian errors (see e.g.,~\cite{dedecker2013minimax, soloff2021multivariate}).

\section{Numerical experiments}\label{sec:numerical}
The first goal of this section is to illustrate the efficiency and scalability of our semismooth Newton based ALM on univariate and bivariate data examples.
We compare the ALM with the state-of-the-art solver \texttt{mix-SQP} \cite{kim2020fast} implemented in the Julia programming language\footnote{The source code is available at \url{https://github.com/stephenslab/mixsqp-paper}.},  and the \texttt{R} package \texttt{REBayes} \cite{koenker2017rebayes}. The \texttt{KWDual} function in the latter package solves the dual formulation \eqref{dual} by an interior point method using the commercial solver Mosek \cite{andersen2000mosek,mosek}.
For the univariate and bivariate data examples, we do not present the PEM algorithm in the comparisons due to three reasons: (1) when $d=1$ or $2$ we are able to choose fine enough support points such that the finite dimensional convex optimization problem \eqref{primal-0} can approximate the NPMLE \eqref{NPMLE1} very well; (2) the convex program \eqref{primal-0} guarantees a globally optimal solution while the PEM only converges to a stationary solution; (3) we observe from our extensive numerical experiments that when $d=1$ or $2$, the empirical Bayes estimates based on the ALM and those based on the PEM are almost indistinguishable; see e.g., the middle and right plots of Figure~\ref{fig-CMD}.

The second goal of this section is to demonstrate the effectiveness of the PEM algorithm; we mainly illustrate this for $d\geq 3$. Here we compare the PEM with the classic EM and the ALM. When $d\geq 3$, the discretization of \eqref{NPMLE1} based on equally spaced grid points in a compact region of $\mathbb{R}^d$ is no longer feasible since the number of grid points then would need to grow exponentially in the number of dimensions $d$. We tackle this difficulty in the ALM by adopting a proposal in~\cite{lashkari2007convex} by taking $m=n$ with $\mu_j = Y_j$, for all $j =1,\ldots, m$, when $d \ge 3$. Note that when $d \ge 3$, we can still implement the PEM and the EM algorithms as it solves~\eqref{adaptive supports 0} that   allows for the update of the locations of the atoms of $\widehat{G}_n$; however, we adopt the above strategy of~\cite{lashkari2007convex} as the starting point of these iterative (non-convex) algorithms.

Most of the numerical experiments reported in this section  are performed in Matlab (version 9.7) on a Windows workstation (24-core, Intel Xeon E5-2680 @ 2.50GHz, 128 GB of RAM), except explicitly mentioned otherwise. Our methods are implemented in Matlab and are available at \url{https://github.com/YangjingZhang/Dual-ALM-for-NPMLE}.

\subsection{Initialization and stopping criteria of the methods}
We first give some details of the stopping conditions for each of the competing methods --- our ALM, the \texttt{mix-SQP} \cite{kim2020fast} solver, the \texttt{R} package \texttt{REBayes}~\cite{koenker2017rebayes}, the EM algorithm and our PEM method. 
\subsubsection{Stopping conditions}
For a given tolerance $\varepsilon > 0$,  the \texttt{mix-SQP} is terminated if
\begin{equation}\label{stop-cond}
  \eta_1:= \max_{1\leq j \leq m} \left[\, \frac{1}{n} L_{\bullet j}^\top ({\bf 1}_n \oslash Lx) - 1 \,\right] \, \leq \, \varepsilon,
\end{equation}
where  $L_{\bullet j}$ represents the $j$-th column of $L$ and $\oslash$ denotes the Hadamard division defined as: $x \oslash y = (x_1/y_1,\dots,x_n/y_n)$, for $x, y\in\mathbb{R}^n$. For \texttt{REBayes}, we adopt its default  termination condition with the relative tolerance of the dual gap `{\tt rtol}' being $10^{-6}$.
We terminate our semismooth Newton based ALM under the following stricter condition
\begin{equation}\label{stop-cond2}
\max(\eta_1,\eta_2) \leq \varepsilon,
\end{equation}
which additionally involves  the KKT residual defined as
\begin{equation*}\
  \eta_2 := \left\|\, x - \max\left(\, x + \frac{1}{n} L^\top ({\bf 1}_n \oslash Lx) - {\bf 1}_m, \, 0 \,\right) \,\right\|_2.
\end{equation*}
In this section and in the reported tables, the residuals  for our ALM and \texttt{REBayes} are computed from $\max(\eta_1, \eta_2)$ given in \eqref{stop-cond2}, while for \texttt{mix-SQP} we use the value $\eta_1$ in \eqref{stop-cond}. We say a solution is more accurate if its residual is smaller. Throughout our numerical experiments,
we set  $\varepsilon = 10^{-6}$ for our ALM and for \texttt{mix-SQP}. In addition, both methods are also terminated if the number of ALM/SQP iterations reaches $100$. The default setting of the maximum number of sub-iterations for the \texttt{mix-SQP} solver is $100$, but we find through our numerical experiments that the overall algorithm may work better if the maximum sub-iteration number allowed is $200$, especially for some difficult instances. So we adopt the latter maximum sub-iteration number in the reported results below.

Both the PEM and EM are terminated at the $k$-th iteration if ${\rm obj}_k - {\rm obj}_{k-1} \leq 10^{-4}$, where ${\rm obj}_k$ denotes the objective function value in \eqref{adaptive supports 0} at the $k$-th iteration. Both the PEM and EM are also terminated if the number of iterations reaches $100$.

\subsubsection{Implementation details of our ALM algorithm}\label{sec:ALM-Implement}
First, we make the following remarks on the implementation of our ALM:

\vskip 0.1in
\noindent
(a) In order to further scale the ALM, 
we  borrow a low rank approximation idea from \cite{kim2020fast}. As shown in Figure~\ref{lowrankL}, the matrix $L$ usually has a lot of singular values close to $0$ when $d=1$ (but not for $d\geq2$).

\vskip 0.1in
\noindent
(b) The solution of \eqref{primal} is invariant to  scaling  each row of $L$ since for any $\alpha_1, \ldots, \alpha_n>0$,
$$
\sum_{i=1}^{n}\log\,(Lx)_i = \sum_{i=1}^{n}\log\, (L_{i\bullet} \, x) = \sum_{i=1}^{n}\log\,(\alpha_i \, L_{i\bullet}\, x) - \sum_{i=1}^{n}\log\,\alpha_i,
$$
where $L_{i\bullet}$ denotes the $i$-th row of $L$. Therefore, we assume without loss of generality that the largest component in each row of $L$ is always $1$ by taking $\alpha_i = 1/\left(\displaystyle\max_{1\leq j\leq m} L_{ij}\right)$.

\vskip 0.1in
\noindent
(c)
Based on the KKT optimality conditions in~\eqref{kkt}, we consider the following initial point for our ALM:
\begin{equation*}
x^0 := \frac{1}{m} {\bf 1}_m, \quad y^0 := L x^0 = \frac{1}{m} L{\bf 1}_m,\quad v^0 = u^0 := {\bf 1}_m \oslash y^0.
\end{equation*}
We find that this  initial point works well for  the synthetic data sets tested in Section~\ref{sec:syn}. For the more challenging real data sets in Section~\ref{sec:real}, we construct the following initial point to fully take advantage of the second order sparsity in the generalized Hessian matrix appearing in the ALM subproblem:
\begin{equation}\label{init-pt-2}
x^0 := \frac{\sigma_0}{2} {\bf 1}_m, \quad y^0 := \frac{1}{m} \, L {\bf 1}_m,\quad u^0 := {\bf 1}_m \oslash y^0,\,\, v^0 := {\bf 0}_n.
\end{equation}
Recall from Table~\ref{table:complexity} that
the computational cost of finding the generalized Hessian in our ALM is $O(ns \min(n,s))$, where $s=\{i:S_{ii}\neq 0\}$ is the number of nonzero entries in the diagonal matrix $S$; see \eqref{eq:hessian}. The idea behind the above initialization  is to make $s$ as small as possible for the first several ALM iterations (and semismooth Newton sub-iterations). Starting from the initial point in \eqref{init-pt-2}, we have that for the first ALM subproblem (i.e., $x = x^0$), it holds that for any dual variable $v\in \mathbb{R}^n$,
\begin{equation*}
S\in\partial  \max\left(\frac{1}{n} L^\top v - {\bf 1}_m + \frac{1}{\sigma_0} x^0, 0\right)
=\partial \max\left(\frac{1}{n} L^\top v - \frac{1}{2}{\bf 1}_m, 0\right).
\end{equation*}
Since we set $v^0 = {\bf 0}_n$, it always holds that $\frac{1}{n} L^\top v^0 - \frac{1}{2}{\bf 1}_m \leq 0$ and $S$ can be taken as a zero matrix for the first semismooth Newton iteration (within the first ALM subproblem). As the algorithm proceeds, the variable $v$ will deviate from the zero vector gradually, and  the number of violated inequalities in $\frac{1}{n} L^\top v - \frac{1}{2}{\bf 1}_m \leq 0$ may increase correspondingly. Hence, the choice of the initial point in \eqref{init-pt-2} would result in  a gradually increasing $s$ (from zero) that helps to reduce the computational cost in the early iterations of the algorithm.

\vskip 0.1in
\noindent
(d) Once we obtain an approximate solution $x$ to problem \eqref{primal} via the ALM, we renormalize $x$ ($x \mapsto x/\sum_{j=1}^{m}x_j$) such that its components add up to one.

\vskip 0.1in
\noindent
(e) We next give the adjustment of the positive scalar $\sigma_k$ in the ALM framework. It follows from Proposition~\ref{prop:rate} that a larger $\sigma_k$ gives rise to a faster local convergence rate $(\kappa/\sqrt{\kappa^2 + \sigma_k^2} + \eta_k )(1-\eta_k)^{-1}$. However, when $\sigma_k$ is very large, the condition number of the the generalized Hessian matrix in \eqref{eq:hessian} will be large since the diagonal entries of the positive definite diagonal matrix $D^k$ will be close to zero. In this case, finding the Newton direction in \eqref{newton-sys} may need more conjugate gradient steps. Therefore, we shall consider the trade-off between the convergence rate of the ALM and the cost of solving the linear systems in the semismooth Newton method. In our implementation, we set $\sigma_0=100$, $\sigma_{k+1} = \sqrt{3}\sigma_k$ if $\chi_k/\chi_{k-1} > 0.6$, and $\sigma_{k+1} = \sigma_k$ otherwise. Here $\chi_k:= \max( \max(\frac{1}{n} L^\top v^k - {\bf 1}_m,0),\|u^k-v^k\|/\|u^k\|)$ characterizes the feasibility of \eqref{dual}. Namely, when the improvement on the feasibility of \eqref{dual} after one iteration is too small, we increase $\sigma_k$. Next, the sequence $\{\varepsilon_k\}_{k \ge 1}$ (the same for $\{\eta_k\}$) in the stopping criteria satisfying \eqref{eq:condition for stop} is chosen as follows: $\varepsilon_0=0.5$, $\varepsilon_{k+1}=\varepsilon_k/\varsigma$. We set $\varsigma = 1.06$ if the $k$-th subproblem has been solved efficiently within 30 semismooth Newton iterations; otherwise, we set
$\varsigma = 1$. Lastly, in the semismooth Newton method (Algorithm~\ref{alg:ssnal}), we set $\bar{\eta}=0.1$, $\tau=0.1$, $\mu=10^{-4}$, and $\beta=0.5$.

\vskip 0.1in
\noindent
(f) The Newton system \eqref{newton-sys} can be solved by direct solvers, for example, via computing the Cholesky factorization of the coefficient matrix, when the size of the coefficient matrix is moderate ($ \leq 5,000$). Alternatively, when the dimension of the linear system is large ($>5,000$), we solve it iteratively by conjugate gradient method.

\subsection{Numerical results when $d=1$}\label{sec:syn}
We first present the numerical results on several one-dimensional synthetic data sets. The main purpose of these experiments is to show the efficiency and scalability of our ALM in terms of $n$ (the number of observations) and $m$ (the number of grid points).

\vskip 0.1in
\noindent
\underline{\bf Example 1}. We replicate the simulation experiment conducted in \cite{brown2009nonparametric,jiang2009general,johnstone2004needles}. Consider $n$ independent observations where  $Y_i \sim {\cal N}(\theta_i,1)$, with each $\theta_i$ taking the value $0$ or $\nu$ with the proportion of $\nu$ being $\tau = 0.5\%n$, $5\%n$, or $50\%n$. We use equally spaced support points on the interval $[\min_i Y_i,\max_i Y_i]$ for a given number of grid points $m$.
Based on the estimated prior $\widehat{G}_n = \sum_{j=1}^m x_j \delta_{\mu_j}$, the empirical Bayes estimator \eqref{EB} reduces to
$ \widehat{\theta}_i= (\sum_{j=1}^{m}\mu_j L_{ij} x_j )/(\sum_{j=1}^{m} L_{ij} x_j )$.

\begin{table}[htbp]\centering
\setlength{\tabcolsep}{1.1mm}
{
\begin{tabular}{lcccccccccc}
\toprule
$\tau$ & $\nu$ & \multicolumn{3}{c}{Time (in sec)} & \multicolumn{3}{c}{Residual} & \multicolumn{3}{c}{$\|\widehat{\theta} - \theta\|^2$} \\
\cmidrule(l){3-5} \cmidrule(l){6-8} \cmidrule(l){9-11}
    &       & ALM & \texttt{mix-SQP} & \texttt{REBayes} & ALM & \texttt{mix-SQP} & \texttt{REBayes}  & ALM & \texttt{mix-SQP} & \texttt{REBayes}  \\
\midrule
\multirow{4}{0.7cm}{5}            & 3 &   0.2 &   4.1 &     0.9 & 5.5e-07 & 1.9e-07        & 2.6e-05 &     34 &     34 &     34 \\
                                  & 4 &   0.2 &   3.7 &     0.9 & 5.0e-07 & 1.4e-07        & 2.8e-05 &     29 &     29 &     29 \\
                                  & 5 &   0.2 &   3.8 &     0.9 & 4.8e-07 & 1.0e-07        & 2.4e-05 &     18 &     18 &     18 \\
                                  & 7 &   0.2 &   4.5 &     0.8 & 3.9e-07 & -9.5e-09       & 4.7e-05 &      7 &      7 &      7 \\[6pt]
\multirow{4}{0.7cm}{50}           & 3 &   0.2 &   3.9 &     0.9 & 4.9e-07 & 1.6e-07        & 3.2e-05 &    156 &    156 &    156 \\
                                  & 4 &   0.2 &   3.9 &     0.8 & 5.0e-07 & 1.4e-07        & 2.0e-05 &    104 &    104 &    104 \\
                                  & 5 &   0.2 &   3.9 &     0.8 & 4.7e-07 & 1.7e-07        & 2.1e-05 &     53 &     53 &     53 \\
                                  & 7 &   0.2 &   23.7* &   0.9 & 5.0e-07 & {\bf 4.3e-01}  & 2.4e-05 &     12 &    {\bf 176} &     12 \\[6pt]
\multirow{4}{0.7cm}{500}          & 3 &   0.2 &   4.0 &     0.7 & 4.9e-07 & 3.1e-07        & 2.8e-05 &    449 &    449 &    449 \\
                                  & 4 &   0.2 &   3.8 &     0.7 & 5.3e-07 & 2.6e-07        & 2.3e-05 &    286 &    286 &    286 \\
                                  & 5 &   0.2 &  9.4* &     0.8 & 5.0e-07 & {\bf 2.8e-02}  & 2.9e-05 &    125 &    {\bf 163} &    125  \\
                                  & 7 &   0.2 & 123.4* &    0.8 & 4.6e-07 & {\bf 8.5e-01}  & 2.6e-05 &     16 &    {\bf 866} &     16  \\
\bottomrule
\end{tabular}
}
\caption{\small Comparison between ALM, \texttt{mix-SQP}, and \texttt{REBayes} for Example 1 with $n=1,000$ and $m=500$ (averaged over $100$ replications).  The sign $*$ indicates that the computational time taken by the \texttt{mix-SQP} solver for these three instances  was excessively long (due to the fact that the solver did not solve the problems successfully and reached the maximal number of iterations).
}
\label{table-sim1-small}
\end{table}

\begin{table}\centering
{
\begin{tabular}{lccccccc}
\toprule
$\tau$ & $\nu$ & \multicolumn{2}{c}{Time (in sec)} & \multicolumn{2}{c}{Residual} & \multicolumn{2}{c}{$\|\widehat{\theta} - \theta\|^2$} \\
\cmidrule(l){3-4} \cmidrule(l){5-6} \cmidrule(l){7-8}
    &       & ALM  & \texttt{REBayes} & ALM  & \texttt{REBayes}  & ALM & \texttt{REBayes}  \\
\midrule
\multirow{4}{1cm}{50}             & 3 &   5.3 & 159.1 & 5.4e-07 & 9.8e-05 &    277 &    277 \\
                                  & 4 &   4.8 & 161.3 & 5.5e-07 & 1.1e-04 &    221 &    221 \\
                                  & 5 &   4.8 & 156.9 & 5.7e-07 & 5.5e-05 &    115 &    115 \\
                                  & 7 &   5.1 & 167.7 & 7.0e-07 & 8.3e-05 &     20 &     20 \\[6pt]
\multirow{4}{1cm}{500}            & 3 &   4.8 & 158.1 & 6.0e-07 & 5.1e-05 &   1476 &   1476 \\
                                  & 4 &   4.7 & 169.8 & 5.8e-07 & 1.5e-04 &    999 &    999 \\
                                  & 5 &   4.9 & 180.8 & 5.9e-07 & 2.3e-04 &    469 &    469 \\
                                  & 7 &   5.0 & 219.5 & 6.1e-07 & 1.1e-04 &     47 &     47 \\[6pt]
\multirow{4}{1cm}{5000}           & 3 &   4.3 & 146.6 & 7.0e-07 & 6.6e-05 &   4437 &   4437 \\
                                  & 4 &   4.4 & 152.0 & 7.1e-07 & 4.7e-05 &   2751 &   2751 \\
                                  & 5 &   4.5 & 176.1 & 7.2e-07 & 5.7e-05 &   1197 &   1197 \\
                                  & 7 &   4.6 & 280.3 & 7.5e-07 & 4.4e-05 &    101 &    101 \\
\bottomrule
\end{tabular}
}
\caption{\small Comparison between ALM  and \texttt{REBayes} for Example 1 with $n=10,000$ and $m=5,000$ (averaged over $100$ replications).
}
\label{table-sim1}
\end{table}

In Table~\ref{table-sim1-small}, we report the numerical performance of our ALM, \texttt{mix-SQP} and~\texttt{REBayes} for relatively small-size instances with $n=1,000$ and $m=500$ (averaged over $100$ replications). We also report the squared error loss $\|\widehat{\theta} - \theta\|_2^2:=\sum_{i=1}^{n}(\widehat{\theta}_i - \theta_i)^2$ between the empirical Bayes estimator $ \widehat{\theta}$ and the true mean $\theta$ for the three different algorithms.
It can be seen from Table~\ref{table-sim1-small} that our ALM outperforms the other two methods for all instances --- the ALM is the fastest algorithm which also produces the smallest `residual' while not compromising on fit (as can be seen from the squared error loss $\|\widehat{\theta} - \theta\|_2^2$). 

In order to illustrate the scalability of our ALM,  we further repeat the experiment on large instances with $n=10,000$ and $m = 5,000$. The corresponding results are recorded in Table~\ref{table-sim1}. We do not include the results for \texttt{mix-SQP} since it cannot handle problems of this scale (where $m>1,000$). It can be seen from Table~\ref{table-sim1} that the computational time of our ALM for each instance is approximately 5 seconds, which is about $30$ times faster than \texttt{REBayes}. In fact, about 4 seconds for each instance of our ALM  are spent for the one-time computation of a low rank approximation of the matrix $L$; the rest of the computation (including the computation of the gradients and the generalized Hessians as well as solving the semismooth Newton equations) is completed in $1$ second.



\vskip 0.2in
\noindent
\underline{\bf Example 2}.
We replicate the experiment conducted in \cite{kim2020fast}, where $50\%$, $20\%$, and $30\%$ of the observations $\{Y_i\}_{i=1}^n$ are draw independently from ${\cal N}(0,1)$, $t_4$, and $t_6$ distributions respectively. Here $t_{\nu}$ denotes Student's $t$-distribution with $\nu$ degrees of freedom. As the observed data can be modeled as a Gaussian {\it scale} mixture, we find $\widehat{G}_n$ by solving a mixture problem of the form $\sum_{j=1}^{m} x_j g_j, \ x_j \geq 0,\  \sum_{j=1}^{m}x_j = 1,$
where $g_j$ is the density of ${\cal N}(0,\sigma_j^2)$ for some given $\sigma_j$, $j = 1,\ldots, m$.
Following \cite{kim2020fast}, we select the grid values $\{\sigma_1^2,\dots,\sigma_m^2\}$ by the method in \cite{stephens2017false}.

\begin{figure}[htbp]
  \centering
  \includegraphics[width=0.4\textwidth, height = 0.22\textheight]{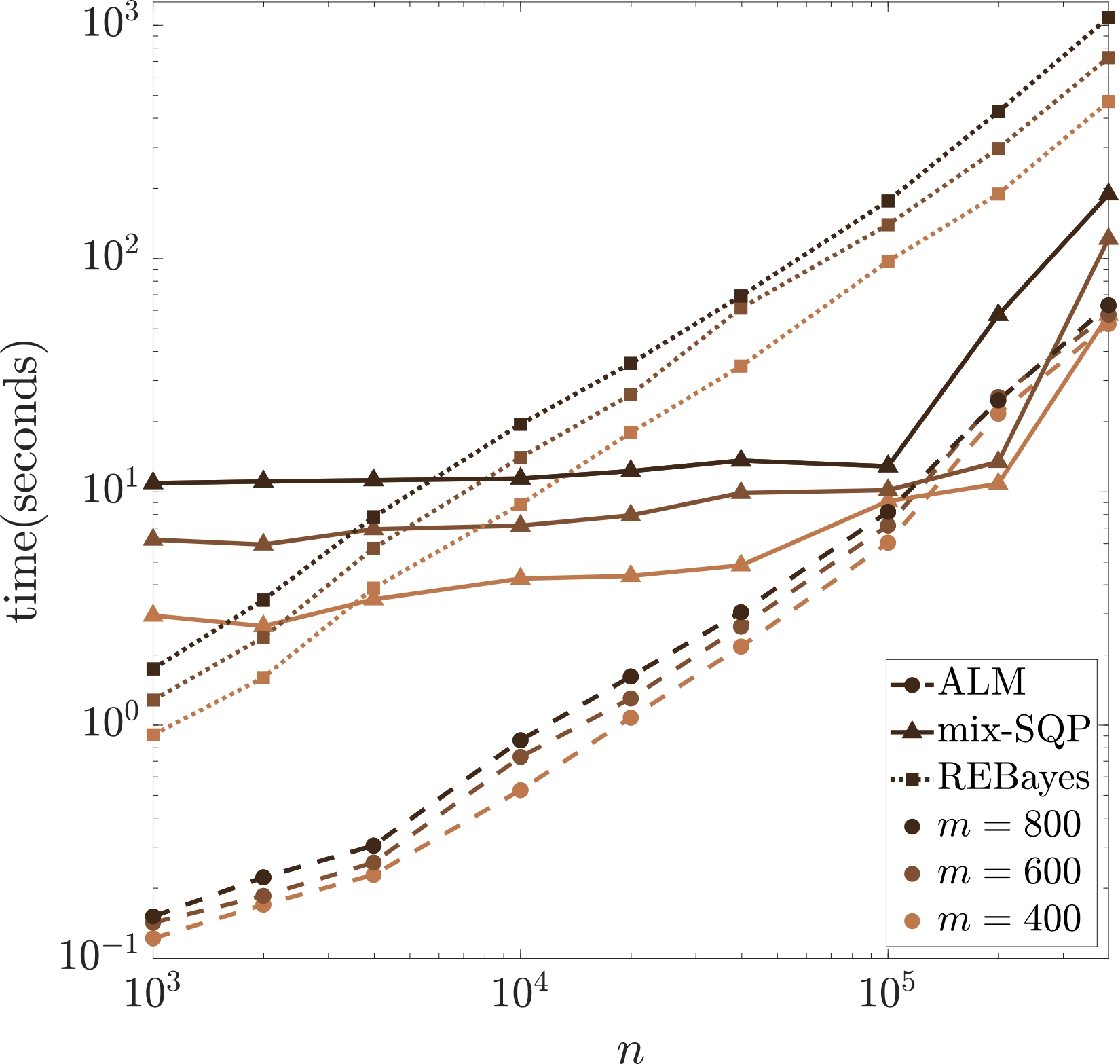}
  \qquad \quad
  \includegraphics[width=0.4 \textwidth, height = 0.22 \textheight]{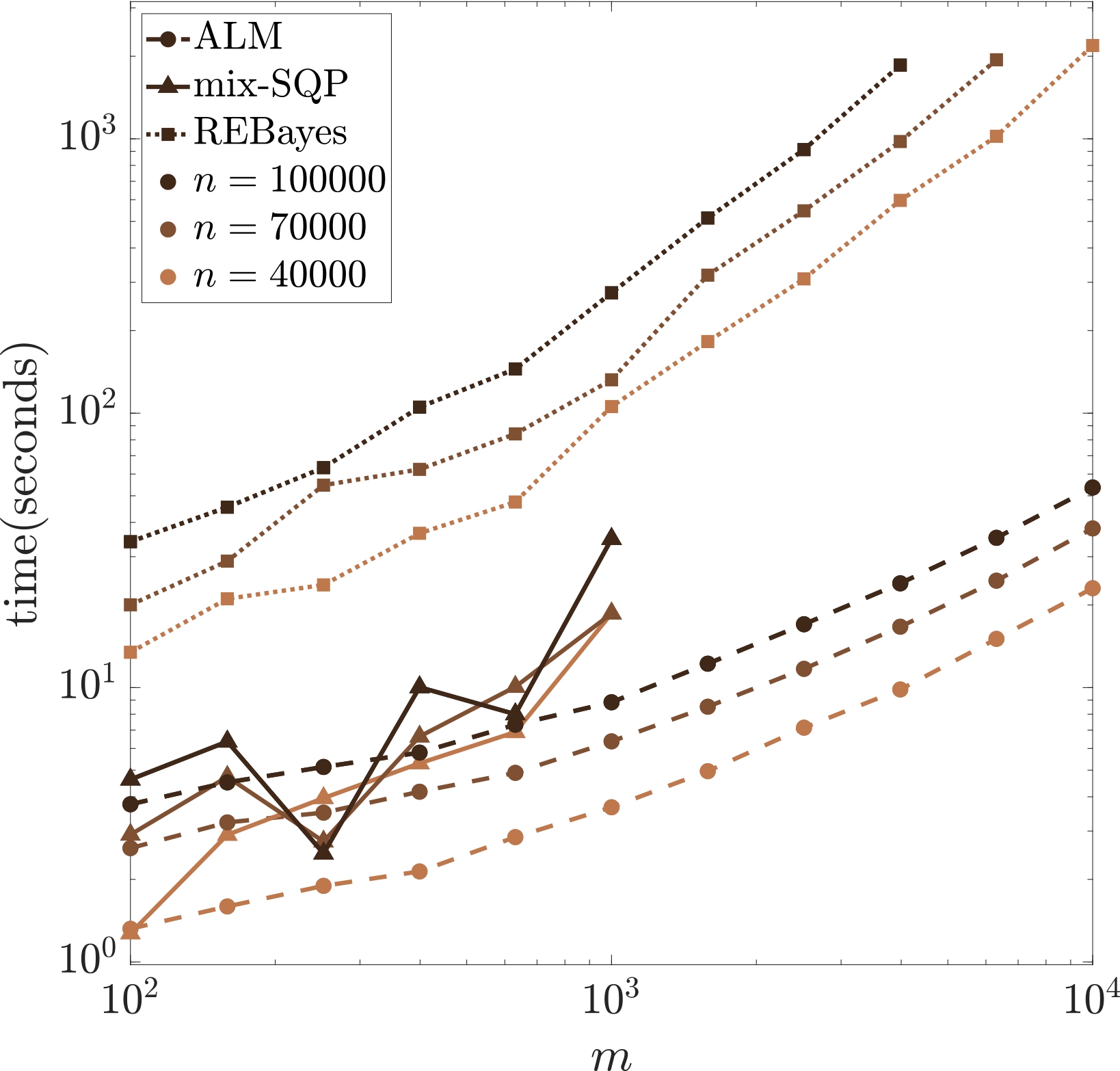}
  \caption{\small Computational time (in seconds) of the ALM, \texttt{mix-SQP}, and \texttt{REBayes} with changing $n$ and $m$ for Example 2 (averaged over $10$ replications).}\label{fig-sim2}
\end{figure}

We test the scalability of the ALM,  \texttt{mix-SQP} and  \texttt{REBayes} for different values of $n$ and $m$, and the results are shown in Figure~\ref{fig-sim2}. On the left panel, we consider
$m  \in\{400,600,800\}$ and $n \in {\rm ceil}\{10^3,10^{3.3}, 10^{3.6}, 10^4, 10^{4.3}, 10^{4.6}, 10^5, 10^{5.3}, 10^{5.6}\}$.
We consider even larger instances with
$n  \in\{4\times 10^4,7\times 10^4,10^5\}$ and
$m \in{\rm ceil}\{10^2,10^{2.2}, 10^{2.4},10^{2.6},10^{2.8}, 10^3, 10^{3.2},10^{3.4},10^{3.6},10^{3.8},10^4\}$ on the right panel of Figure~\ref{fig-sim2}.
We found that when $m>1,000$,  \texttt{mix-SQP} usually fails to solve the instances within $100$ iterations under the stopping criterion $\varepsilon = 10^{-6}$;
as a result,  we have not included the results for  \texttt{mix-SQP} when $m > 1,000$ in the plot. Although the \texttt{REBayes} solver is able to solve most instances, it takes about $100$ times more computational time compared to our ALM. In particular, for the largest test instance with $n=10^5$ and $m=10^4$, it only takes the ALM $53.61$ seconds to get a highly accurate solution. However,  \texttt{REBayes} failed to solve this instance.

\subsection{Numerical results when $d=2$}\label{sec:real}

To further show the effectiveness and scalability of our proposed ALM, we analyze 2 two-dimensional astronomy data sets obtained from Gaia-TGAS~\cite{brown2016gaia} and APOGEE \cite{majewski2017apache}.

\begin{figure}[!h]
  \centering
  \subfloat[][\texttt{mix-SQP} with $m=20^2$]{ \includegraphics[width=0.48\textwidth]{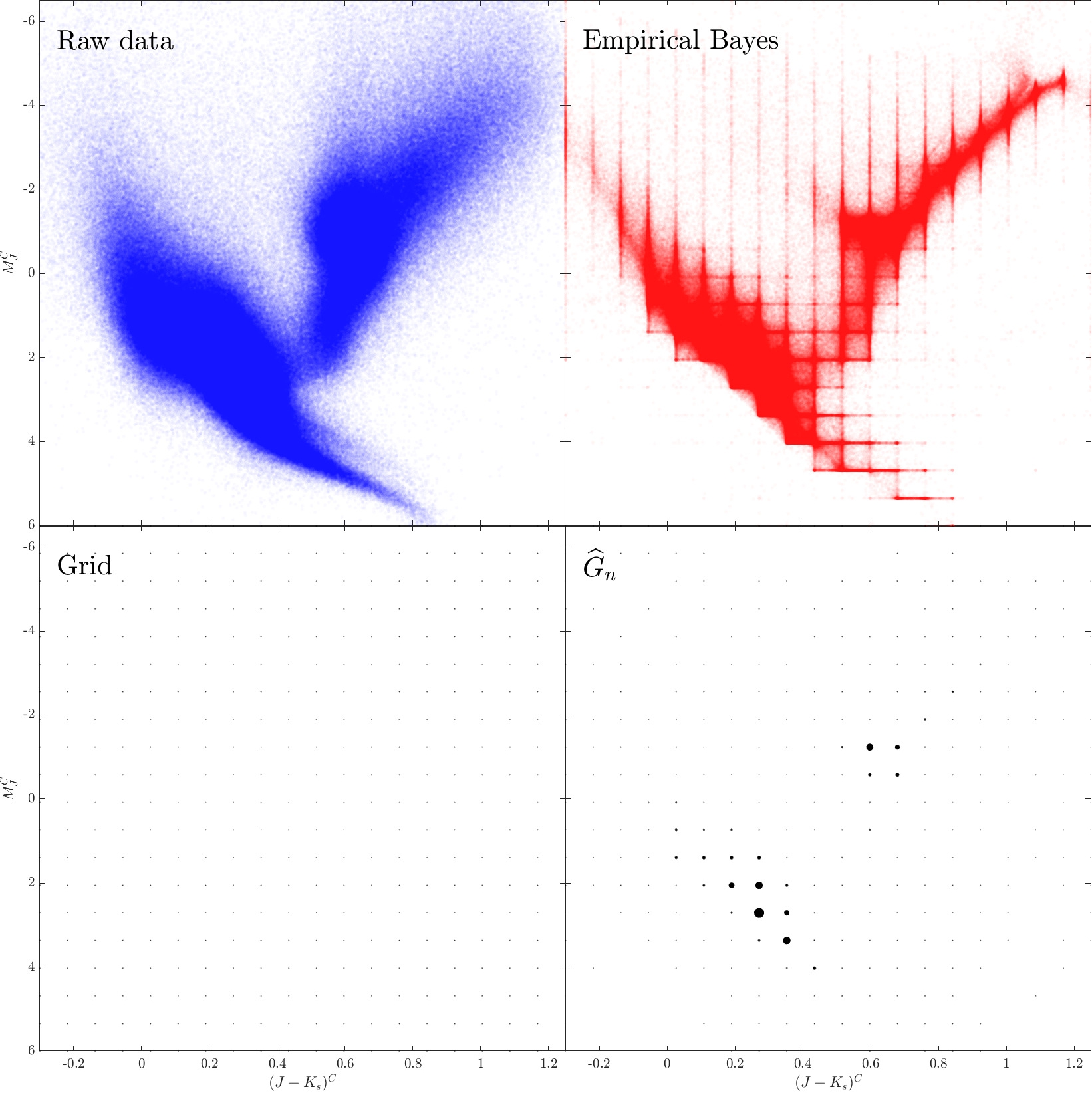}} \,
   \subfloat[\small ALM with $m=100^2$]{  \includegraphics[width=0.48\textwidth]{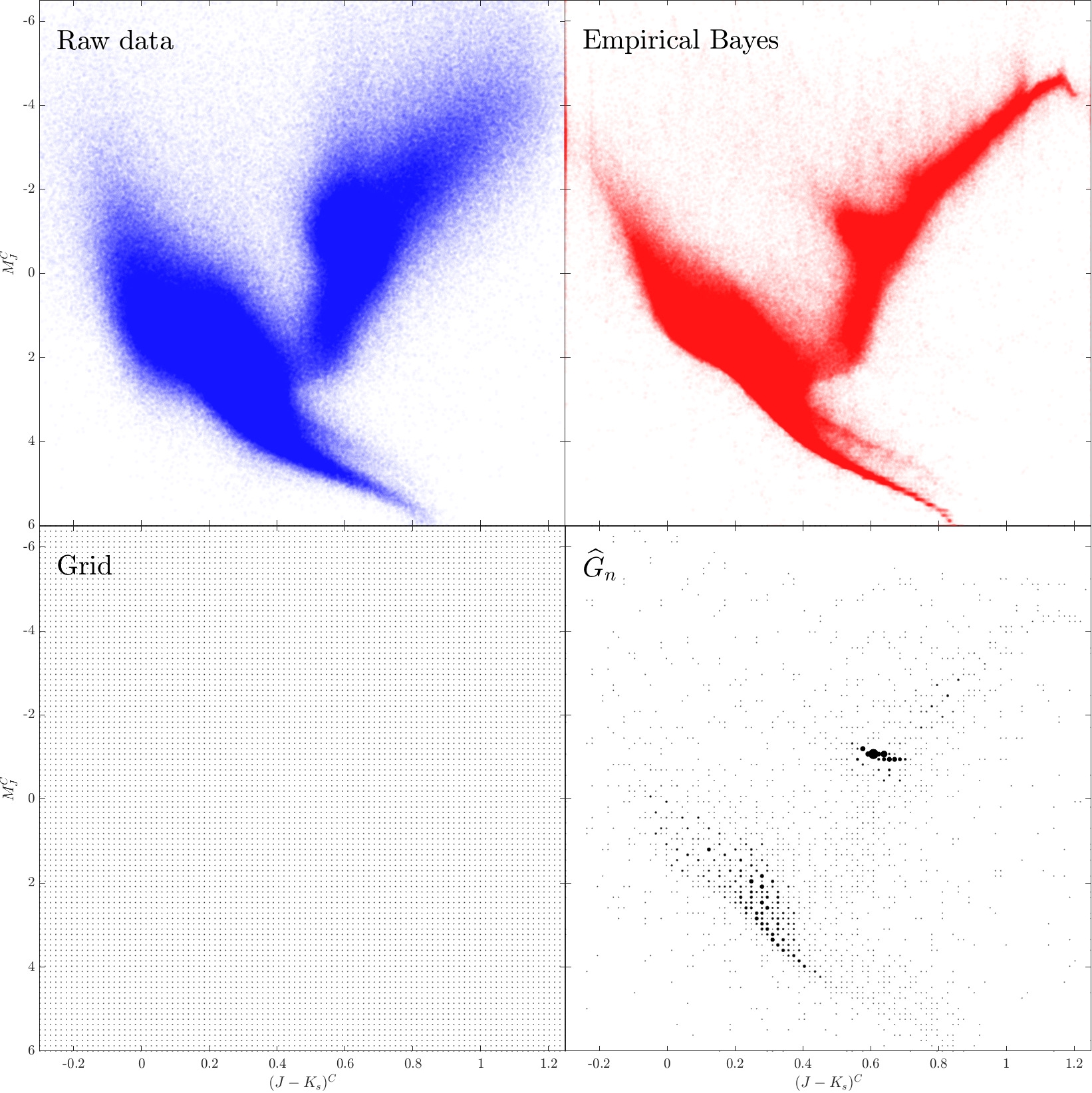}}
  \caption{\small Results for the $d=2$ dimensional Gaia-TGAS data obtained from: (a) \texttt{mix-SQP} (with $m = 20^2$),  (b) ALM (with $m = 100^2$). The number of support points of $\widehat{G}_n$ (see the (2,2) subplots) are:  (a) $307$,  (b) $1,677$, where the size of each support point plotted is proportional to its weight. The run times are:  (a) 28 minutes (failed to converge within 100 iterations), (b) 472 minutes. {With denser grid points we are able to obtain sharper denoised estimates that reveal finer details of the CMD.}
  }\label{fig-Gaia}
\end{figure}

\vskip 0.1in
\noindent
\underline{\bf Data set 1 (Gaia-TGAS)}.
We first consider  the astronomy data Gaia-TGAS \cite{brown2016gaia} that has been studied in  \cite{anderson2018improving}, where the extreme deconvolution algorithm \cite{bovy2011extreme} was used to estimate the true parallax and photometry of every star.
This data set contains $n=1,363,432$ observations $\{Y_i\}_{i=1}^n \subset \mathbb{R}^2$, which can be modeled as a two-dimensional Gaussian location mixture where $Y_i$ is assumed to have density $f_{G^*,\Sigma_i}$ with a known diagonal covariance matrix $\Sigma_i$; see \eqref{mixture-dis}.
We plot the raw data $\{Y_i\}_{i=1}^n$, the empirical Bayes estimates $\{\widehat\theta_i\}_{i=1}^n$, the initial grid, and the estimated prior $\widehat{G}_n$  in Figure~\ref{fig-Gaia}. 
With such a large $n\approx 10^6$, we found that \texttt{mix-SQP} can only handle this problem with small $m$, up to several hundreds. Thus, in Figure~\ref{fig-Gaia}(a), we use a $20\times 20$ grid points (i.e., $m=400$) when solving it by \texttt{mix-SQP}. In contrast, we display in Figure~\ref{fig-Gaia}(b) the solution obtained from our ALM with $100\times 100$ grid points.  We can see from the  empirical Bayes estimates given by \texttt{mix-SQP} in Figure~\ref{fig-Gaia}(a) that the $20\times 20$ grid points are not fine enough to denoise this data properly. In contrast, the empirical Bayes estimates obtained by ALM in Figure~\ref{fig-Gaia}(b) show more shrinkage overall; there are sharp tails in the bottom (i.e., the main sequence) and top right (i.e., the red clump) of the plot of the empirical Bayes estimates. One can easily see the benefits of working with a large $m$ here --- with denser grid points we are able to obtain sharper denoised estimates (i.e., posterior means) that reveal finer details of the CMD\footnote{In particular, we point out that 128 GB of RAM was not enough for solving the ALM problem shown in Figure~\ref{fig-Gaia}(b), since  the matrix $L\in\mathbb{R}^{1,363,432\times 10,000}$ alone consumes approximately 109 GB of storage. The results in Figure~\ref{fig-Gaia}(b) were obtained in Matlab (version 9.5) on a Windows workstation (32-core, Intel Xeon Gold 6130 CPU @ 2.10 GHz (2 processors), {\bf 256 GB of RAM}) in approximately $472$ minutes.}.
Note that \texttt{REBayes} cannot handle $n \approx 10^6$ and $m \approx 10^4$. To illustrate the performance of \texttt{REBayes}, we draw a subsample of size $n=100,000$ from the original data set (with $n = 1,363,432$). For this subsampled data set (not plotted here) with $m=10^4$ grid points, our ALM takes approximately 5 minutes, while \texttt{REBayes} takes approximately 73 minutes.

\vskip 0.2in
\noindent
\underline{\bf Data set 2 (APOGEE)}. Our second real data set is taken from the Apache Point Observatory Galactic Evolution Experiment survey (APOGEE) \cite{majewski2017apache}. Following the pre-processing in \cite{ratcliffe2020tracing} to remove the outliers with anomalous abundance measurements, the data set contains $n=27,135$ observations in $\mathbb{R}^{19}$.

We first analyze $d=2$ features picked from the 19 dimensions. For $d=2$ we always use $m=100\times100$ equally spaced grid points inside the minimum axis-aligned bounding box  of the raw data (i.e., the smallest rectangle that contains all the data points), that is known to contain all the support points of $\widehat{G}_n$ (see \cite{soloff2021multivariate}). To compare the empirical performance of the ALM and \texttt{REBayes}, we first select 5 pairs of features from the 19 dimensions and run both algorithms. Table~\ref{table-APOGEE-time} shows that for all instances the ALM is faster than \texttt{REBayes} and the solutions returned by the ALM are more accurate (i.e., have smaller KKT residuals). For this real data set with both large $n=27,135$ and $m=10,000$,  \texttt{mix-SQP} is not applicable.
We remind the reader that we have not incorporated a low rank approximation of the matrix $L$ here since it does not work well for multivariate data as discussed in the Introduction; see Figure~\ref{lowrankL}. Therefore, the second order sparsity in the generalized Hessian mostly contributes to the efficiency of our ALM.

\begin{figure}[!h]
\centering
\includegraphics[width=0.99\textwidth]{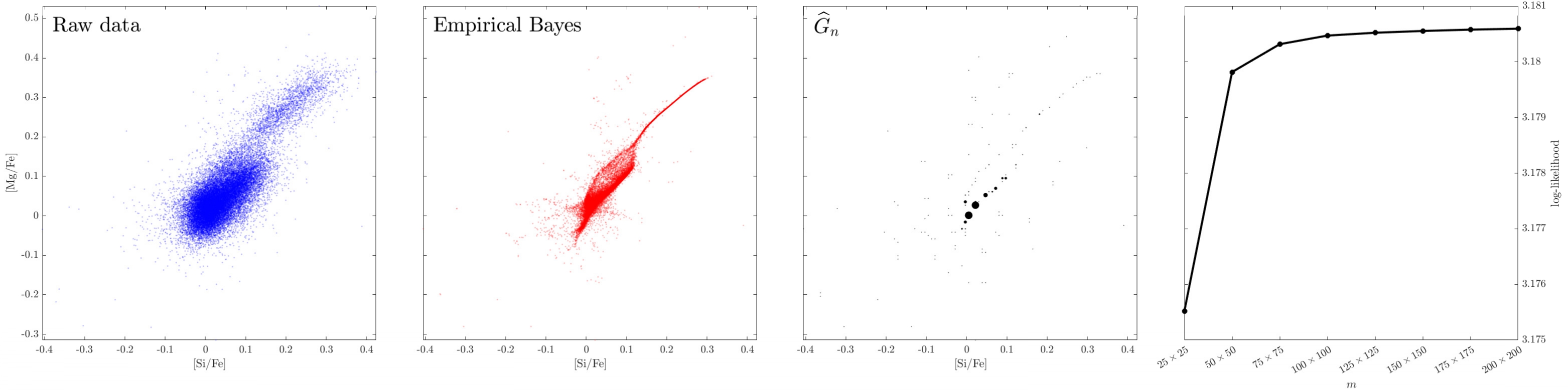}
\caption{\small Results for the $d=2$ dimensional APOGEE data in the [Mg/Fe]-[Si/Fe] plane with $m=100^2$ grid points. The empirical Bayes estimates (2nd plot from left) show strong association and a manifold-like structure in the upper right region, and the fitted $\widehat{G}_n$ is very sparse (3rd plot from left). The rightmost plot gives the log-likelihood value against the number of grid points.  Observe that the log-likelihood value improves a lot as the number of grid points increases from $25\times 25$ and attains a plateau near  $100\times 100$. 
}
\label{fig-APOGEE-1-10}
\end{figure}

We next illustrate the performance of our ALM on the $2$-dimensional plane [Mg/Fe]-[Si/Fe]. The first three plots (from the left) in Figure~\ref{fig-APOGEE-1-10} show the raw data $\{Y_i\}_{i=1}^n \subset \R^2$ for $n = 27,135$, the empirical Bayes estimates $\{\widehat{\theta}_i\}_{i=1}^n$, and the estimated prior $\widehat{G}_n$. From the plot of the empirical Bayes estimates in Figure~\ref{fig-APOGEE-1-10} we infer the strong association (in fact, a `curve'/manifold-like structure) between the variables [Mg/Fe] and [Si/Fe], especially in the upper right corner. In order to get a sense of how dense the grid points should be to obtain a good approximation of~\eqref{NPMLE1} for this data set, we plot the log-likelihood value $\frac{1}{n}\sum_{i=1}^{n}\log\big(\sum_{j=1}^{m} x_j\phi_{\Sigma_i}(Y_i-\mu_j)\big)$ against the number of grid points $m\in\{ 25^2,50^2,75^2,100^2,125^2,150^2,175^2,200^2 \}$; see the rightmost plot of Figure~\ref{fig-APOGEE-1-10}. We see that the objective value improves a lot as the number of grid points increases from $25\times 25$ and attains a plateau near  $100\times 100$. This justifies our choice of taking a set of $100\times 100$ grid points for denoising this data set.

\begin{figure}[!h]
\centering
\includegraphics[width=0.48\textwidth]{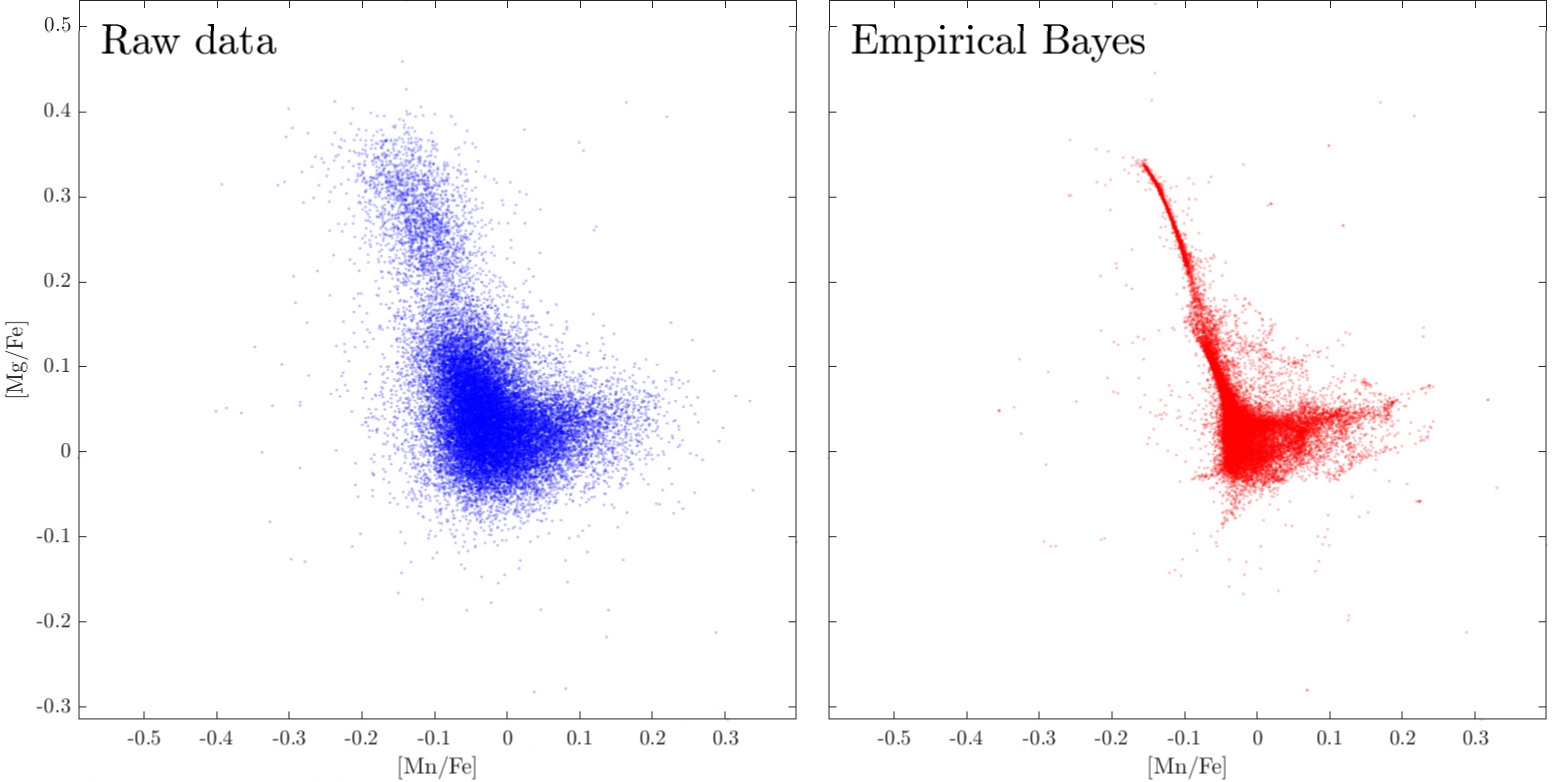}\,\,
\includegraphics[width=0.48\textwidth]{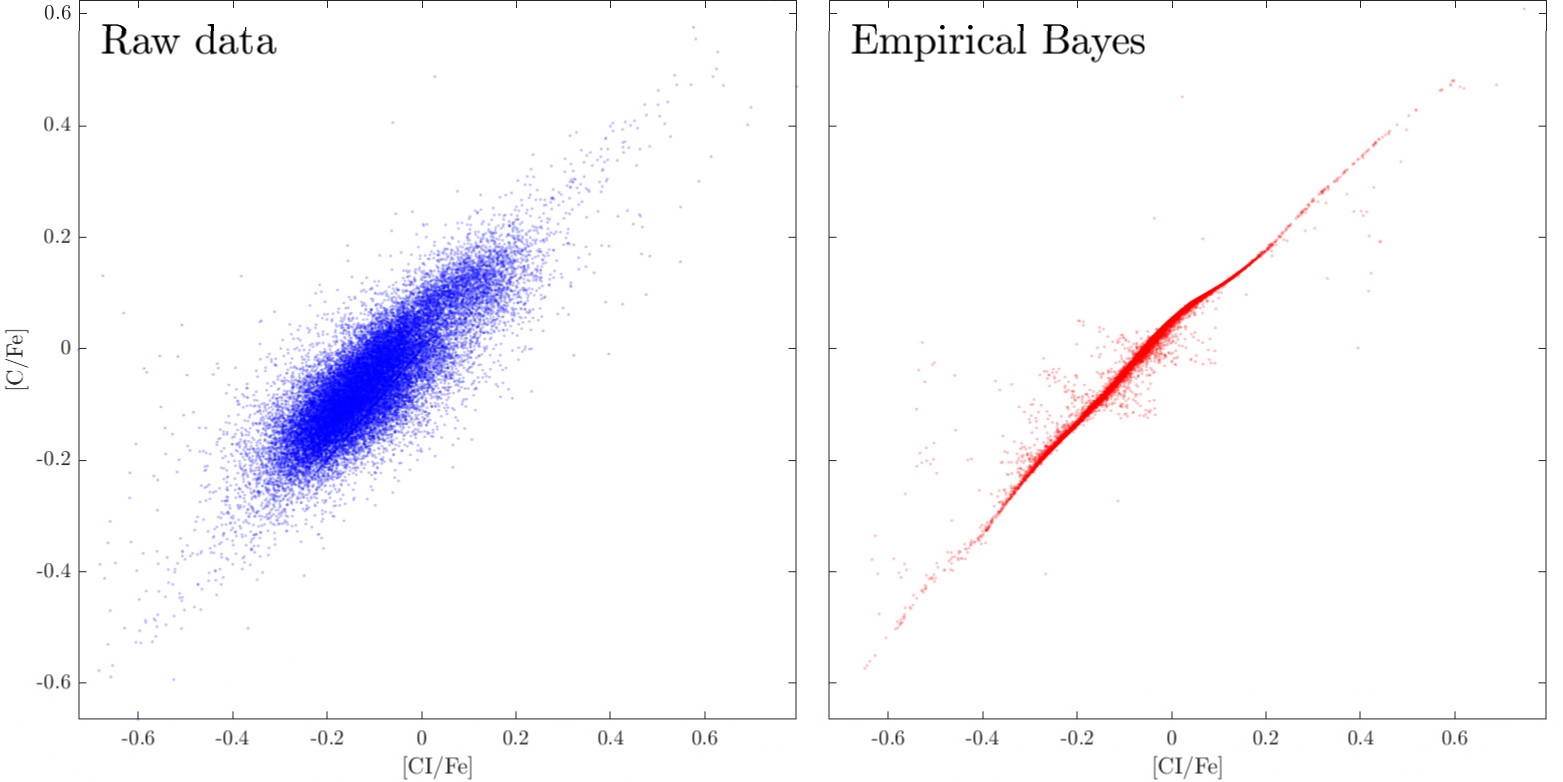}\\
\includegraphics[width=0.48\textwidth]{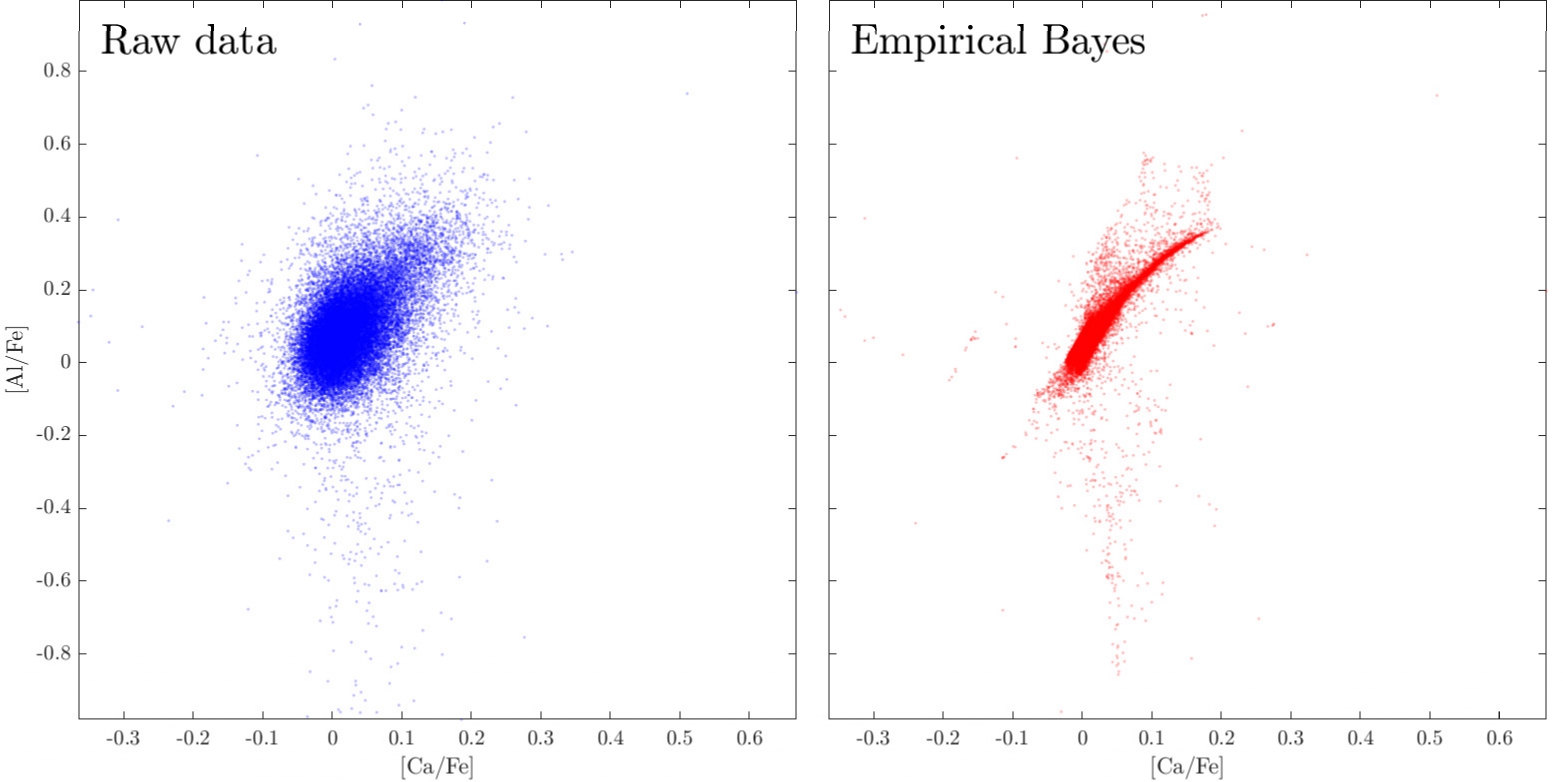}\,\,
\includegraphics[width=0.48\textwidth]{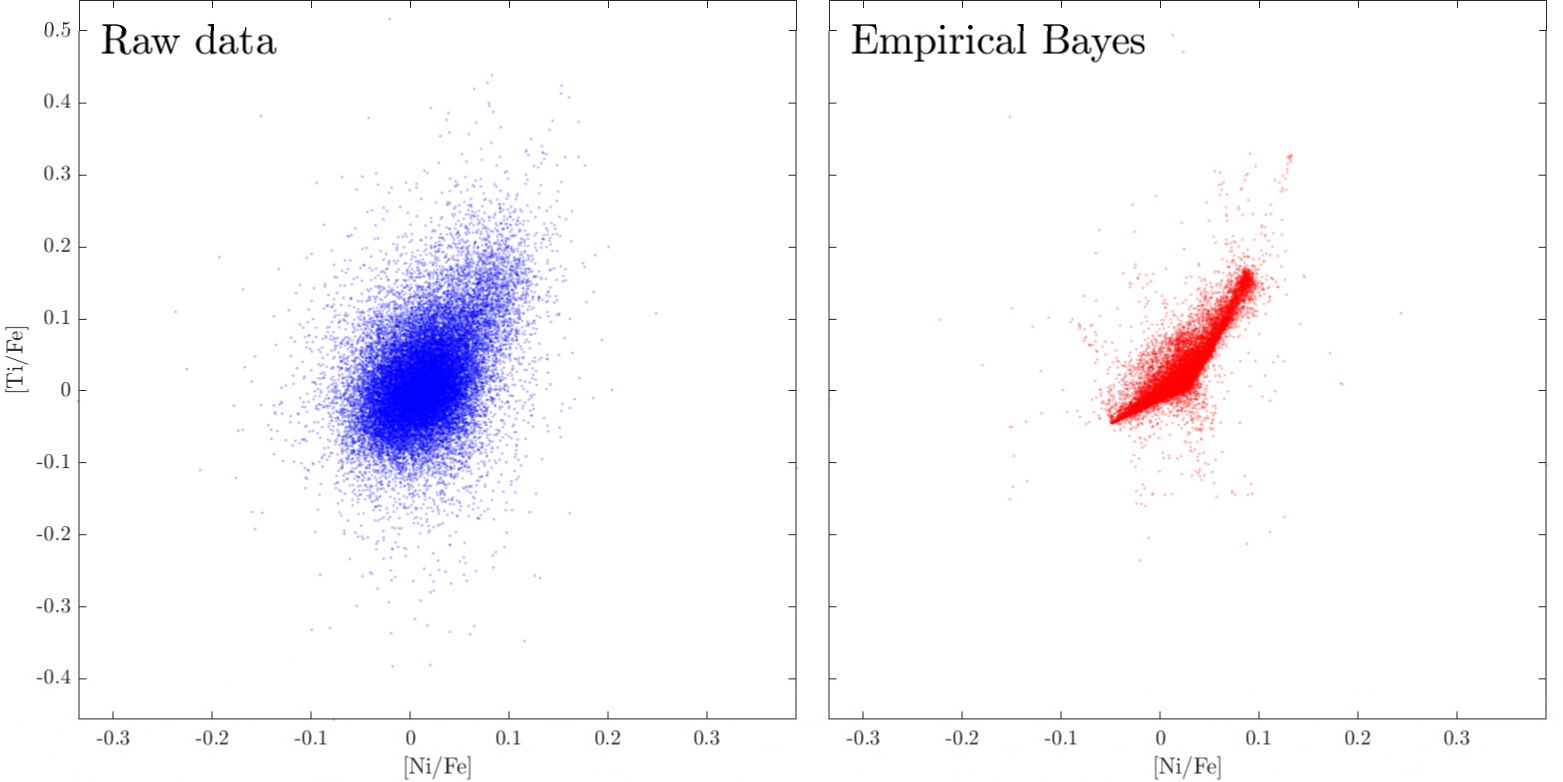}
\caption{\small Results from 4 abundance-abundance data sets (with $d=2$) from the APOGEE survey.  {Observe that the denoised estimates reveal interesting structure not  visible in the raw data plots.}}
\label{fig-APOGEE-add}
\end{figure}

We provide additional two-dimensional abundance-abundance plots along with their denoised empirical Bayes estimates (obtained from a choice of $100\times 100$ grid points) in Figure~\ref{fig-APOGEE-add}. In all the examples we see that the denoised estimates reveal interesting structure not  visible in the raw data plots. In particular, the empirical Bayes estimates of the data in the [C/Fe] and [CI/Fe] plane (see the top right plots in Figure~\ref{fig-APOGEE-add}) reveal very strong (almost linear) association between the two variables.
\begin{table}[htbp]\centering
{
\begin{tabular}{lcccc}
\toprule
Plane & \multicolumn{2}{c}{Time (in sec)} & \multicolumn{2}{c}{Residual} \\
\cmidrule(l){2-3} \cmidrule(l){4-5}
      & ALM  & \texttt{REBayes} & ALM  & \texttt{REBayes}  \\
\midrule
{\rm [Mg/Fe]-[Fe/H]}    &  64.9 & 157.5 & 9.4e-07 & 2.7e-04 \\
{\rm [Mg/Fe]-[Si/Fe]}   &  32.0 & 540.8 & 1.0e-06 & 5.2e-05 \\
{\rm [Mg/Fe]-[Ca/Fe]}   &  27.3 & 526.0 & 1.3e-06 & 6.2e-05 \\
{\rm [Mg/Fe]-[Ti/Fe]}   &  32.6 & 2070.8 & 1.0e-06 & 2.3e-05 \\
{\rm [Mg/Fe]-[Till/Fe]} &  44.4 & 2262.3 & 8.6e-07 & 5.0e-05 \\
\bottomrule
\end{tabular}
}
\caption{\small Numerical performance of ALM  and \texttt{REBayes} for 5 abundance-abundance data sets (with $d=2$) from the APOGEE survey.}
\label{table-APOGEE-time}
\end{table}

\subsection{Numerical results when $d\geq 3$}\label{sec:d3}
In this subsection we aim to investigate the effectiveness of our ALM and the PEM algorithm in approximating the NPMLE when $d\geq 3$. We first introduce our simulation settings.
\begin{figure}
\centering
\includegraphics[width=0.48\textwidth]{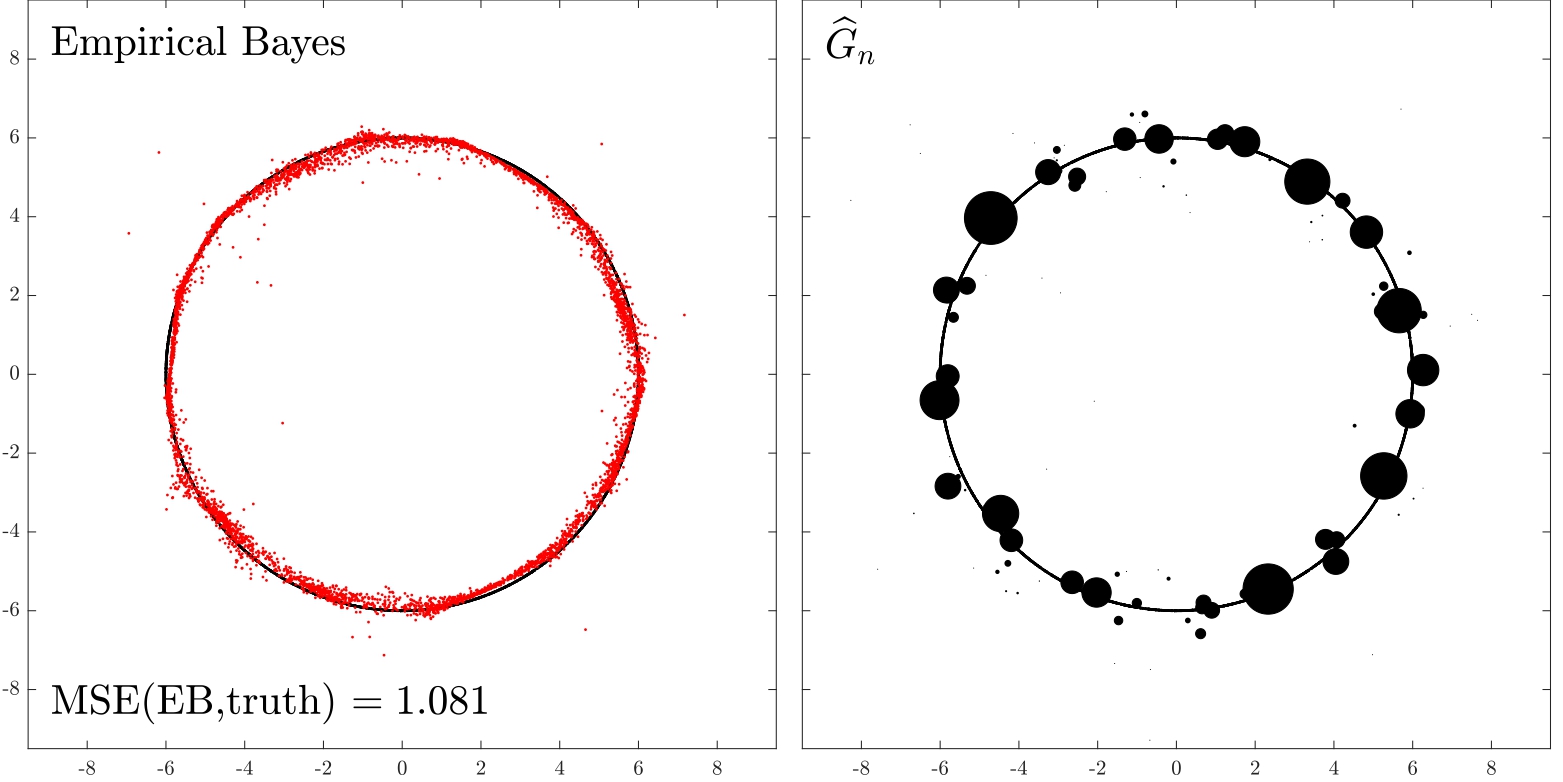}	
\includegraphics[width=0.48\textwidth]{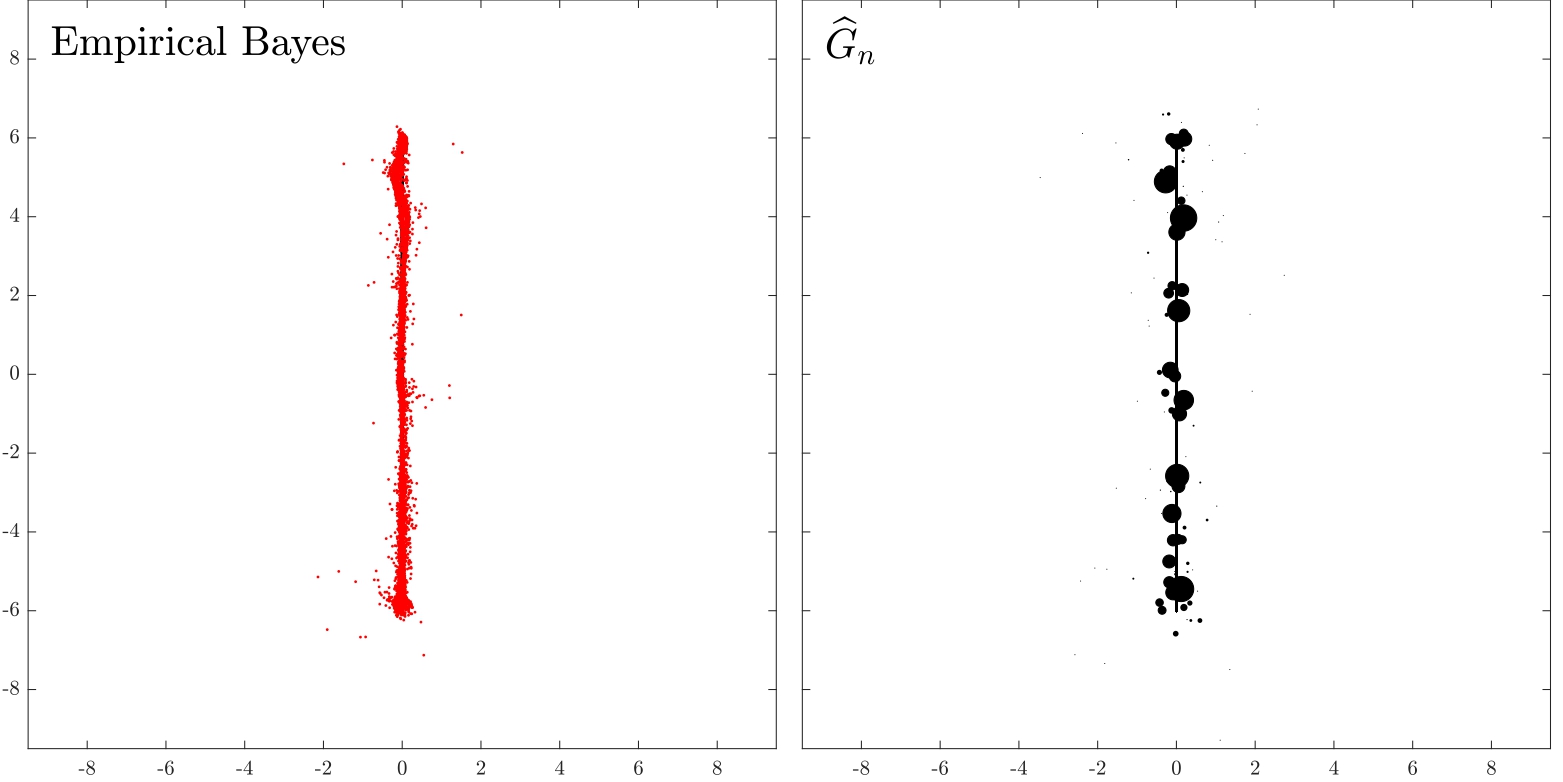}\\

\includegraphics[width=0.48\textwidth]{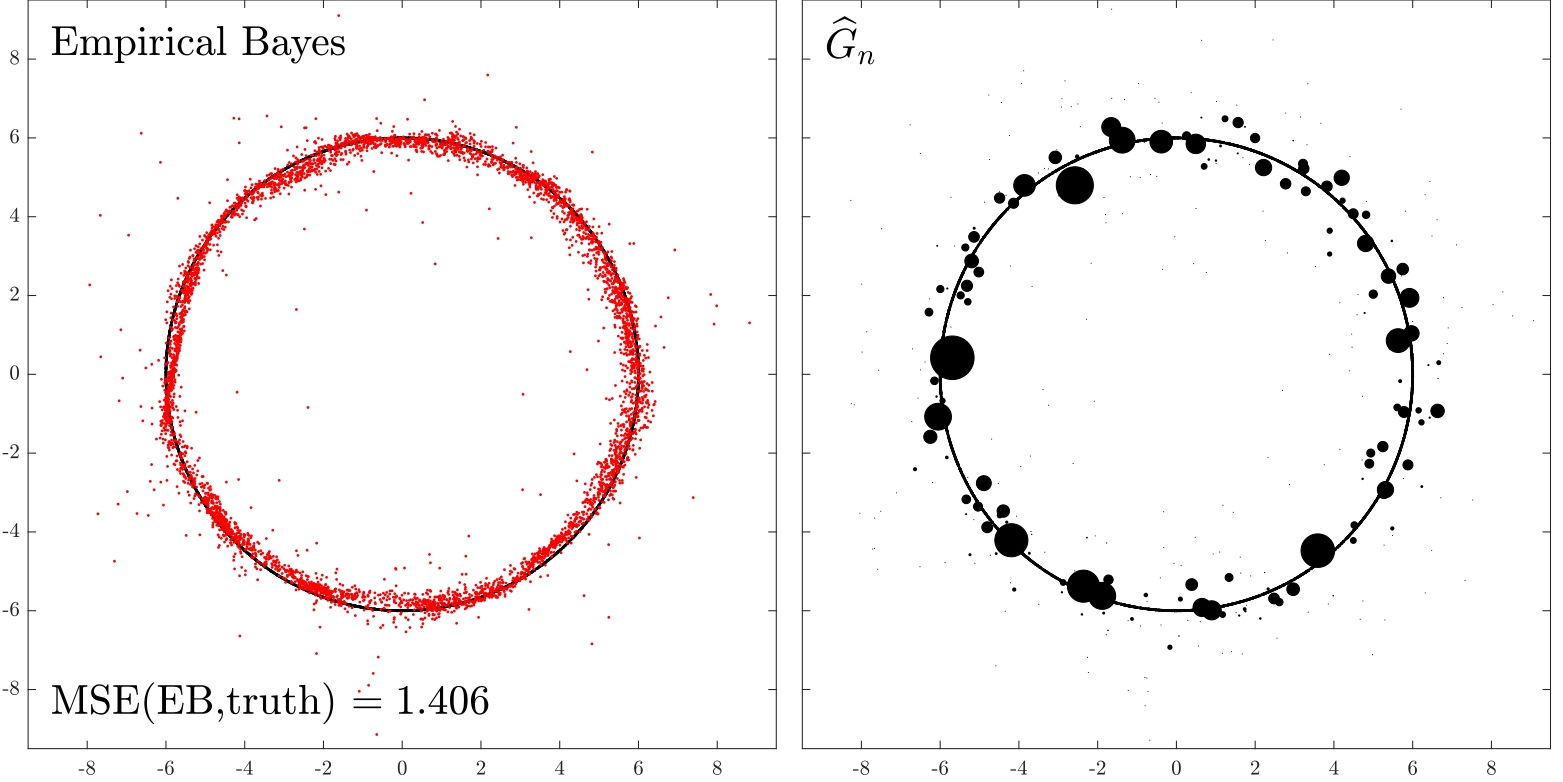}	
\includegraphics[width=0.48\textwidth]{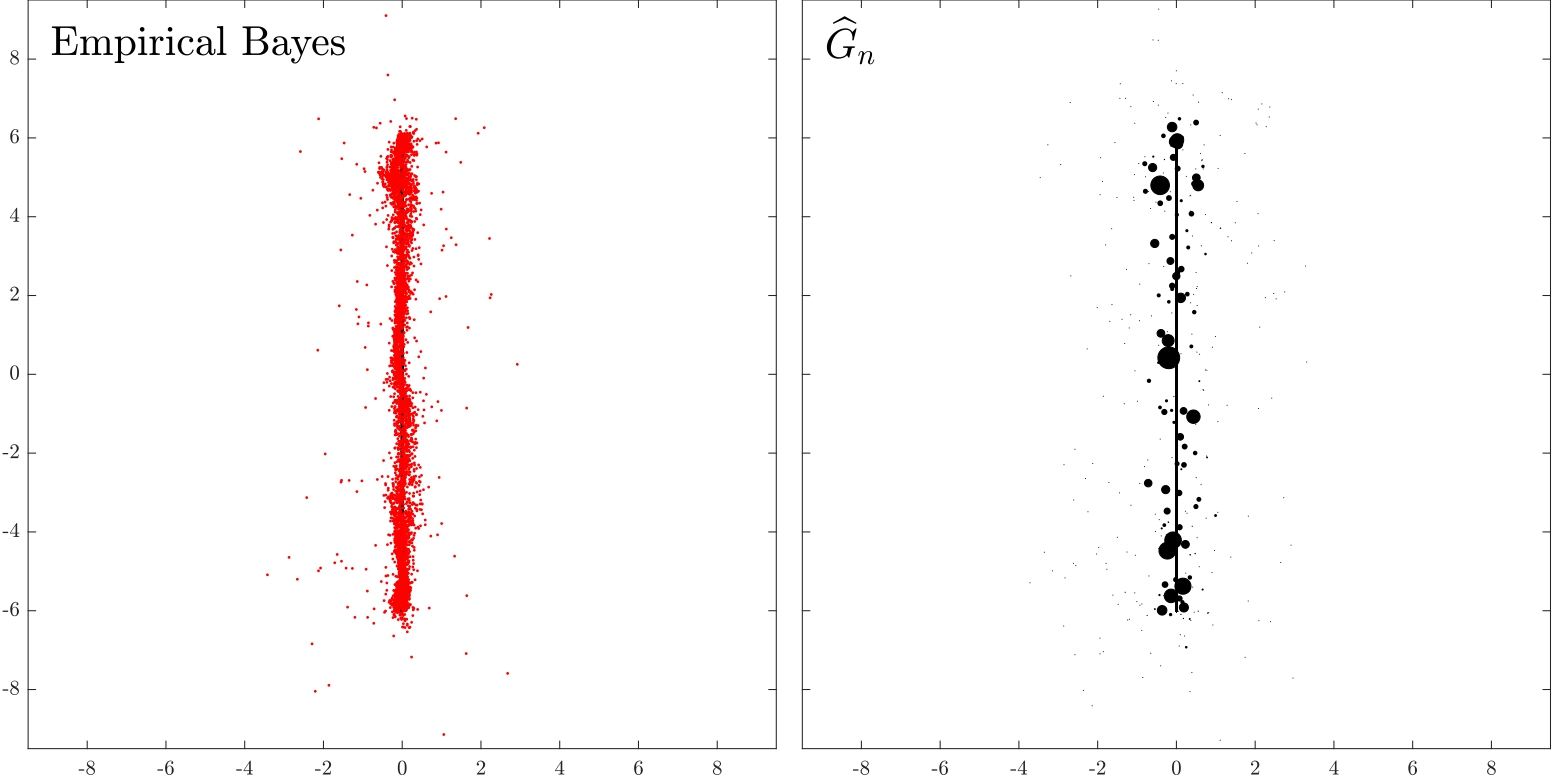}	\\

\includegraphics[width=0.48\textwidth]{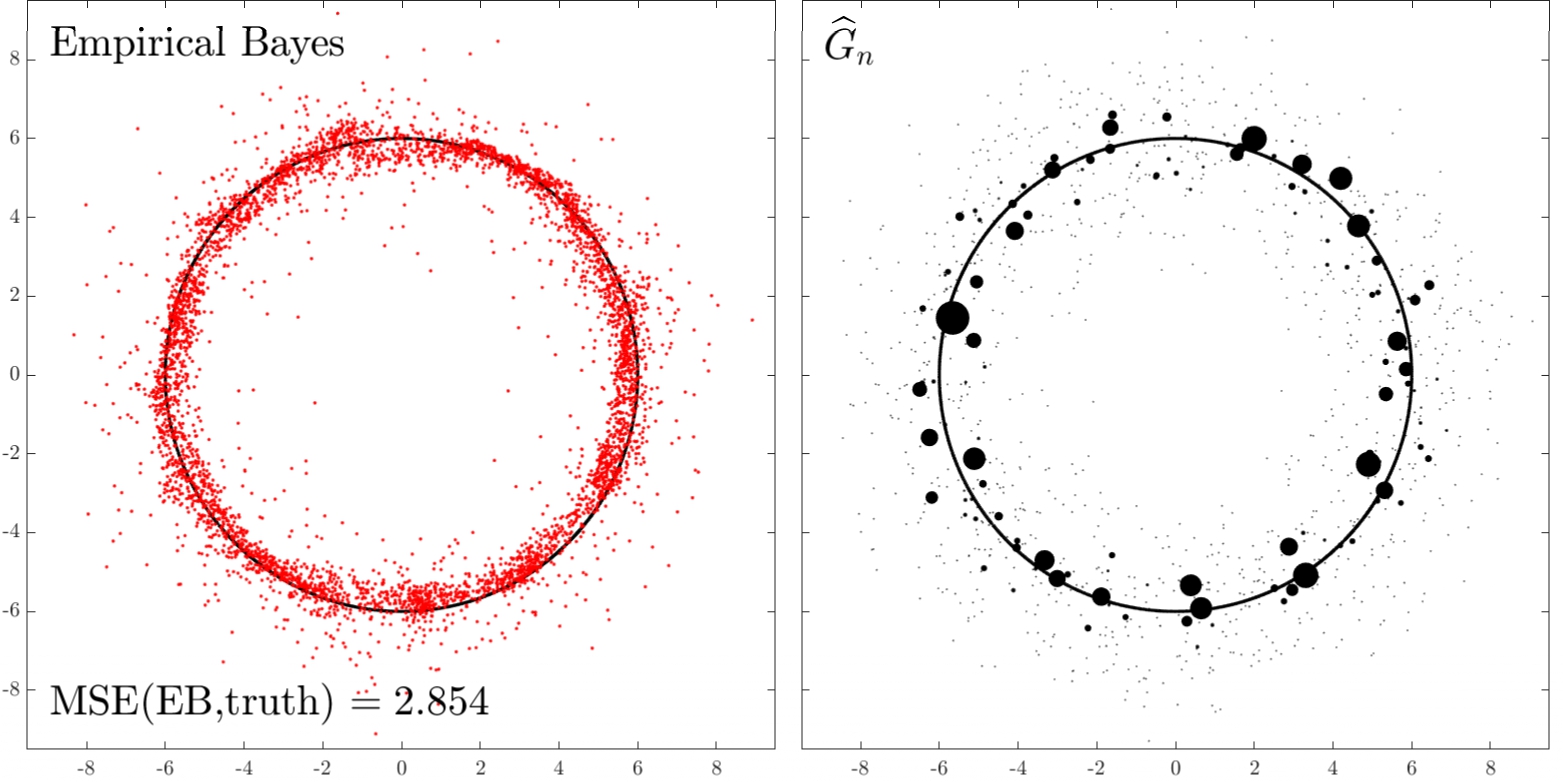}
\includegraphics[width=0.48\textwidth]{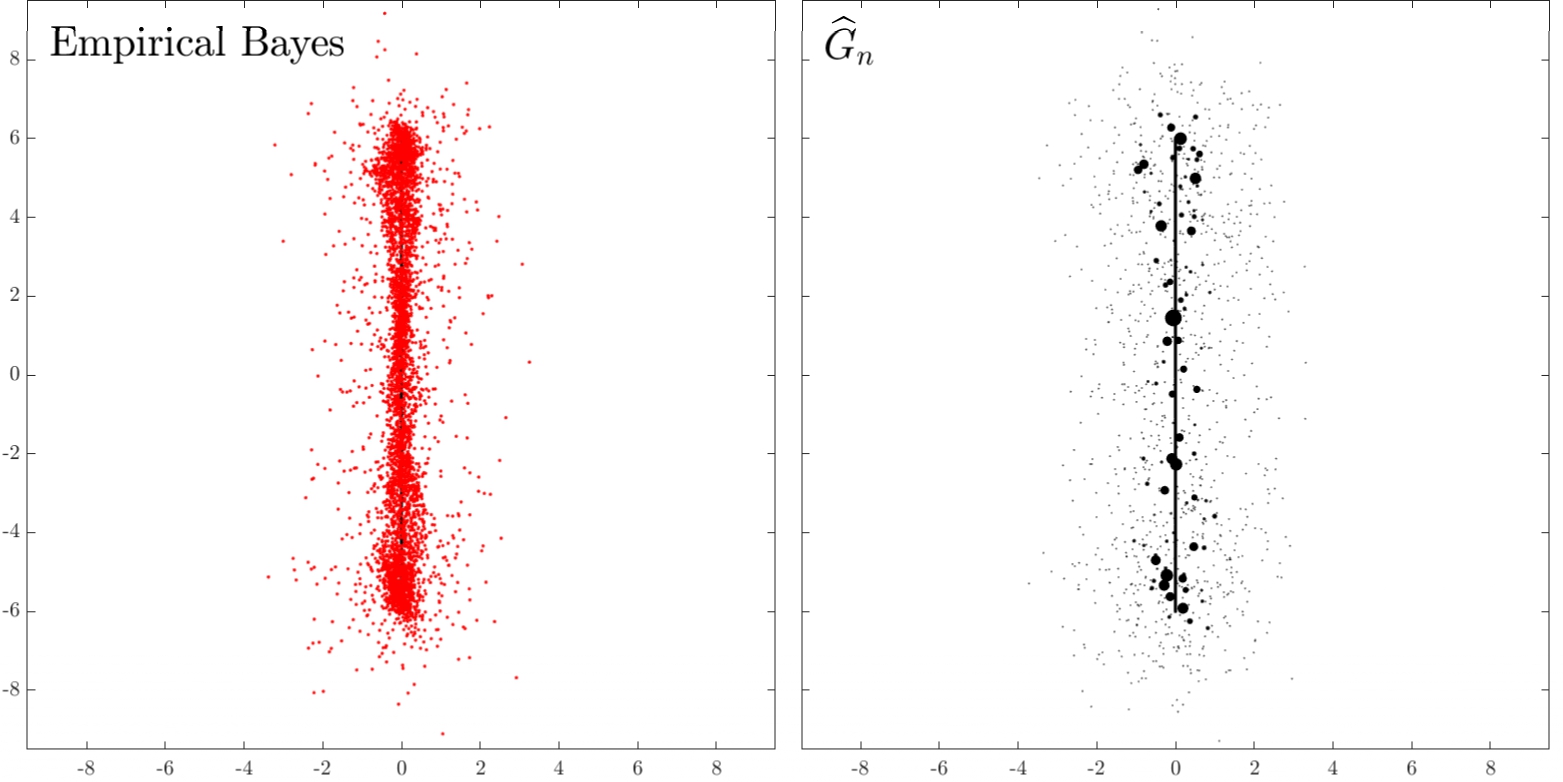}\\

\includegraphics[width=0.48\textwidth]{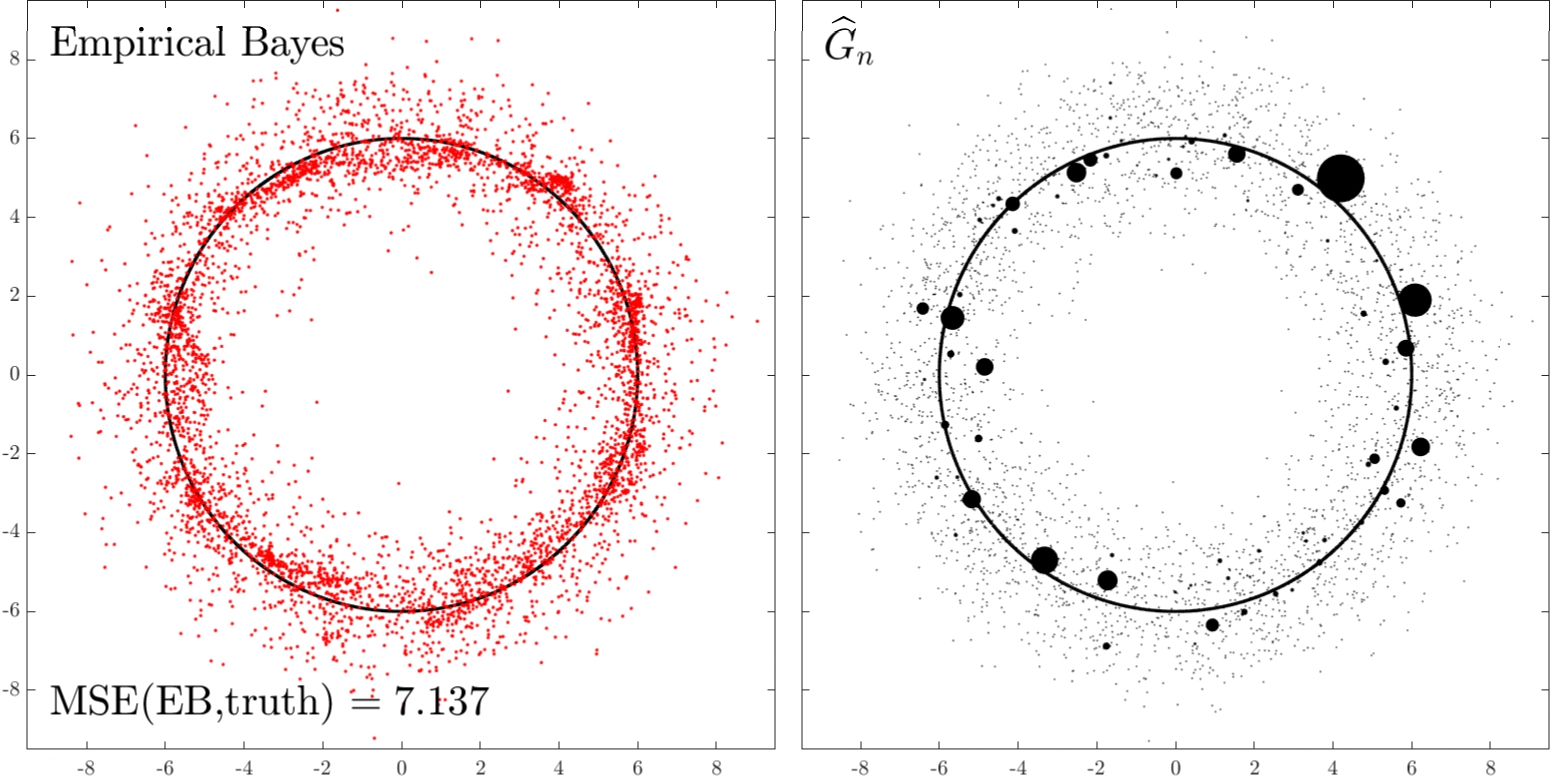}
\includegraphics[width=0.48\textwidth]{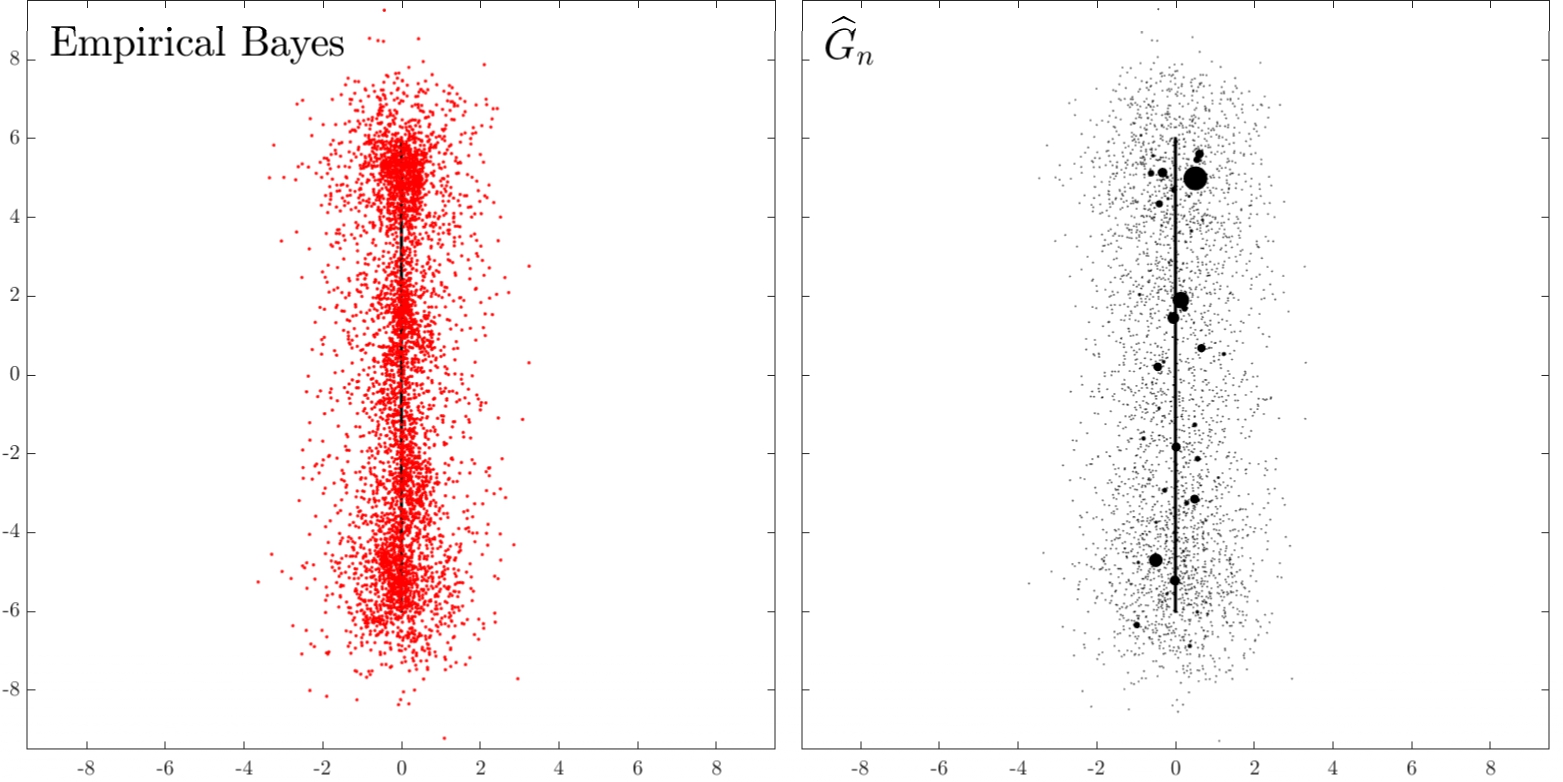}	\\

\caption{\small Plots of the projections of the empirical Bayes estimates (in red), the true $G^*$ (in black), and $\widehat{G}_n$ (in black dots) obtained from our ALM onto 1-2 plane (see columns 1 and 2) and 2-3 plane (see columns 3 and 4) for data obtained from Example 3(a). The four rows correspond to $d=3,6,9$ and $12$ (from top to bottom). Here we take $n= m = 5,000$ and $\mu_i = Y_i$ for all $i=1,\ldots,m$. {Observe that the quality of the empirical Bayes estimates deteriorates as $d$ increases.}}
\label{fig-sim5a-L}
\end{figure}

\vskip 0.2in
\noindent
\underline{\bf Example 3}. We consider the following four simulation settings for model \eqref{denoising data} where $\{\theta_i\}_{i=1}^n \subset \mathbb{R}^d$ is generated as:
\begin{enumerate}
  \item[3(a)] The first two coordinates of $\theta_i \in \R^d$ are drawn uniformly at random from the circle of radius 6 (centered at $(0,0) \in \R^2$), and the remaining entries are set to zero;
  \item[3(b)] $\theta_i = {\bf 0} \in \R^d$, for all $i=1,\dots,n$;
  \item[3(c)] Each $\theta_i$ is generated independently from the discrete distribution $G^*= \frac{1}{3}(\delta_{{\bf e}_1} + \delta_{{\bf e}_2} + \delta_{{\bf e}_3})$, where ${\bf e}_1 = (0,  \dots ,0)\in\mathbb{R}^d$, ${\bf e}_2 = (6, 0, 0, \dots ,0)\in\mathbb{R}^d$,  and ${\bf e}_3 = (0, 6, 0, \dots, 0)\in\mathbb{R}^d$;
  \item[3(d)] $\theta_i \overset{iid}{\sim} {\cal N}({\bf 0}, I_d)$, for $i=1,\dots,n$;
\end{enumerate}
Given the $\theta_i$'s, the observed data are generated independently according to $Y_i \sim {\cal N}(\theta_i, I_d)$, i.e., we consider the homoscedastic setting $\Sigma_i \equiv I_d$ (for simplicity). We set $n=5,000$ and the dimension of the problem $d$ is varied within the set $\{3,4,\dots,12\}$.

\subsubsection{Performance of the ALM when $d \ge 3$}
As we have mentioned earlier, in this multivariate setting the discretization of \eqref{NPMLE1} based on equally spaced grid points in a compact region of $\mathbb{R}^d$ is no longer feasible since the number of such grid points grow exponentially in the number of dimensions $d$, e.g., $m=100^d$. Therefore, when $d\geq 3$ for our ALM we simply take all observations as grid points, i.e., we take $m=n$ and $\mu_i = Y_i$ for all $i=1,\ldots,n$; this approach was advocated in~\cite{lashkari2007convex}. 
From these simulation experiments we make the following observations.

\begin{figure}
\centering
\includegraphics[width=0.48\textwidth]{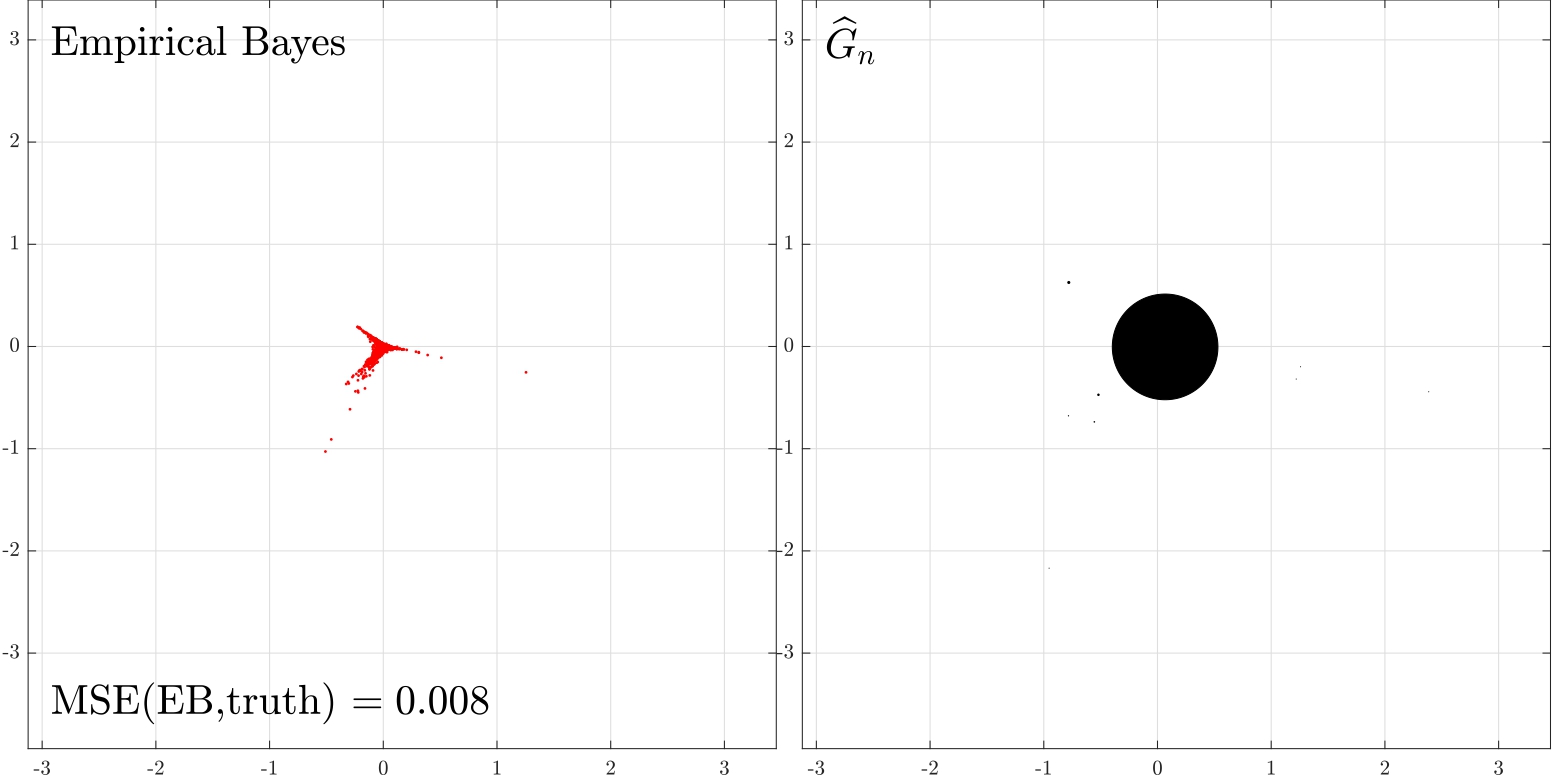}	\,
\includegraphics[width=0.48\textwidth]{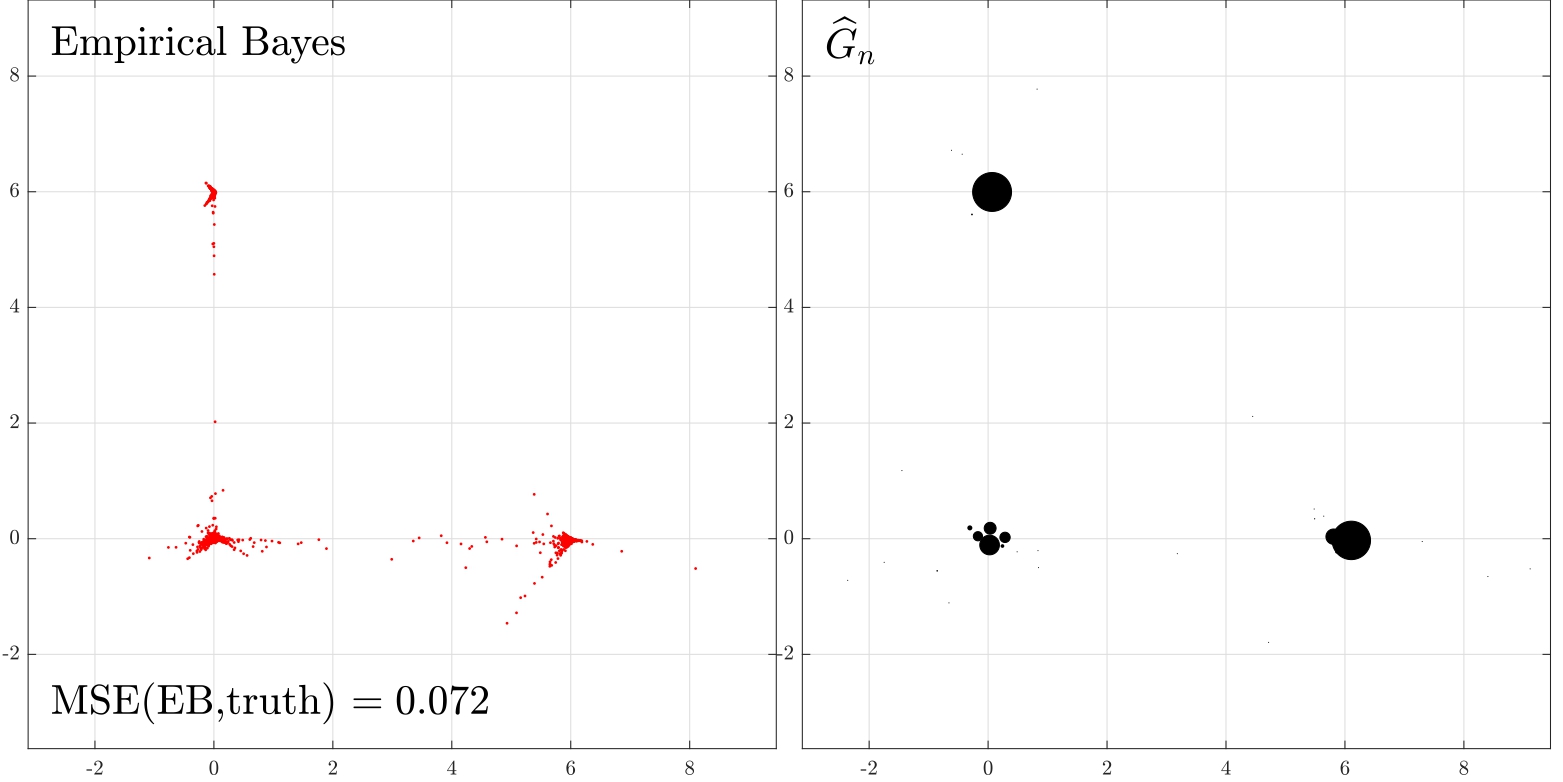}
\\

\includegraphics[width=0.48\textwidth]{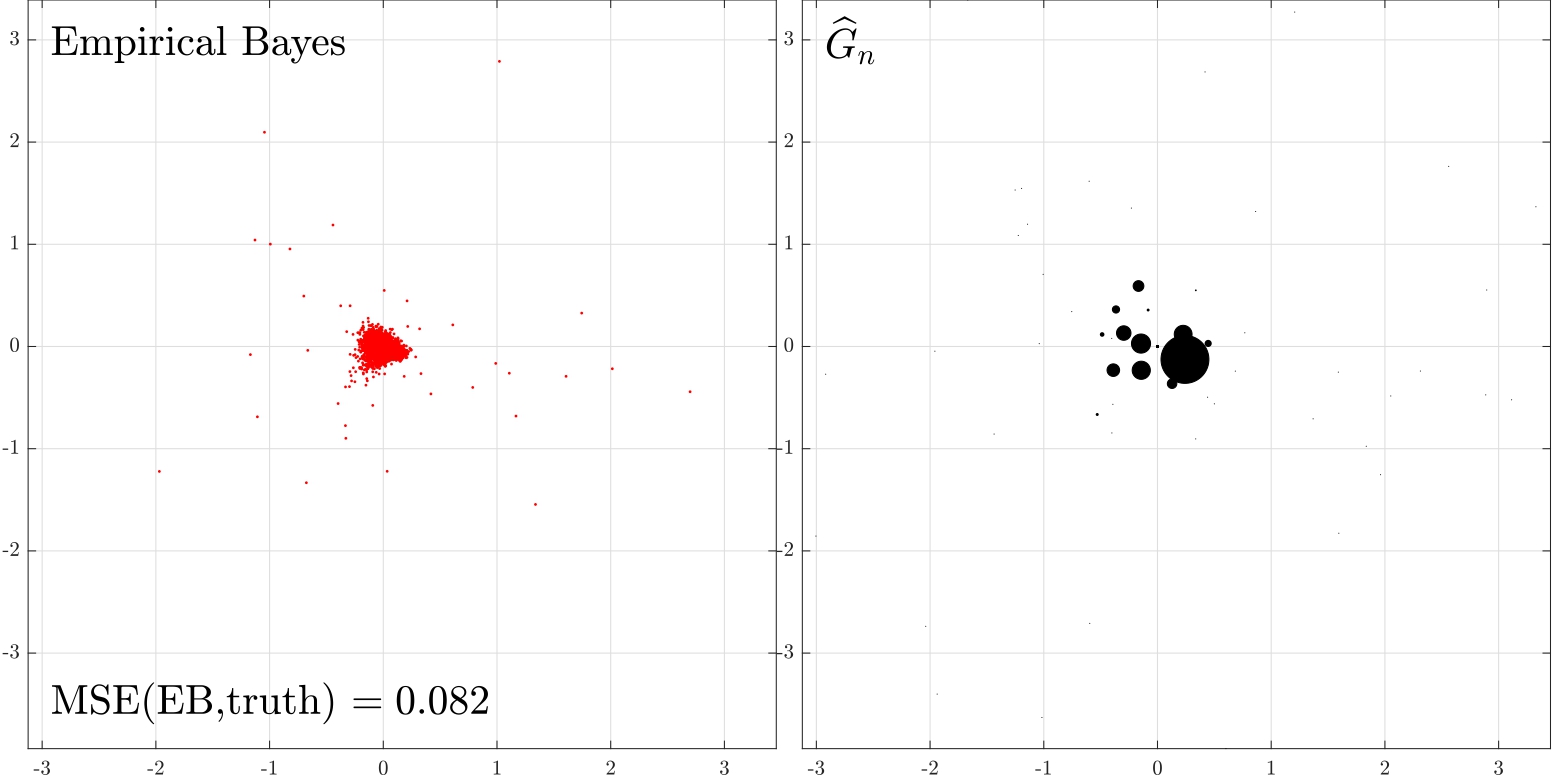}	\,
\includegraphics[width=0.48\textwidth]{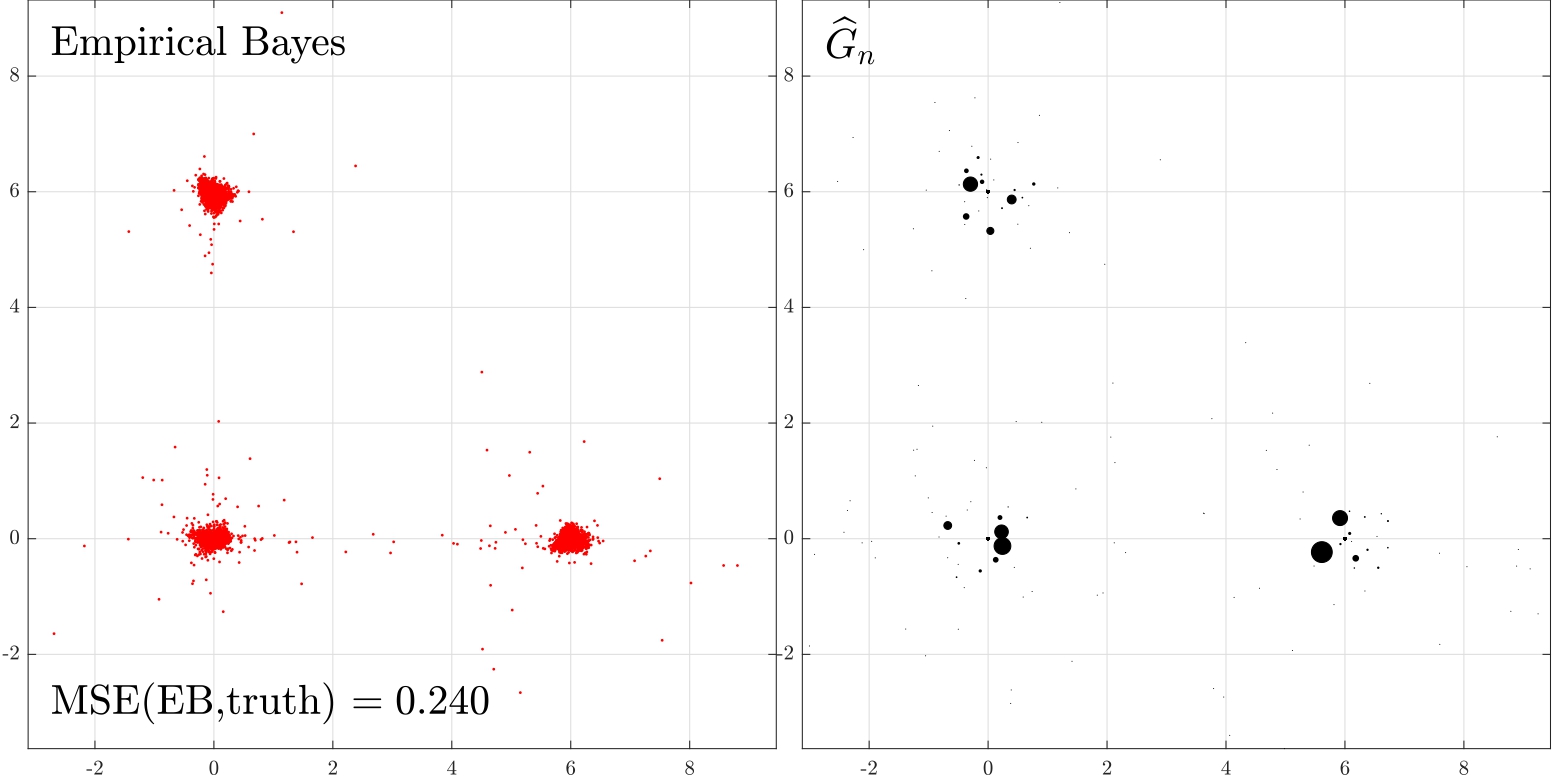}	
\\

\includegraphics[width=0.48\textwidth]{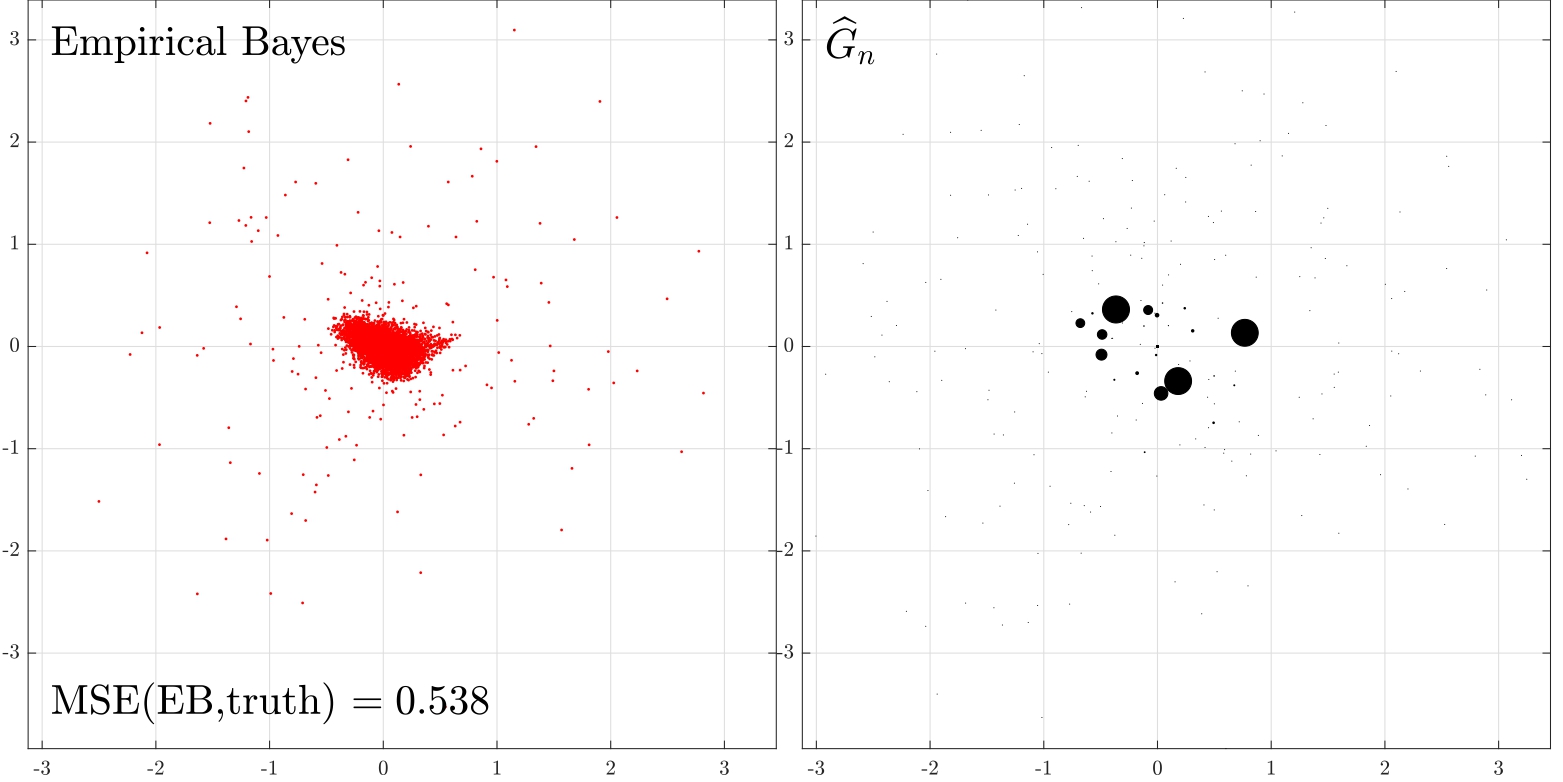} \,
\includegraphics[width=0.48\textwidth]{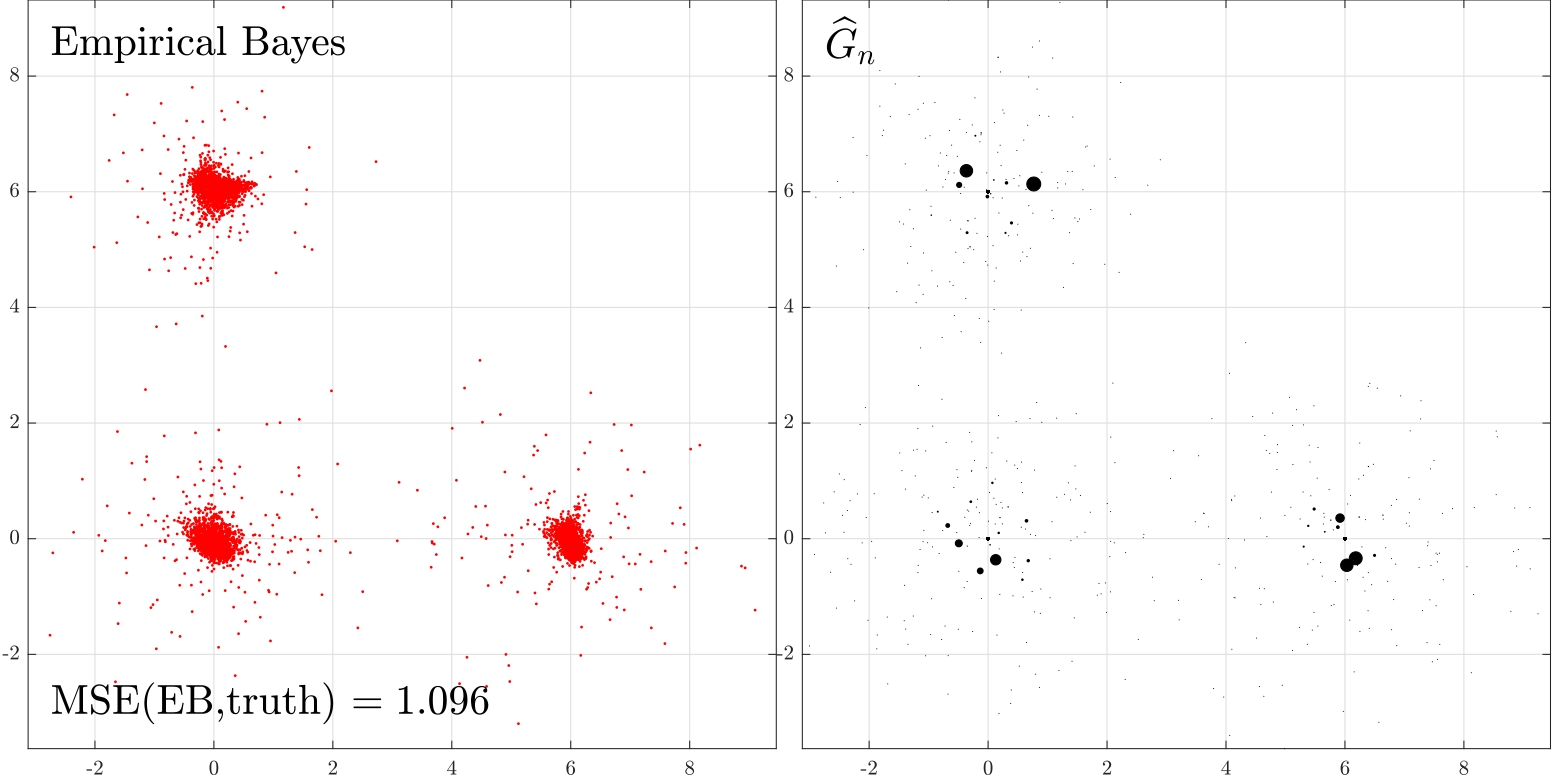}
\\

\includegraphics[width=0.48\textwidth]{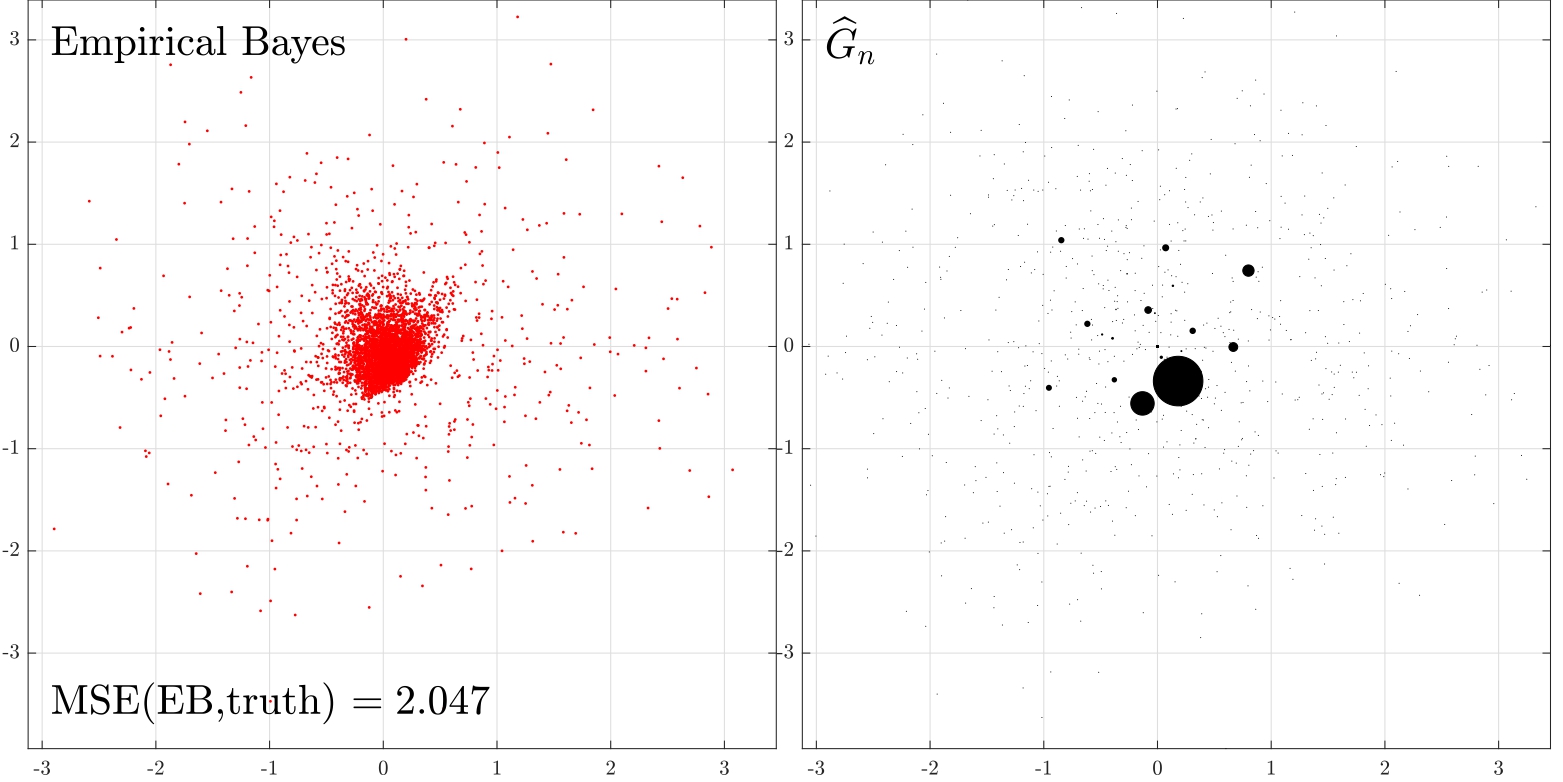} \,
\includegraphics[width=0.48\textwidth]{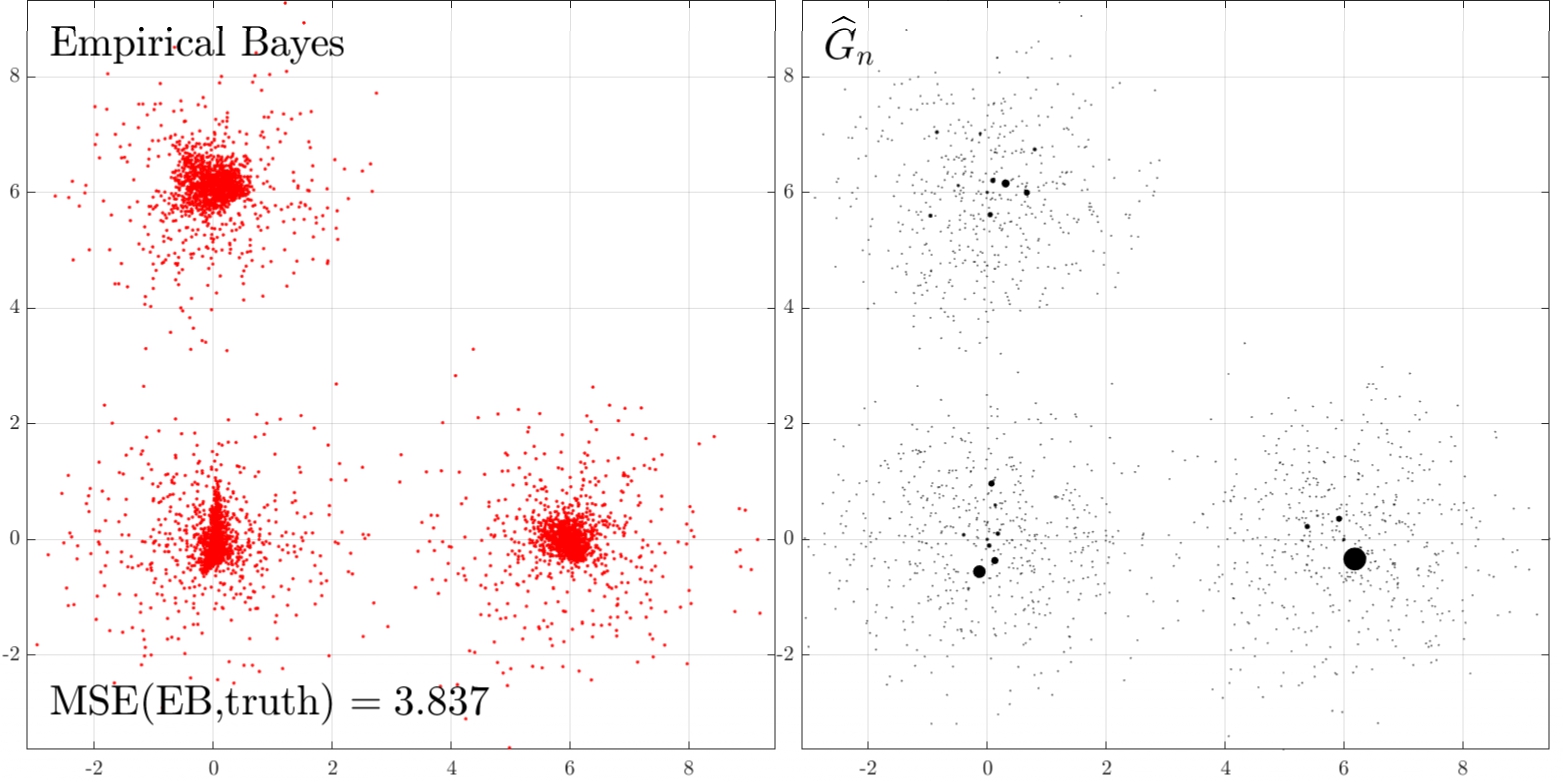}
\\

\caption{\small Plots of the projections of the empirical Bayes estimates (in red) and $\widehat{G}_n$ (in black dots) obtained from our ALM  for data obtained from  Example 3(b) (see columns 1 and 2) and Example 3(c) (see columns 3 and 4). The four rows correspond to $d=3,6,9$ and $12$ (from top to bottom). Here we take $n= m = 5,000$ and $\mu_i = Y_i$ for all $j=1,\ldots,n$.
}
\label{fig-sim5bc-L}
\end{figure}
\vskip 0.1in
\noindent {\bf Behavior of the empirical Bayes estimates as $d$ increases}:  We first illustrate the results for data generated from Example 3(a) using our ALM with the grid points chosen as our data points (here $n=m = 5,000$). Figure~\ref{fig-sim5a-L} displays the projected empirical Bayes estimates onto the first two dimensions and the second and third dimensions for $d \in\{3,6,9,12\}$. The plots indicate that the quality of the empirical Bayes estimates deteriorates as $d$ increases. This phenomenon is also observed in the other simulation settings (see e.g., Figure~\ref{fig-sim5bc-L}) and is intuitively expected since the task of Gaussian denoising gets more difficult as $d$ grows.

\begin{figure}[!h]
\centering
\subfloat[]{\includegraphics[width=0.24\textwidth]{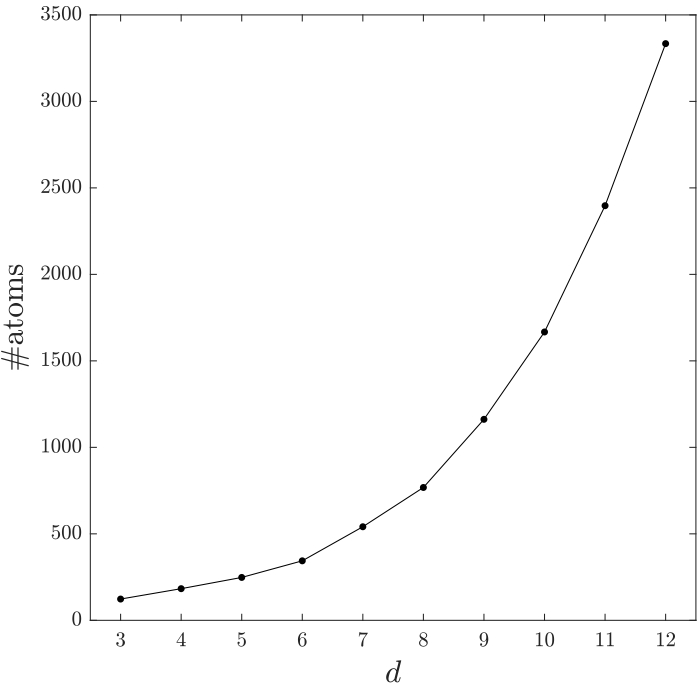}}	
\subfloat[]{\includegraphics[width=0.24\textwidth]{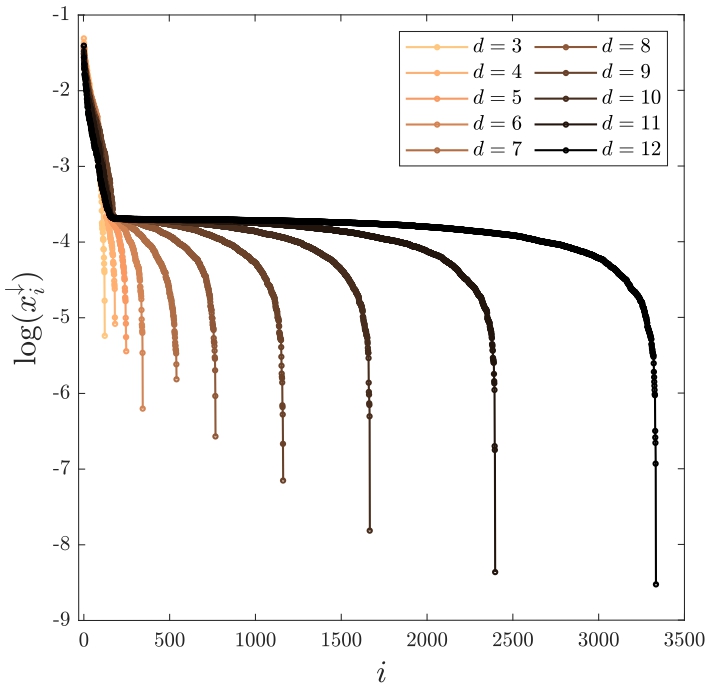}}	
\subfloat[]{\includegraphics[width=0.24\textwidth]{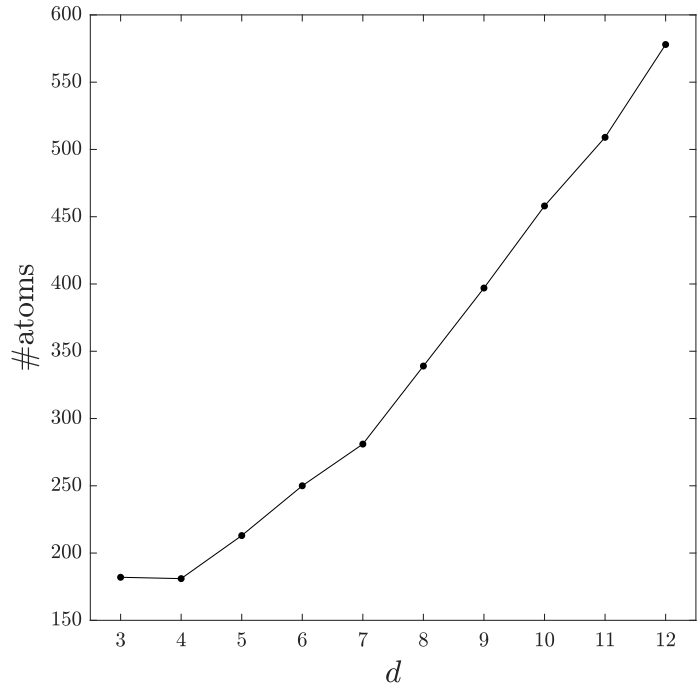}}	
\subfloat[]{\includegraphics[width=0.24\textwidth]{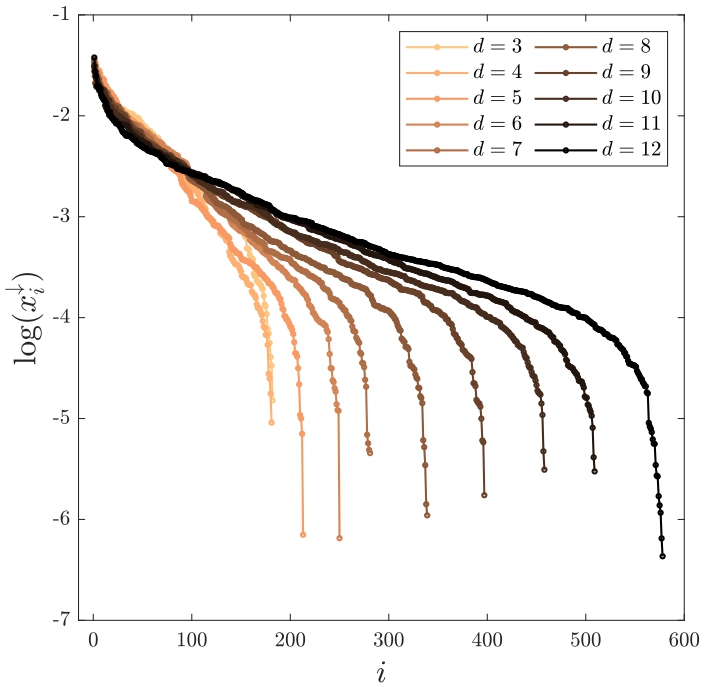}	}
\caption{\small (a) The number of atoms of the estimated $\widehat{G}_n$ (obtained via our ALM) against the dimension $d$; (b) plot of $\log(x^{\downarrow}_{i})$ for different $d$ where $x^{\downarrow}$ consists of the sorted elements of the vector $x$ (see~\eqref{primal-0}) in descending order; (c)-(d) depict similar plots when we incorporate $L_{ii}=0$, for all $i$, in our ALM. All the plots here are obtained from one run from Example 3(a) with $n= m = 5,000$. {Observe that the fitted $\widehat{G}_n$ has more atoms as $d$ grows and that when $d=12$  about 3,400 atoms (out of 5,000 grid points) have nonzero mass and most of them (excluding the largest/smallest 200) have mass approximately $2\times10^{-4}$.}}
\label{fig-sim5-atoms}
\end{figure}

\vskip 0.1in
\noindent {\bf Behavior of $\widehat{G}_n$ as $d$ increases}: Additionally, we observe from the plots in Figure~\ref{fig-sim5a-L} that the estimated $\widehat{G}_n$ (obtained via our ALM) has more atoms (support points) as $d$ grows. To further illustrate this phenomenon we plot (for this data example) the number of atoms, i.e., the number of nonzero entries of the solution $x$ to problem \eqref{primal-0}, against the dimension $d$ in Figure~\ref{fig-sim5-atoms}(a) and the weights on the atoms in Figure~\ref{fig-sim5-atoms}(b). In particular, we observe that when $d=12$  about 3,400 atoms (out of 5,000 grid points) have nonzero mass and most of them (excluding the largest/smallest 200; see Figure~\ref{fig-sim5-atoms}(b)) have mass approximately $2\times 10^{-4}$.

\begin{figure}[!h]
\centering
\includegraphics[width=0.947\textwidth]{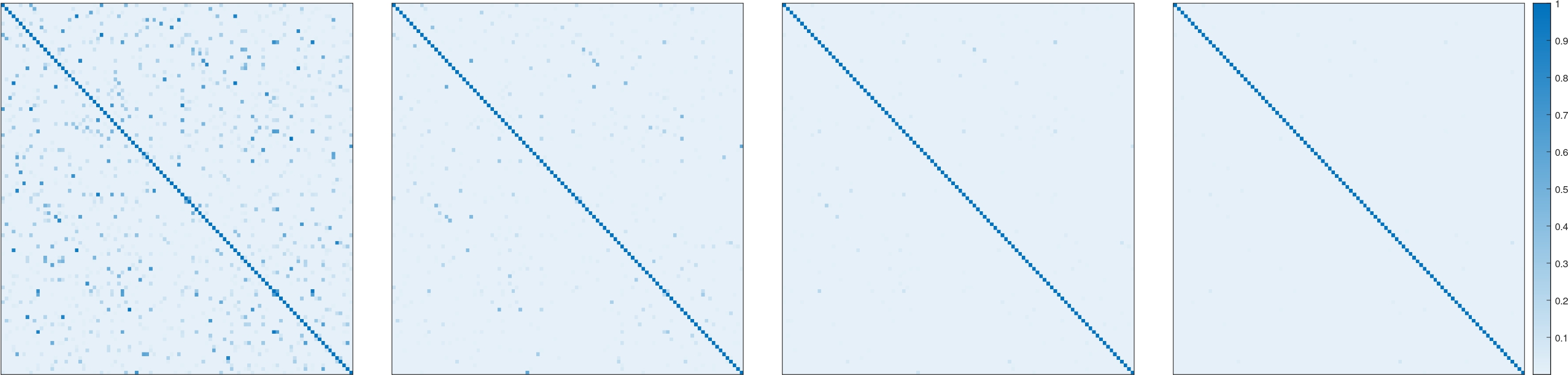}	\\[0.5cm]
\includegraphics[width=0.95\textwidth]{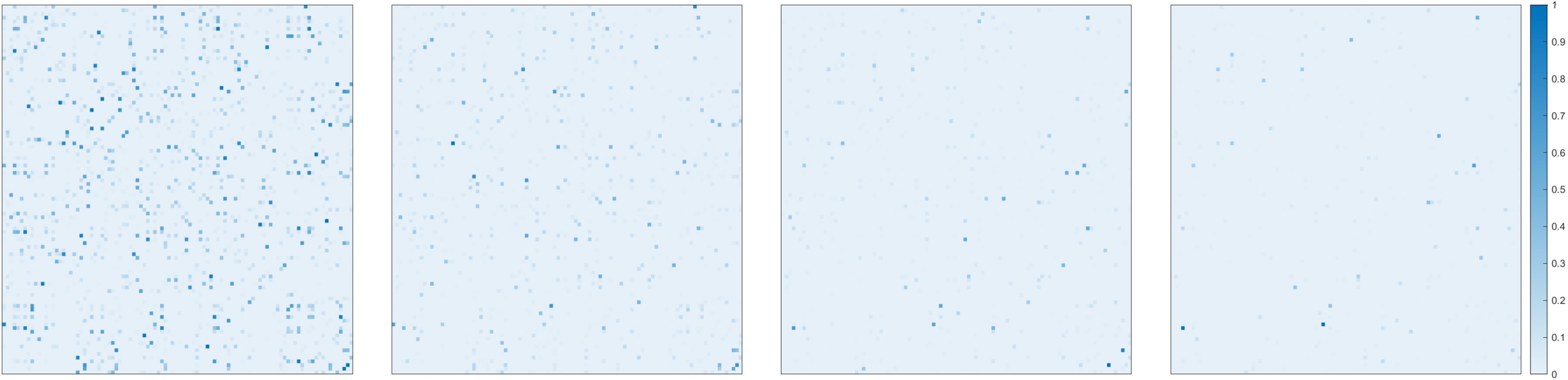}	
\caption{\small The plots in the first row show the heatmap of the sub-matrix $L_{\mathcal{J},\mathcal{J}}$, obtained from data from Example 3(a), as $d$ varies in $\{3,6,9,12\}$ (from left to right). Here $\mathcal{J}$ is subsampled randomly from $\{1,2,\dots,n\}$ with $| \mathcal{J}|=100$. {The plots in the first row show that the matrix $L$ is approaching the identity matrix as $d$ increases. The plots in the second row are obtained after enforcing $L_{ii}=0$ for all $i$.}}
\label{fig-sim5-heatmap}
\end{figure}

To explain this behavior of the estimated $\widehat{G}_n$, as $d$ increases, we plot the heatmap of the scaled\footnote{We scale the matrix $L$ such that the maximum entry in each row is one; see Section~\ref{sec:ALM-Implement}.} matrix $L$ computed from data obtained from Example 3(a) for $d=3,6,9,12$ in the first row of Figure~\ref{fig-sim5-heatmap}. In the heatmap\footnote{For better visualization we only show the heatmap of $L_{\mathcal{J},\mathcal{J}}$ --- the submatrix of $L$ with rows and columns restricted to a randomly sampled index set $\mathcal{J} \subseteq \{1,\ldots, 5000\}$ with $|\mathcal{J}|=100$.}, the values of entries in the matrix $L$ are represented by colors in each square. We can infer from the first row of Figure~\ref{fig-sim5-heatmap} that the matrix $L$ is approaching the identity matrix as $d$ increases. This is because the diagonal entries $L_{ii} = \phi_{\Sigma_i}(Y_i - \mu_i) = \phi_{\Sigma_i}({\bf 0})$ dominate the off-diagonal entries $L_{ij}=\phi_{\Sigma_i}(Y_i - \mu_j)$ (for $i\neq j$) which are typically much smaller, as most points are far from each other in high dimensions. Note that when $L=I_m$, we know that $x=\frac{1}{m}{\bf 1}_m$ is the solution to \eqref{primal-0} and thus the number of support points of $\widehat{G}_n$ should be $n = m$. This explains why the number of atoms of $\widehat{G}_n$ is increasing with $d$, and each (non-zero) weight is approaching a fixed value; cf.~Figure~\ref{fig-sim5-atoms}(b).

\begin{figure}
\centering
\includegraphics[width=0.48\textwidth]{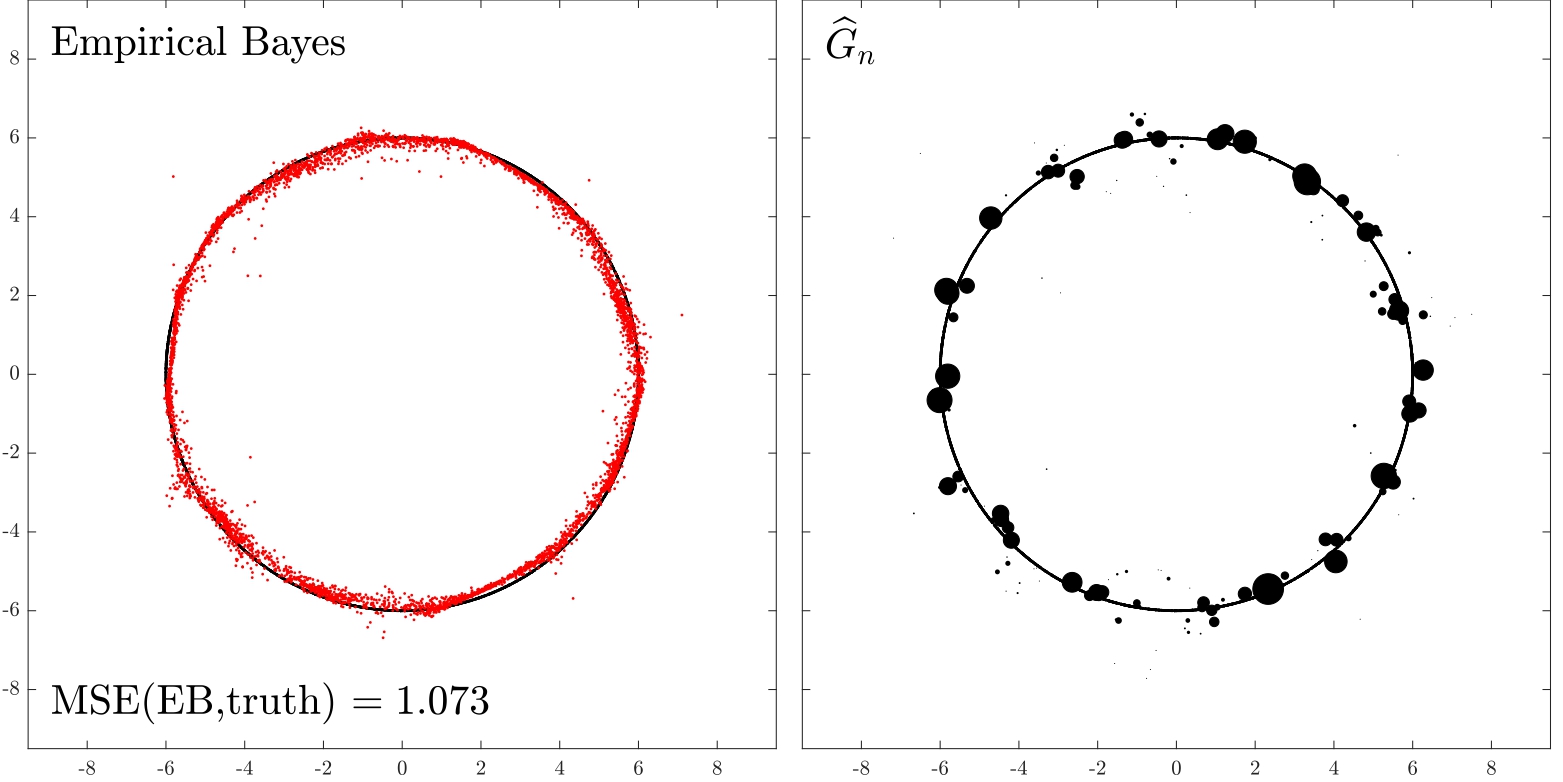}	
\includegraphics[width=0.48\textwidth]{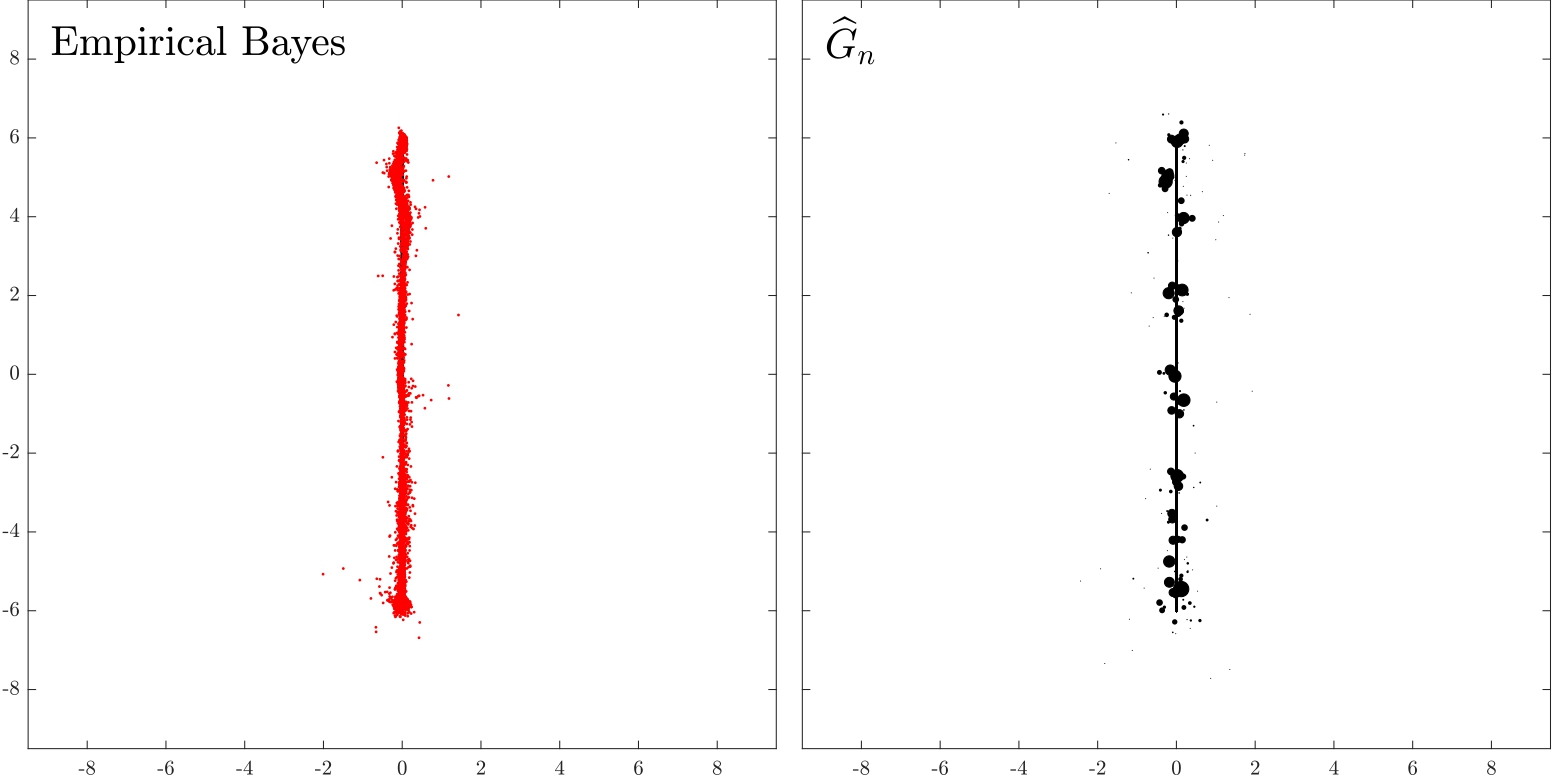}\\

\includegraphics[width=0.48\textwidth]{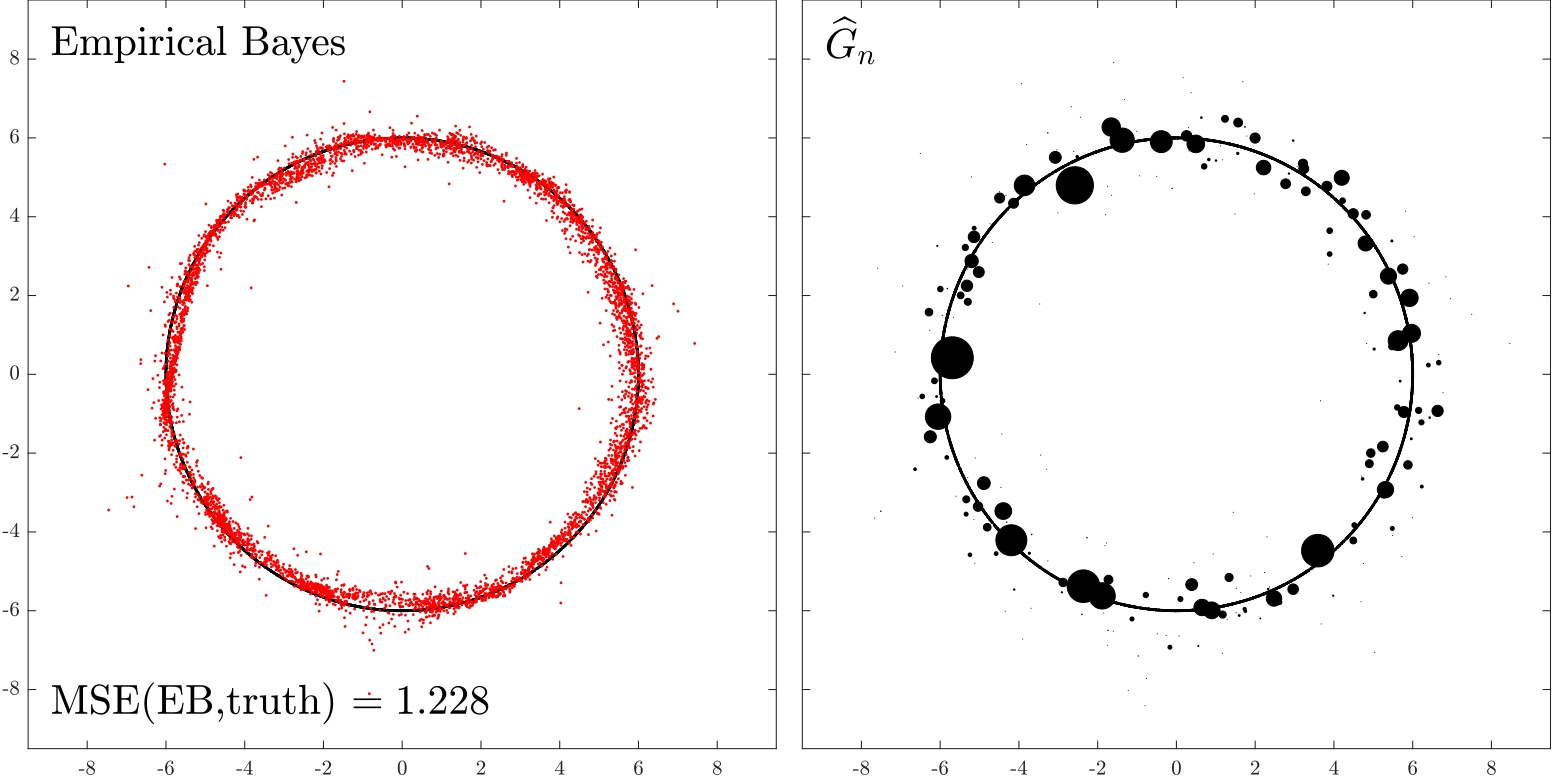}	
\includegraphics[width=0.48\textwidth]{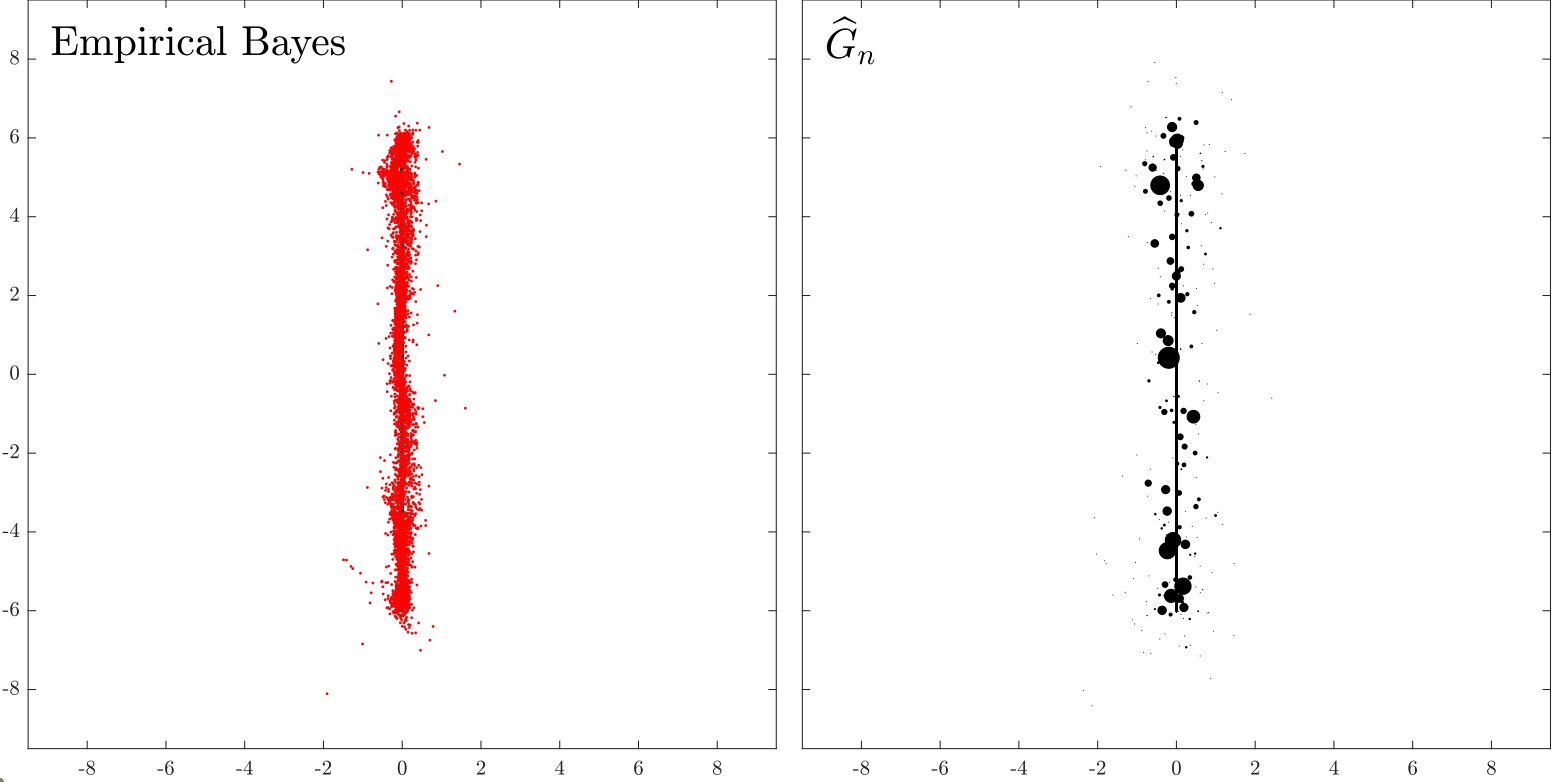}	\\

\includegraphics[width=0.48\textwidth]{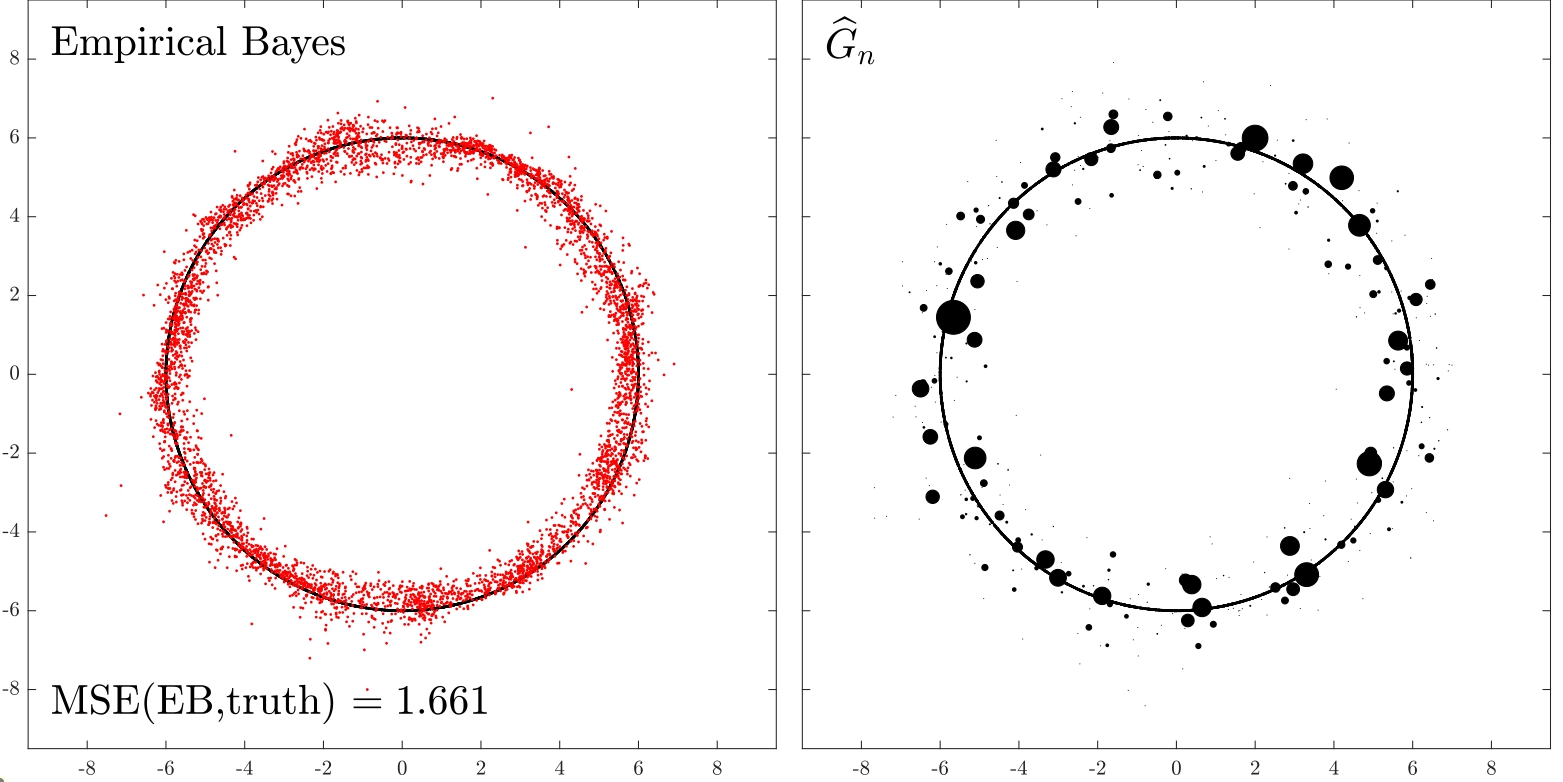}
\includegraphics[width=0.48\textwidth]{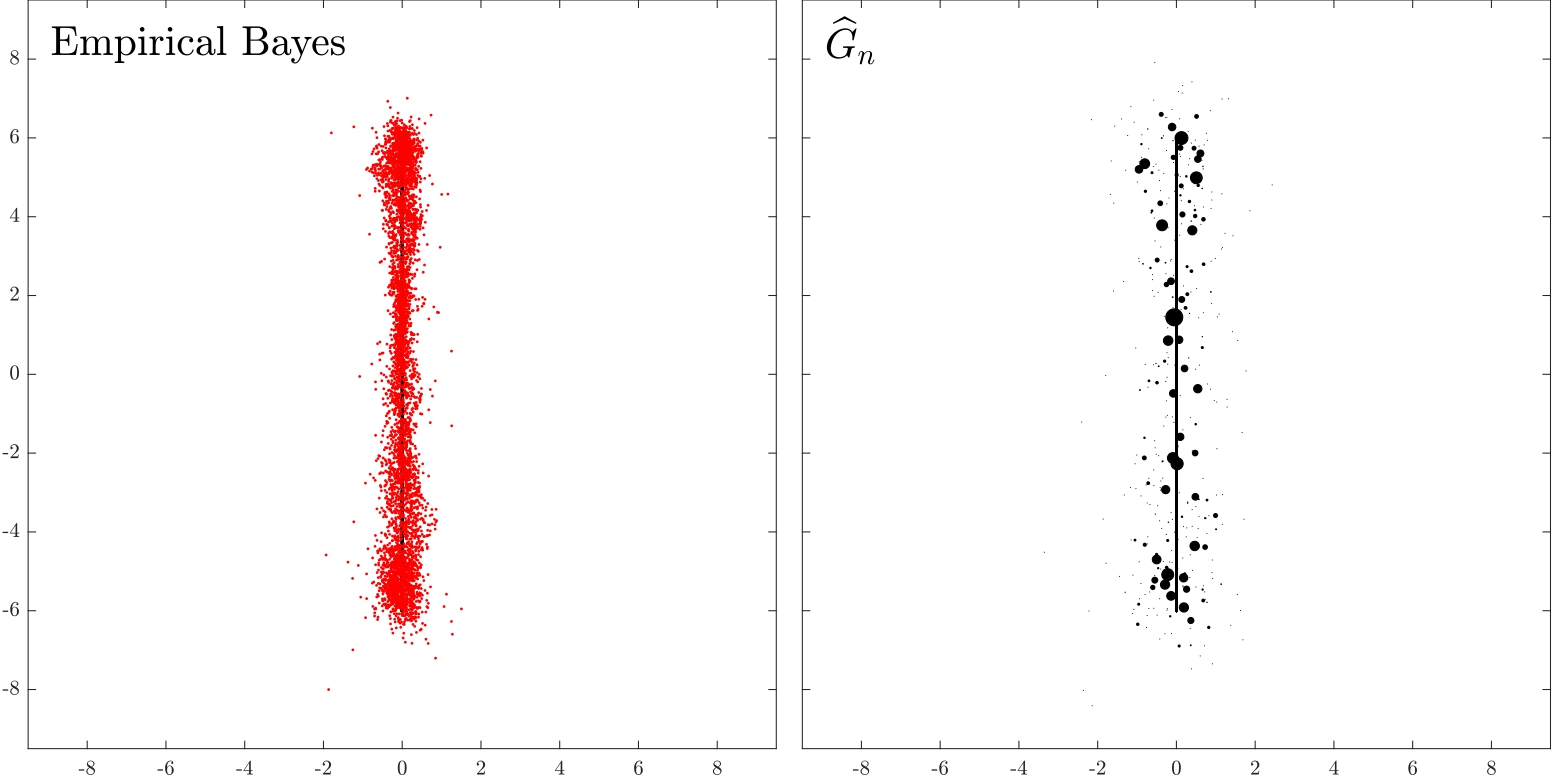}\\

\includegraphics[width=0.48\textwidth]{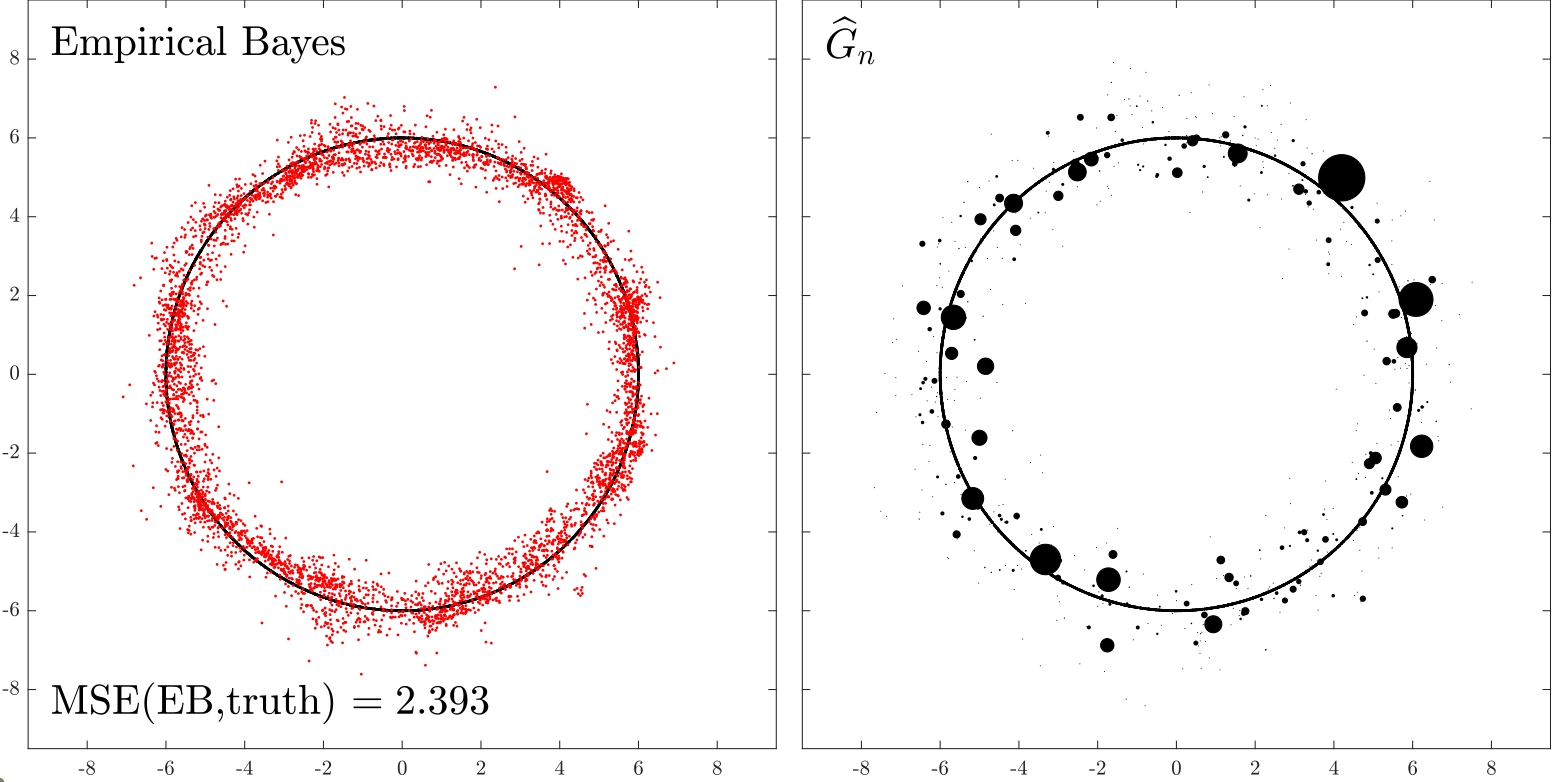}
\includegraphics[width=0.48\textwidth]{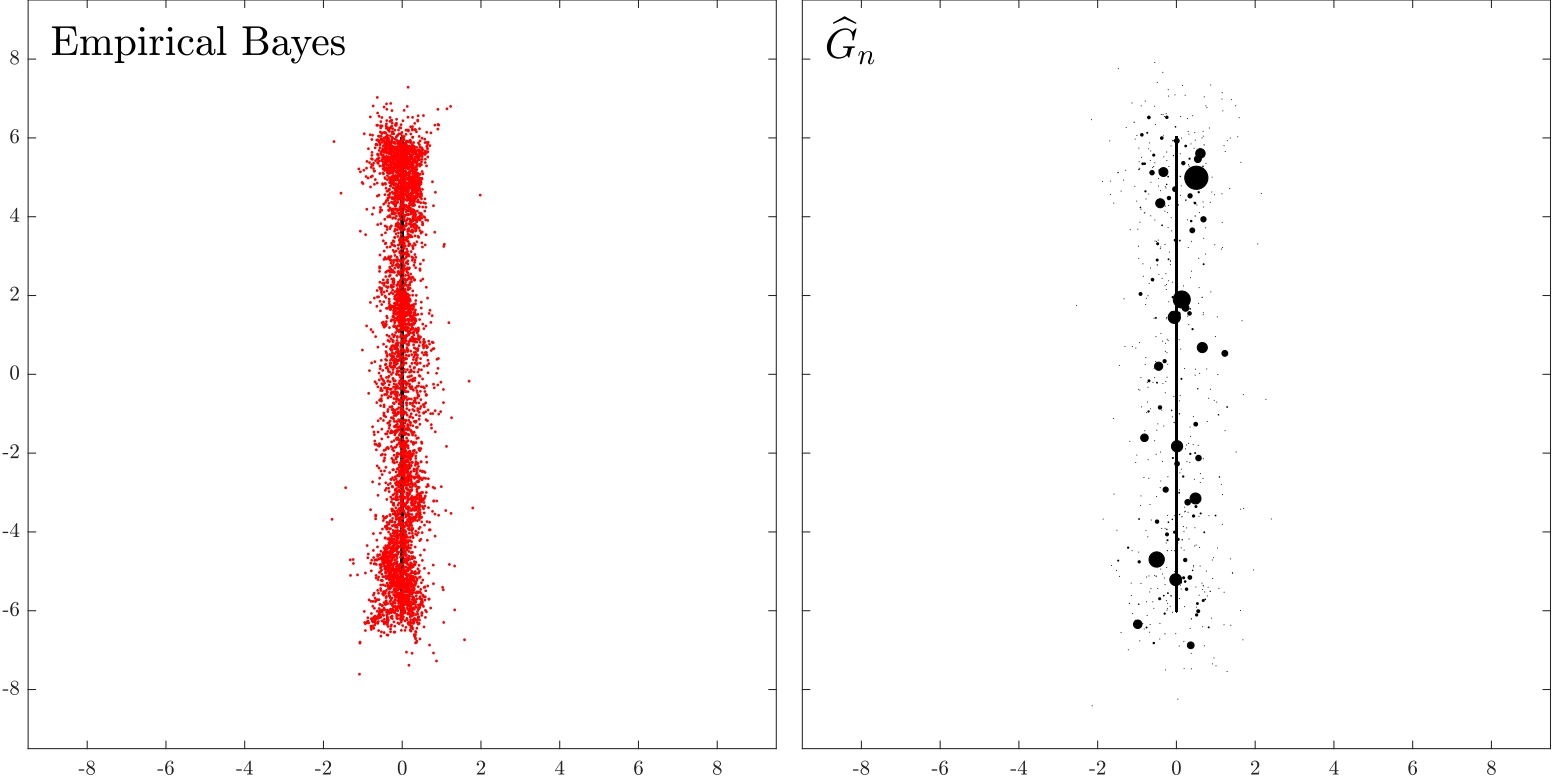}	\\

\caption{\small Plots of the projections of the empirical Bayes estimates (in red), the true $G^*$ (in black), and $\widehat{G}_n$ (in black dots) obtained from our ALM onto 1-2 plane (see columns 1 and 2) and 2-3 plane (see columns 3 and 4) for data obtained from Example 3(a) where we enforce $L_{ii}=0$, for all $i=1,\ldots, n$. The four rows correspond to $d=3,6,9$ and $12$ (from top to bottom). Here we take $n= m = 5,000$ and $\mu_i = Y_i$ for all $i=1,\ldots,n$.}
\label{fig-sim5a-L0}
\end{figure}

\vskip 0.1in
\noindent {\bf An effective strategy to mitigate this curse of dimensionality}: We introduce the following strategy to slightly modify $L$ when $d\geq 3$ and the support points are taken to be exactly the data points, i.e., $\mu_i = Y_i$ for $i = 1,\ldots,n$. This modification enhances the performance of the obtained $\widehat{G}_n$ and the resulting empirical Bayes estimates. The strategy is to set all the diagonal entries of $L$ to zero, but keep all  the off-diagonal entries of $L$ intact. Namely, we redefine $L$ as:
\begin{equation}\label{set-L}
L_{ij} =
\begin{cases}
\phi_{\Sigma_i}(Y_i - \mu_j)  , & \mbox{if } i\neq j, \\
0, & \mbox{if } i=j.
\end{cases}
\end{equation}
This adjustment mitigates the domination of the diagonal entries over the off-diagonal entries in $L$ when $d$ is large (as can be seen from the plots in the second row of Figure~\ref{fig-sim5-heatmap}). We found from our extensive simulation experiments that this simple modification can generally improve the denoising results, especially when $d$ is moderately large (e.g., when $d$ varies between 5 and 15); see Figure~\ref{fig-sim5a-L0} and compare with Figure~\ref{fig-sim5a-L}. Figure~\ref{table-5c-d9} also shows that the ALM, with this adjustment, performs surprisingly well in terms of denoising, compared to the EM and the PEM.

This strategy also yields a $\widehat{G}_n$ with less support points that provides a better estimator of $G^*$; see Figure~\ref{fig-sim5-atoms}(c)-(d) and Figure~\ref{fig-sim5a-L0}. We can see from Figure~\ref{fig-sim5-atoms}(c) that most of the grid points  have zero mass (e.g., only about 600 of the 5,000 grid points have nonzero mass when $d=12$). By comparing Figures~\ref{fig-sim5a-L} and \ref{fig-sim5a-L0}, we can see that for small dimensions (e.g., $d=3$) this tweak does not have much effect, whereas for moderate dimensions (e.g., $d=12$) the effect can be substantial. This strategy dramatically enhances the performance of the obtained empirical Bayes estimates, e.g., when $d=12$ the mean squared error (MSE) between $\widehat{\theta}$ and ${\theta}$, defined as $\frac{1}{n}\sum_{i=1}^{n}\|\widehat{\theta}_i - \theta_i\|^2_2$, equals $2.393$ with this adjustment whereas MSE $= 7.137$ without this tweak (compare Figures~\ref{fig-sim5a-L} and \ref{fig-sim5a-L0}).

\begin{table}[!h]
  \centering
  \setlength{\tabcolsep}{5mm}
\begin{tabular}{llccc}
  \toprule
  Example   &   & ALM & EM & PEM \\
  \midrule
    3(a) & MSE & {\bf  1.0577 } &  1.0639 &  1.0705 \\
 $d=3$ & Log-likelihood & -6.4556 &  -6.4551 & {\bf  -6.4542 } \\
 & Time &{\bf  2.2123 } &  24.6584 &  8.7832 \\
 & Iterations & 16.0    &  18.5    &  5.3    \\
\hline

  3(b) & MSE & {\bf  0.0084 } &  0.0099 &  0.0099 \\
 $d=3$ & Log-likelihood & -4.2673 &  -4.2680 & {\bf  -4.2673 } \\
 & Time &{\bf  3.0083 } &  22.1838 &  5.1697 \\
 & Iterations & 27.9    &  16.9    &  2.0    \\
\hline

  3(c) & MSE & {\bf  0.0667 } &  0.0699 &  0.0719 \\
 $d=3$ & Log-likelihood & -5.3589 &  -5.3592 & {\bf  -5.3586 } \\
 & Time &{\bf  2.1694 } &  23.6469 &  5.1121 \\
 & Iterations & 21.5    &  18.1    &  2.6    \\
\hline

  3(d) & MSE &  1.5413 & {\bf  1.5361 } &  1.5481 \\
 $d=3$ & Log-likelihood & -5.2917 &  -5.2929 & {\bf  -5.2911 } \\
 & Time &{\bf  3.2420 } &  19.1286 &  13.9843 \\
 & Iterations & 22.9    &  14.8    &  3.0    \\

   \hline
     3(a) & MSE & {\bf  1.0875 } &  1.1165 &  1.1236 \\
 $d=4$ & Log-likelihood & -7.8713 &  -7.8649 & {\bf  -7.8643 } \\
 & Time &{\bf  2.0612 } &  31.6009 &  16.9594 \\
 & Iterations & 13.8    &  21.2    &  11.5    \\
\hline

  3(b) & MSE & {\bf  0.0188 } &  0.0239 &  0.0256 \\
 $d=4$ & Log-likelihood & -5.6851 &  -5.6850 & {\bf  -5.6845 } \\
 & Time &{\bf  2.6203 } &  28.1069 &  8.5390 \\
 & Iterations & 23.3    &  19.6    &  3.8    \\
\hline

  3(c) & MSE & {\bf  0.0852 } &  0.0986 &  0.1018 \\
 $d=4$ & Log-likelihood & -6.7763 &  -6.7737 & {\bf  -6.7732 } \\
 & Time &{\bf  1.9809 } &  29.8528 &  13.5334 \\
 & Iterations & 21.1    &  20.9    &  8.8    \\
\hline

  3(d) & MSE & {\bf  2.0773 } &  2.0893 &  2.1022 \\
 $d=4$ & Log-likelihood & -7.0410 &  -7.0380 & {\bf  -7.0365 } \\
 & Time &{\bf  2.2863 } &  35.2624 &  24.0392 \\
 & Iterations & 19.2    &  24.7    &  7.5    \\
%
%
%
\bottomrule
\end{tabular}
  \caption{\small Results from the ALM, EM, PEM on Example 3 (over 10 replications).}
\label{table-EM1}
\end{table}

\subsubsection{Comparison of the ALM, EM, and PEM when $d \ge 3$}\label{subsec: numerical PEM}
In this subsection, we compare the PEM algorithm proposed in Section~\ref{sec:adaptive supports} with the classical EM algorithm and our ALM. Here we adopt the strategy \eqref{set-L} (i.e., setting $L_{ii}=0$ for all $i$) when implementing the ALM for solving \eqref{primal-0}. Table~\ref{table-EM1} compares the performance of the ALM, EM, PEM on Examples 3(a)-(d) with $d=3,4$ and $n =5,000$. The table shows that the PEM always achieves the highest objective value (i.e., largest log-likelihood\footnote{For our ALM the log-likelihood value is computed as $\frac{1}{n}\sum_{i=1}^{n}\log\big(\sum_{j=1}^{m} x_j\phi_{\Sigma_i}(Y_i-\mu_j)\big)$ without setting $L_{ii}=0$.}), and the ALM is the fastest in terms of computational time.

\begin{figure}
\centering
\includegraphics[width=0.3\textwidth]{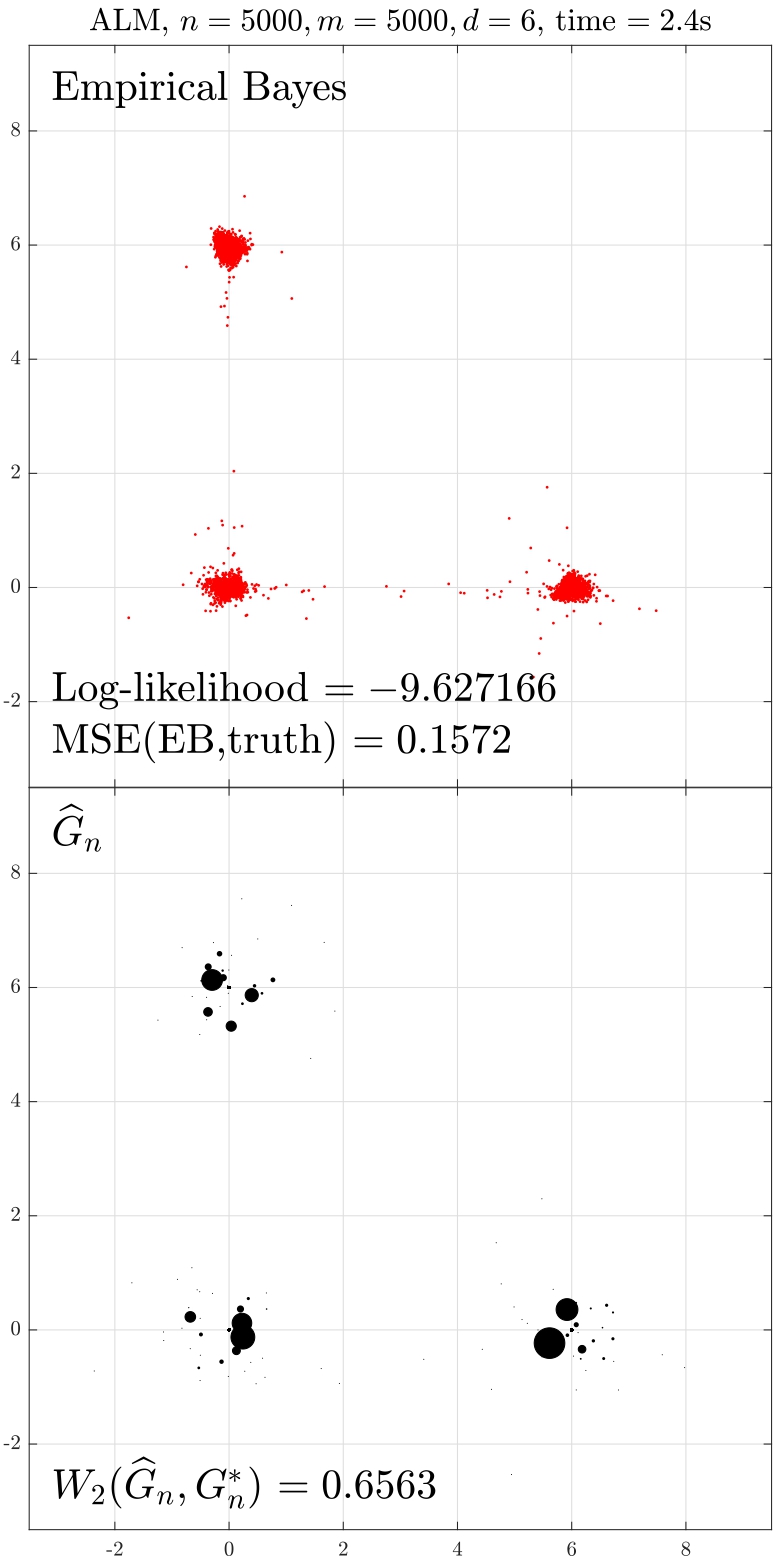}
\includegraphics[width=0.3\textwidth]{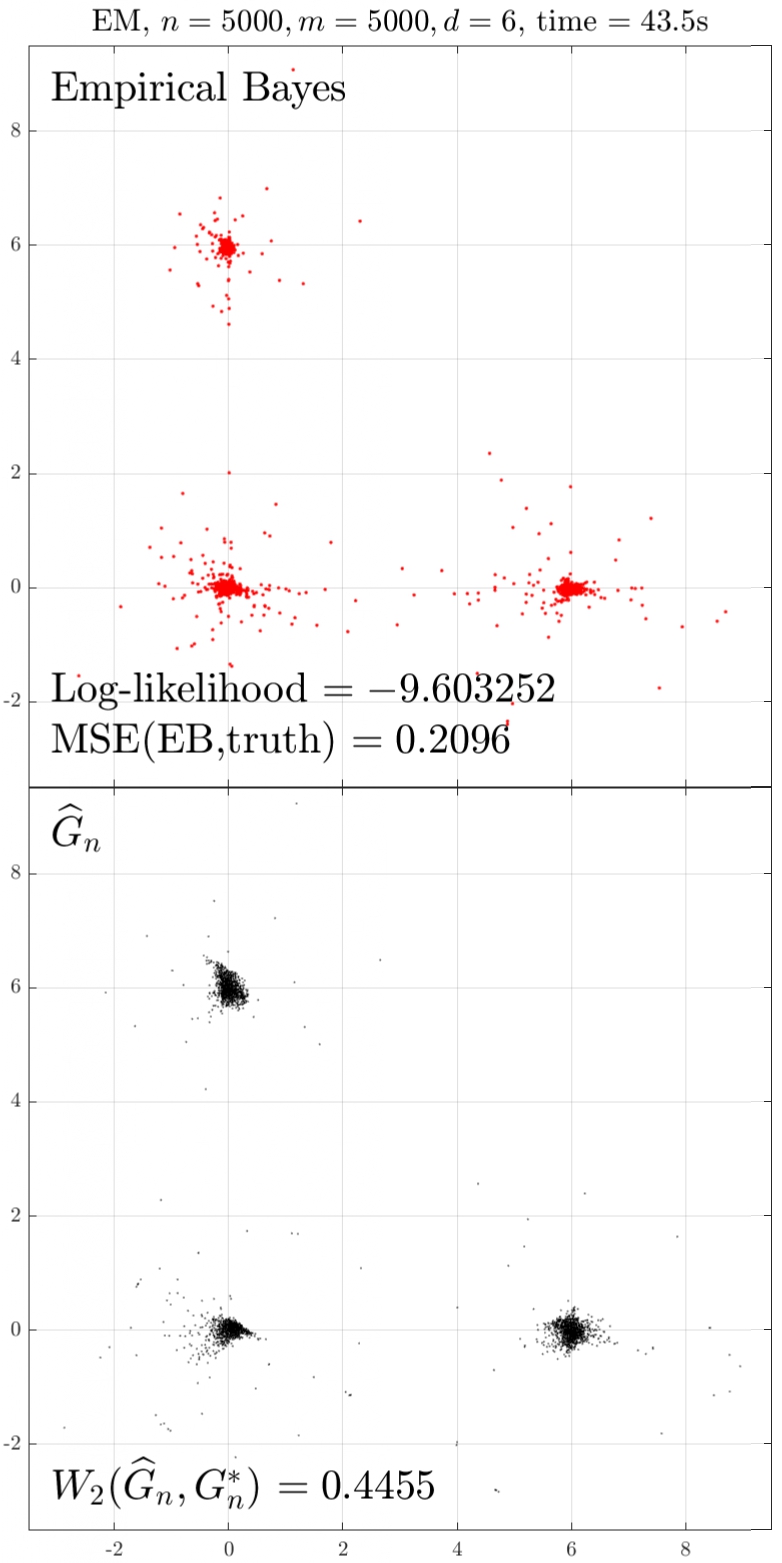}
\includegraphics[width=0.3\textwidth]{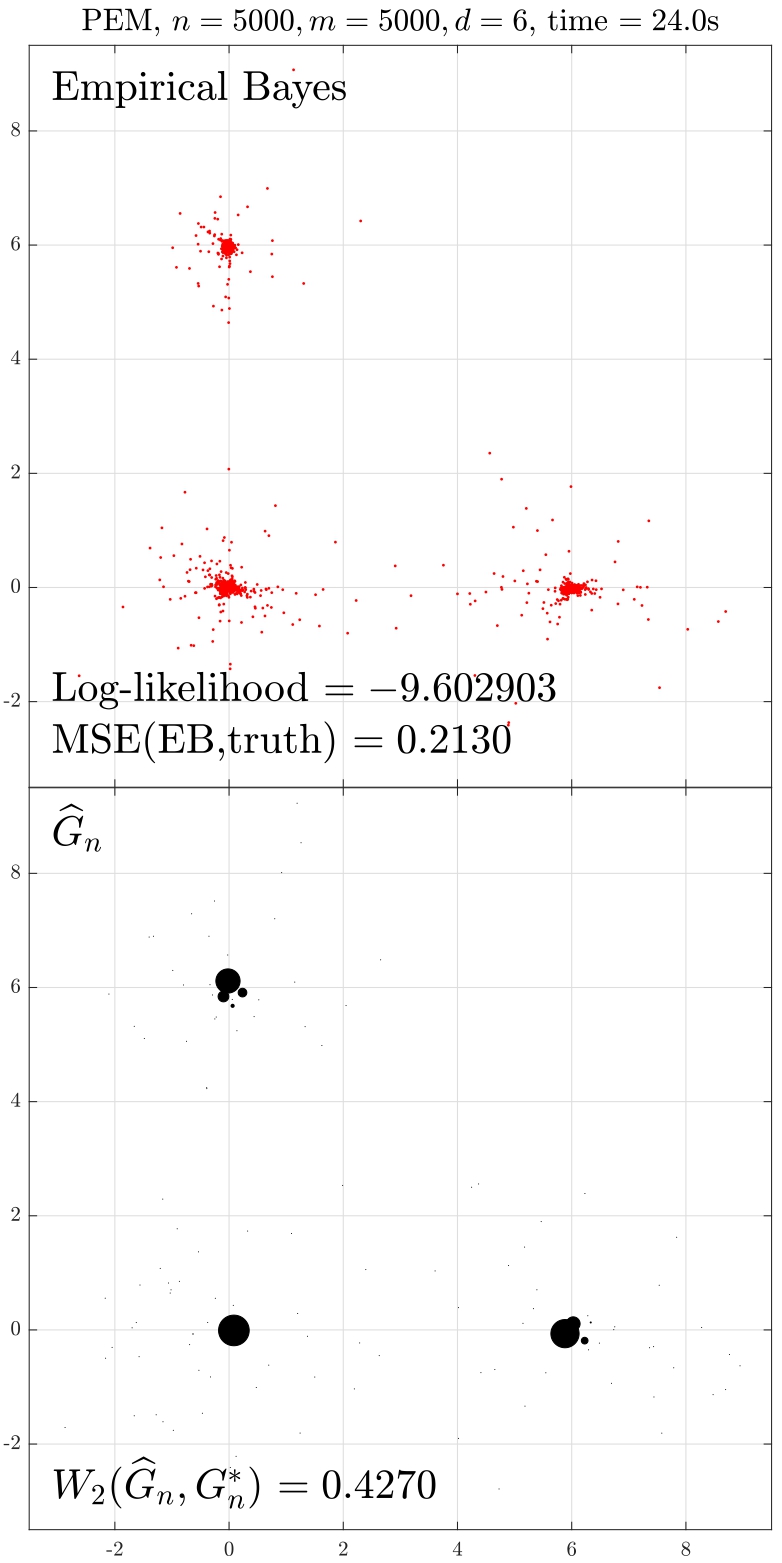}\\
\caption{\small Plots of the empirical Bayes estimates and $\widehat{G}_n$ obtained from our ALM (first column), the EM algorithm (second column), and the PEM (third column) for Example 3(c) with $d=6$.} \label{table-5c-d6}
\includegraphics[width=0.3\textwidth]{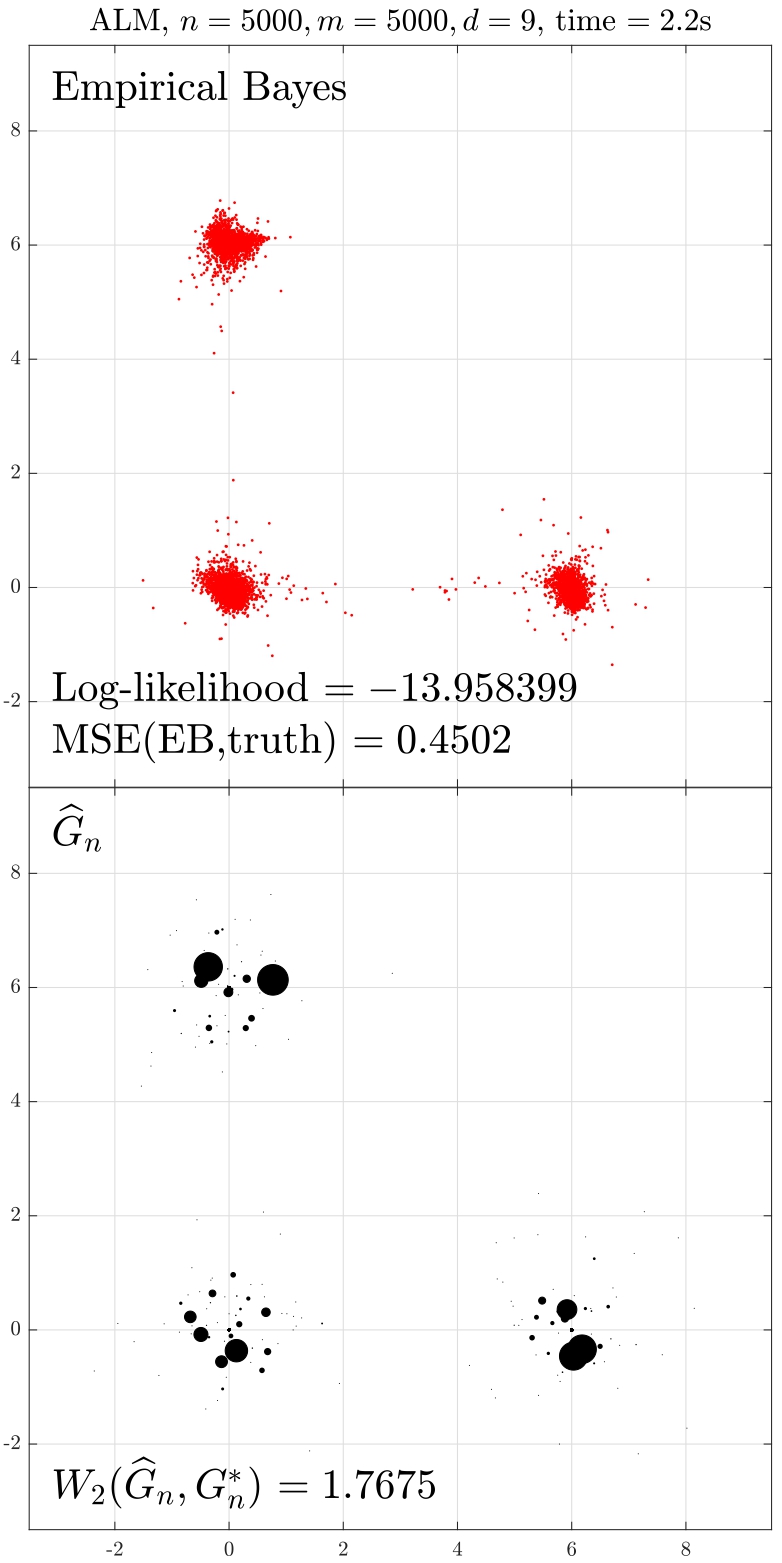}
\includegraphics[width=0.3\textwidth]{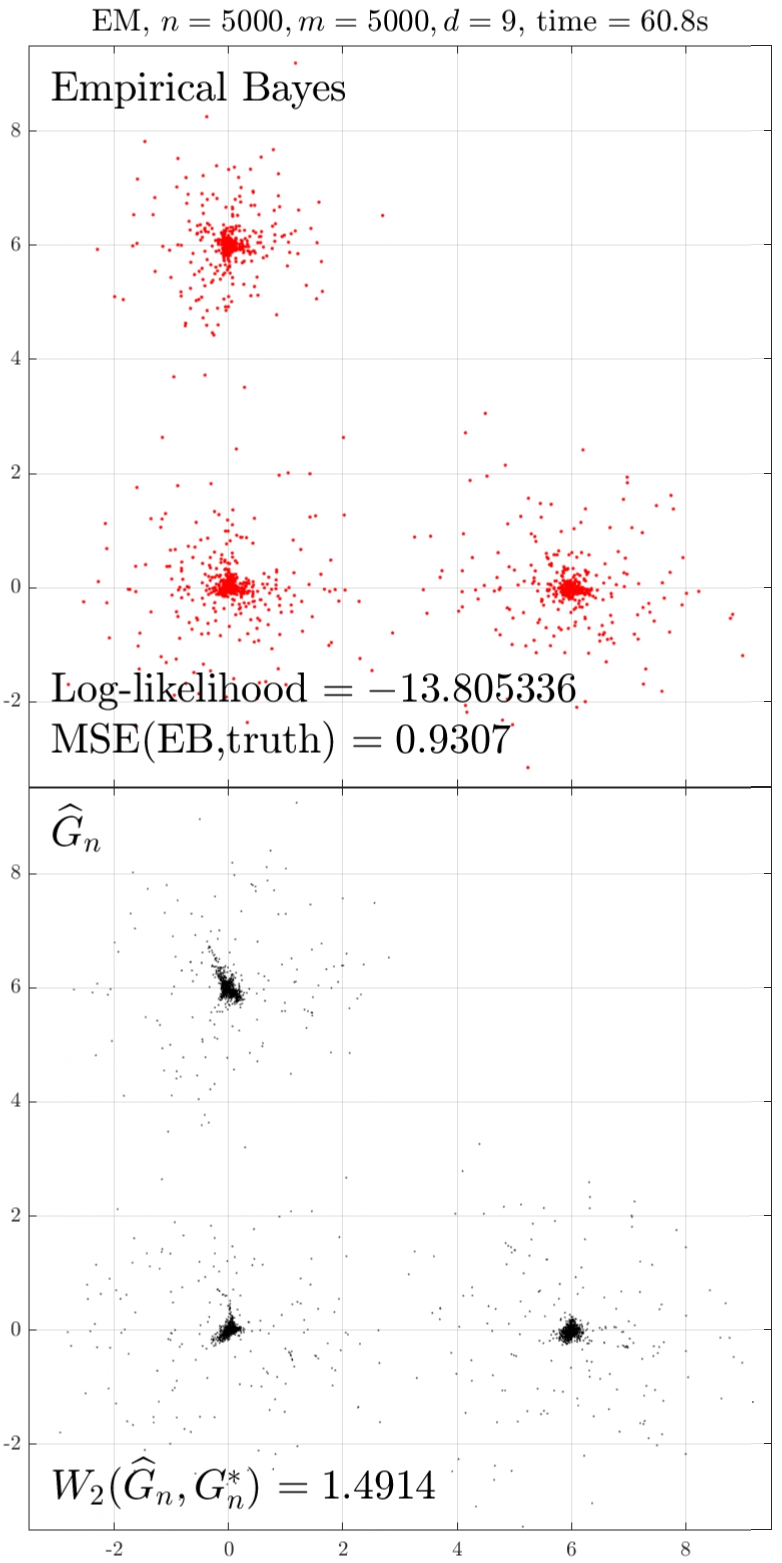}
\includegraphics[width=0.3\textwidth]{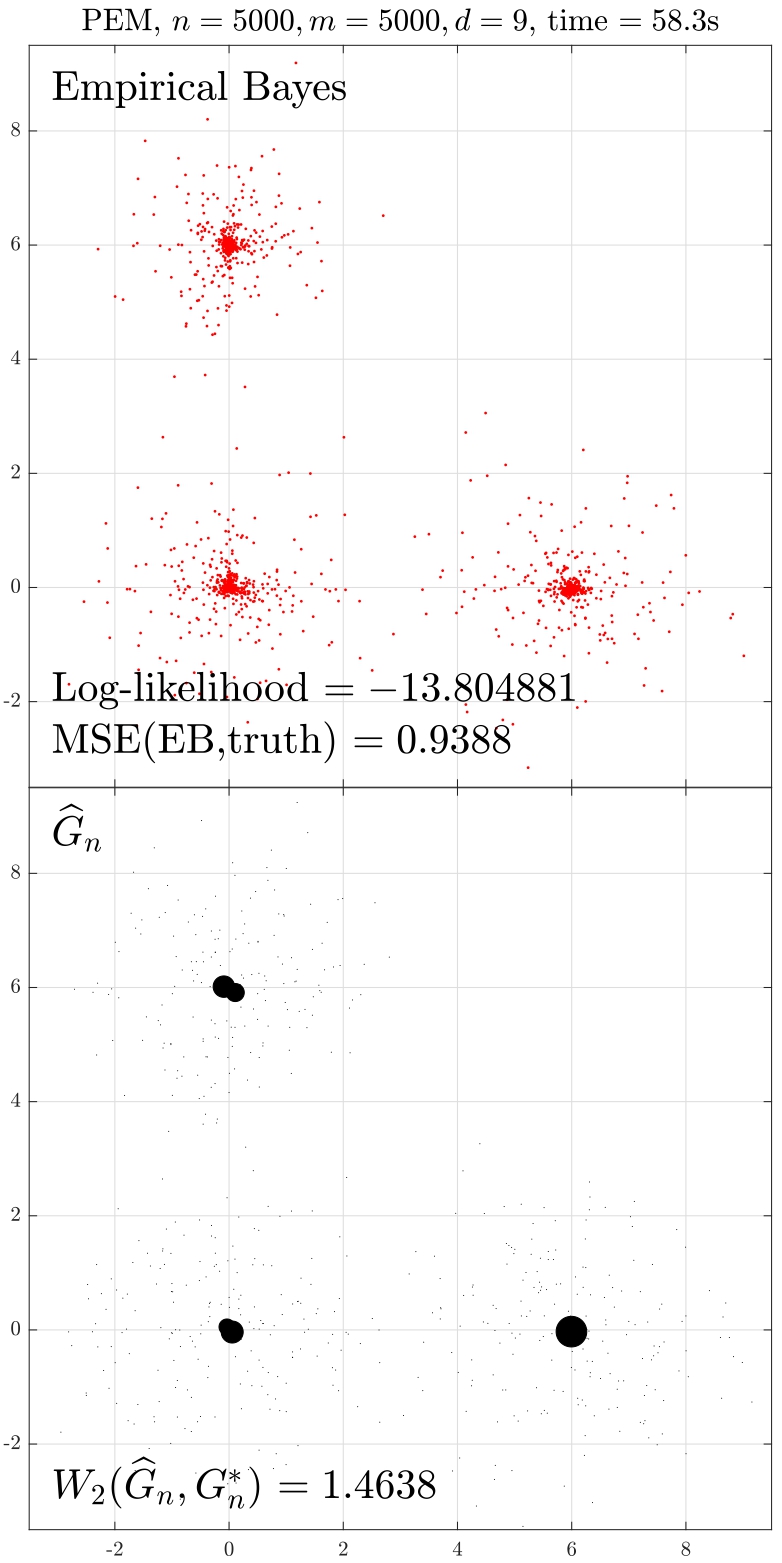}
\\
\caption{\small Plots of the empirical Bayes estimates and $\widehat{G}_n$ obtained from our ALM (first column), the EM algorithm (second column), and the PEM (third column) for Example 3(c) with $d=9$. {The EM and the PEM perform similarly in terms of denoising; however the PEM algorithm leads to a more accurate estimator of $G^*$. The ALM performs surprisingly well in terms of denoising here, although the resulting $\widehat{G}_n$ is not very accurate.} } \label{table-5c-d9}
\end{figure}

We further increase the dimension to $d=6$ and $9$, and plot the results for Example 3(c) in Figures~\ref{table-5c-d6} and \ref{table-5c-d9} respectively. We observe that the support points of $\widehat{G}_n$ obtained from the PEM algorithm are much sparser compared to that of the EM. More importantly, the resulting $\widehat{G}_n$, obtained from the PEM algorithm, is much closer to the true unknown distribution $G^*$. Although we do not show the corresponding plots for the other examples in this paper, our extensive simulations confirm that these general conclusions carry over to other settings. 
\begin{figure}[htbp]
\centering
\includegraphics[width=0.3\textwidth]{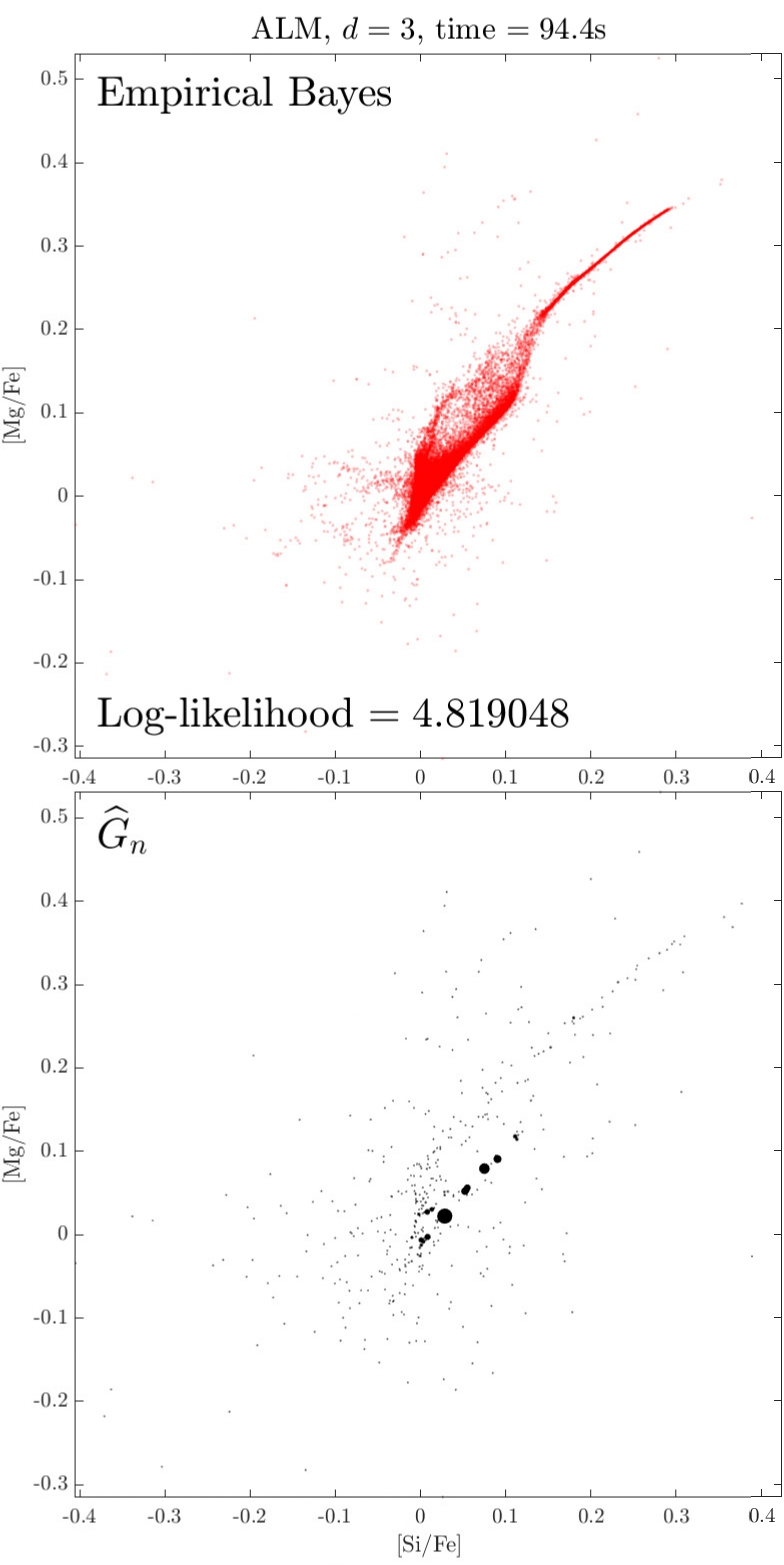}\,
\includegraphics[width=0.3\textwidth]{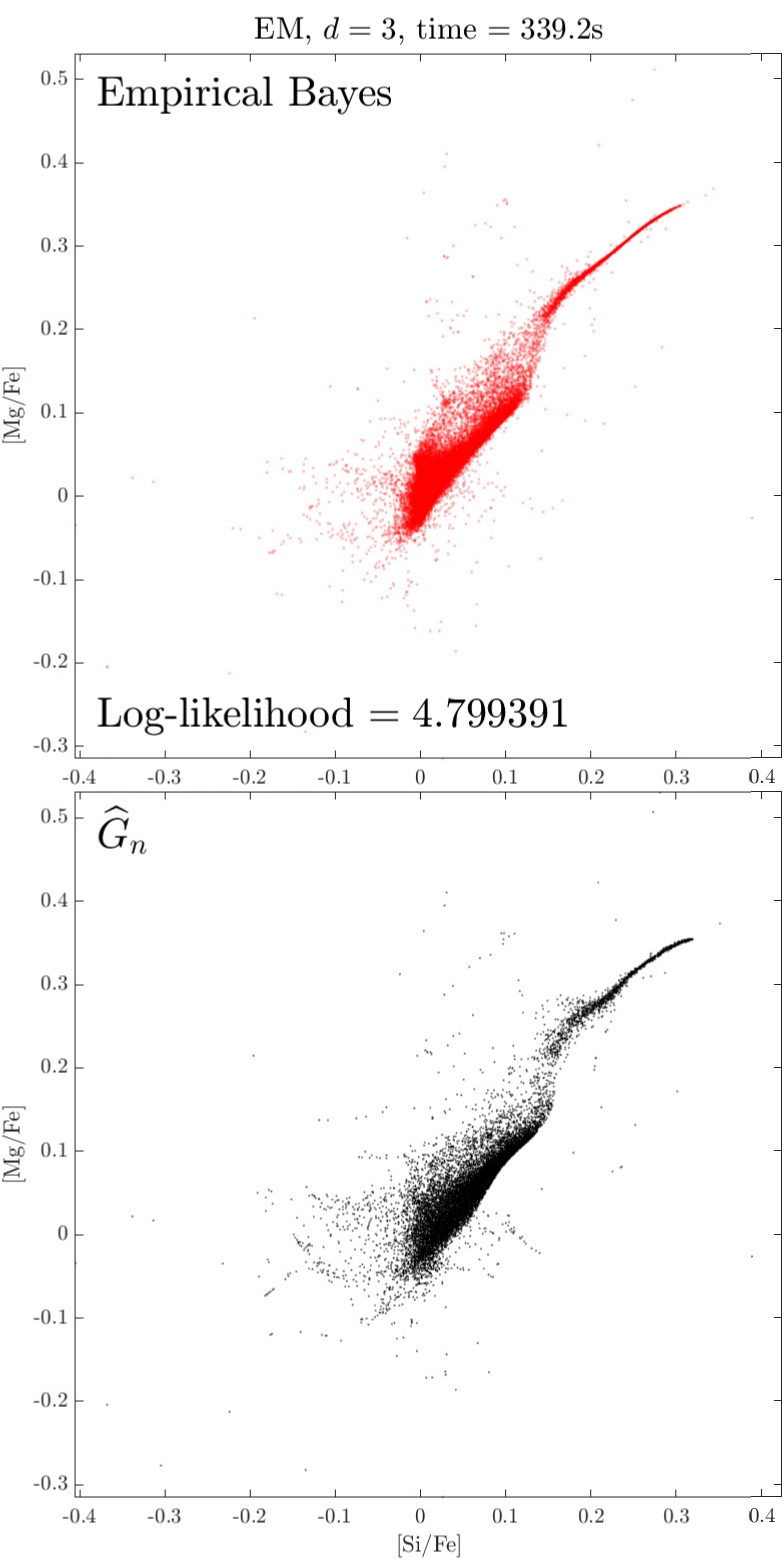}\,
\includegraphics[width=0.3\textwidth]{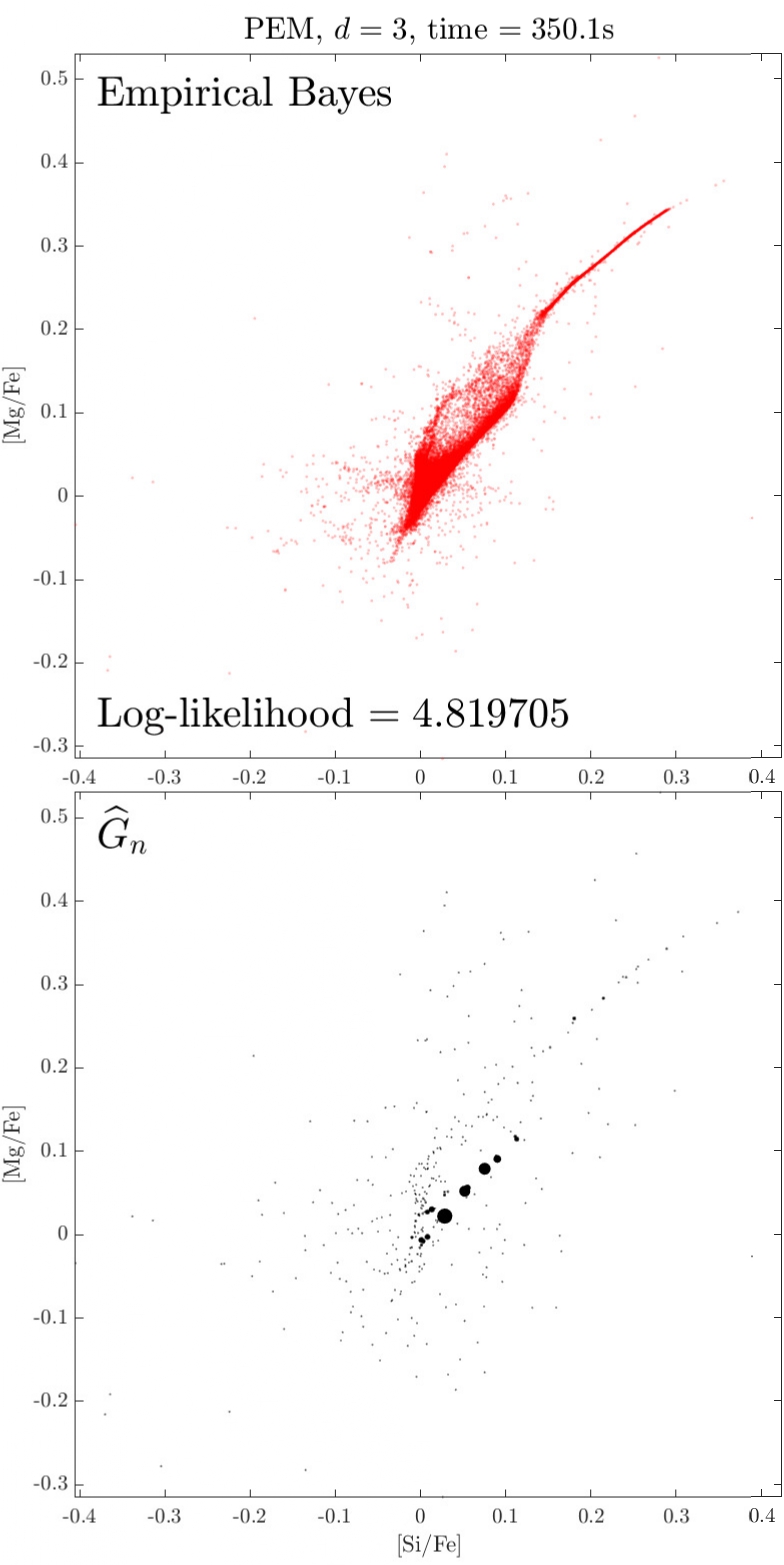}	
\caption{\small Projected results of the APOGEE data in the [Mg/Fe]-[Si/Fe] plane, solved in [Mg/Fe]-[Si/Fe]-[Mn/Fe] spaces ($d=3$). Left/Middle/Right plot is solved by ALM/EM/PEM, respectively.
}
\label{fig-APOGEE-PEM-d3}
\includegraphics[width=0.3\textwidth]{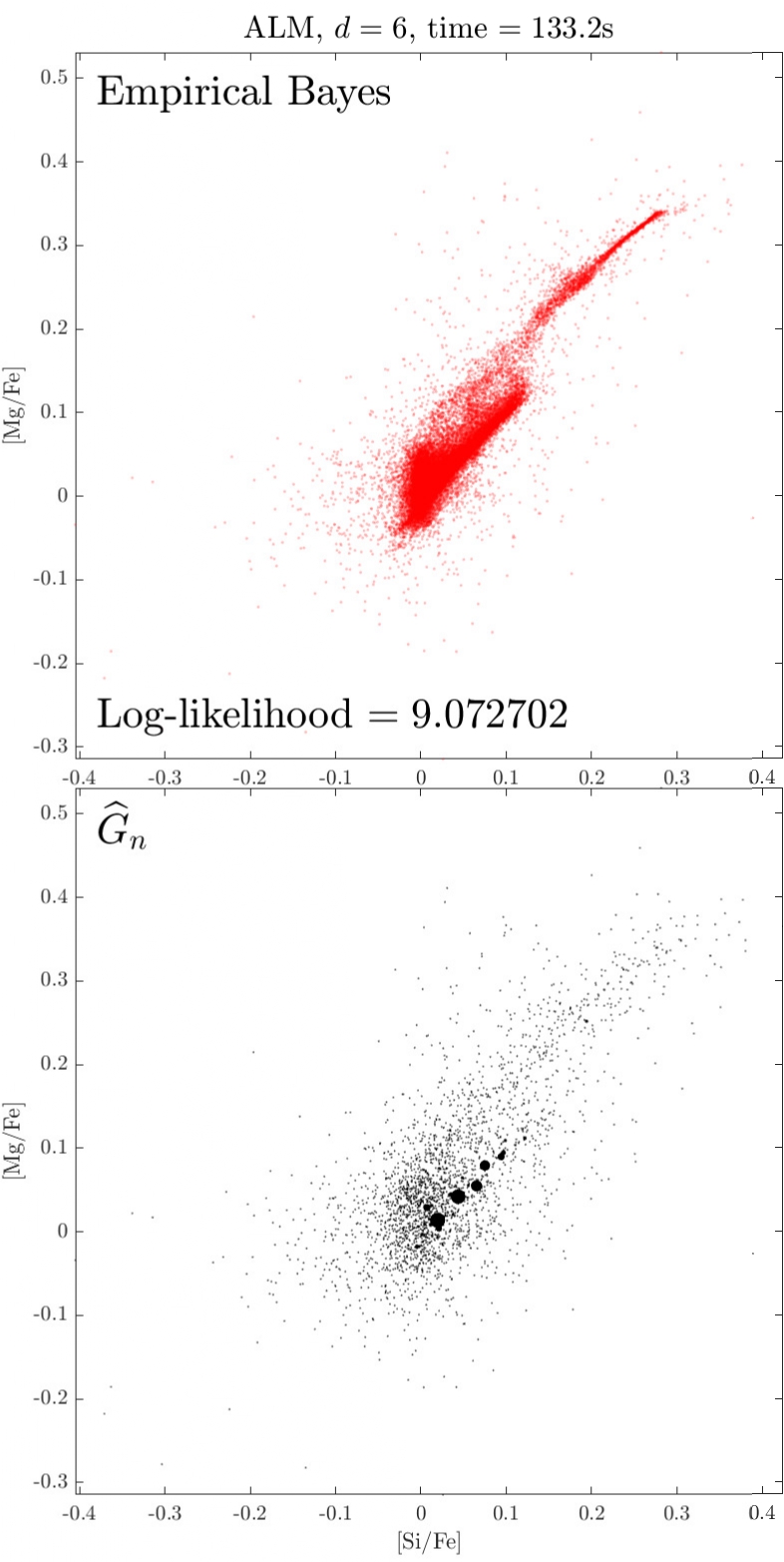}\,
\includegraphics[width=0.3\textwidth]{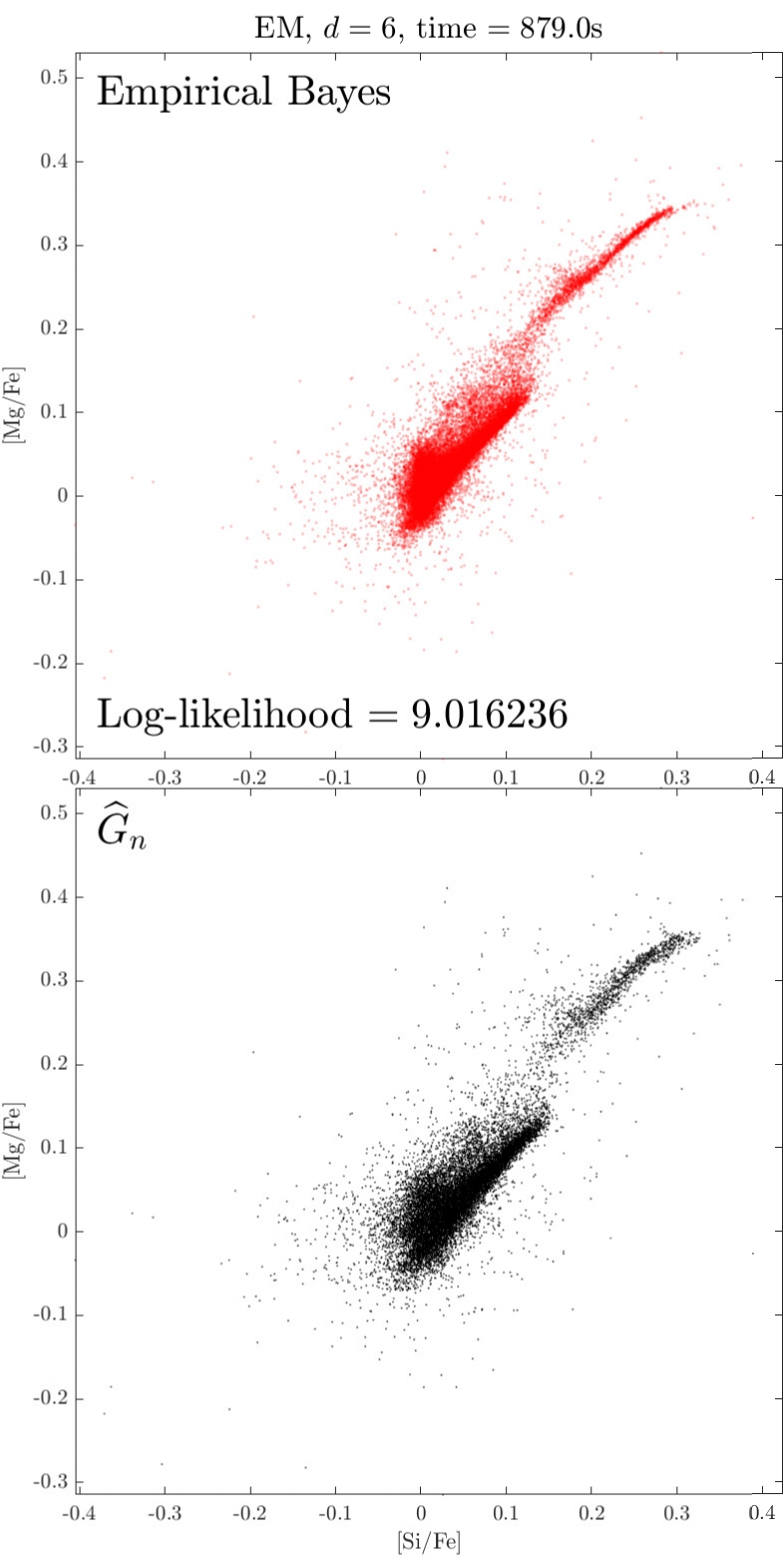}\,
\includegraphics[width=0.3\textwidth]{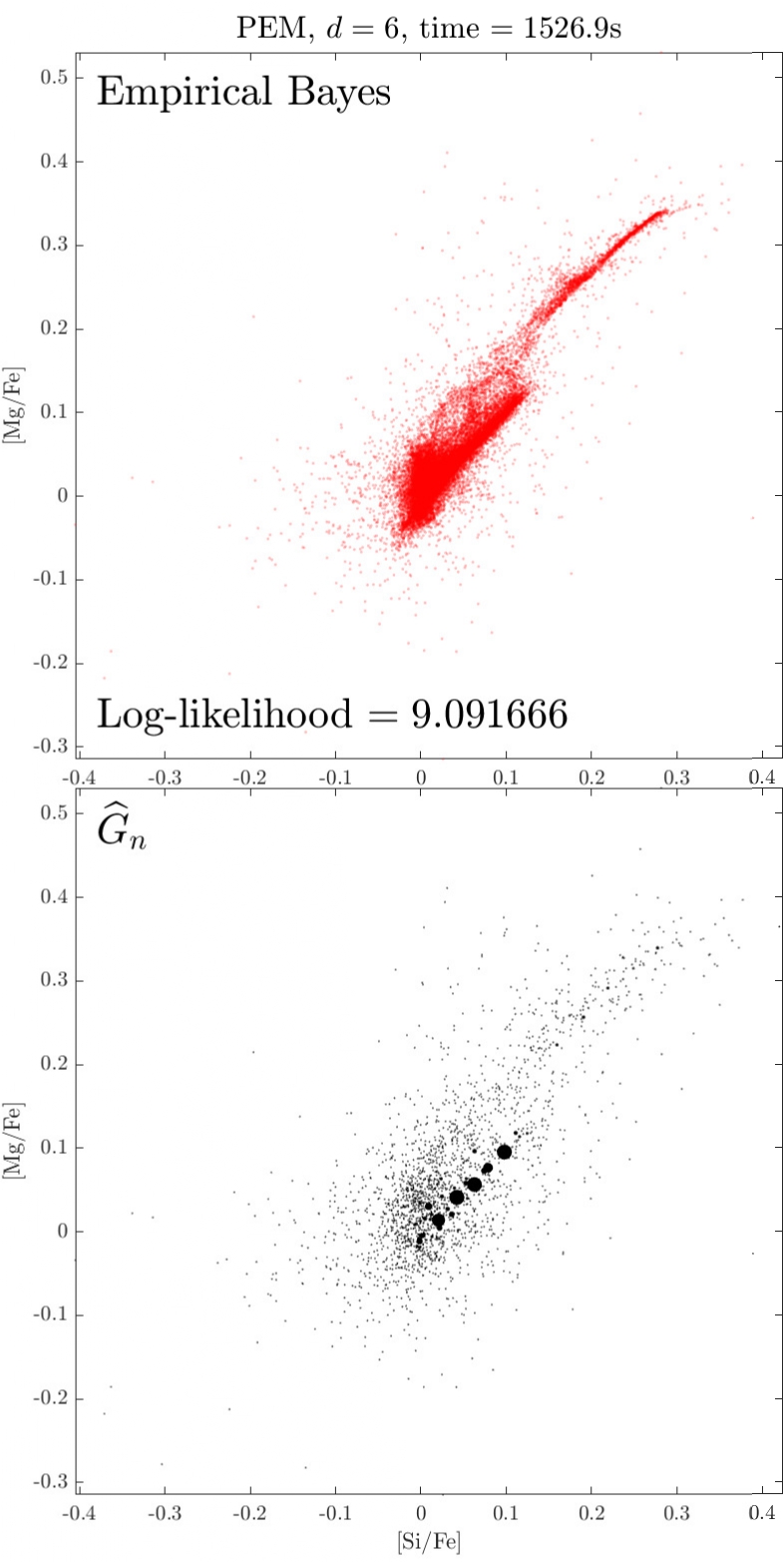}
\caption{\small Projected results for the APOGEE data in the [Mg/Fe]-[Si/Fe] plane, solved in [Mg/Fe]-[Si/Fe]-[Mn/Fe]-[N/Fe]-[Ti/Fe]-[K/Fe] spaces ($d=6$). Left/middle/right plots are obtained from the ALM/EM/PEM methods. { The plots reveal that all the three competing methods perform similarly in denoising the observations; however, the EM algorithm cannot yield a sparse solution for $\widehat{G}_n$.}
}
\label{fig-APOGEE-PEM-d6}
\end{figure}
\begin{figure}[htbp]
  \centering
  \includegraphics[width=0.30\textwidth]{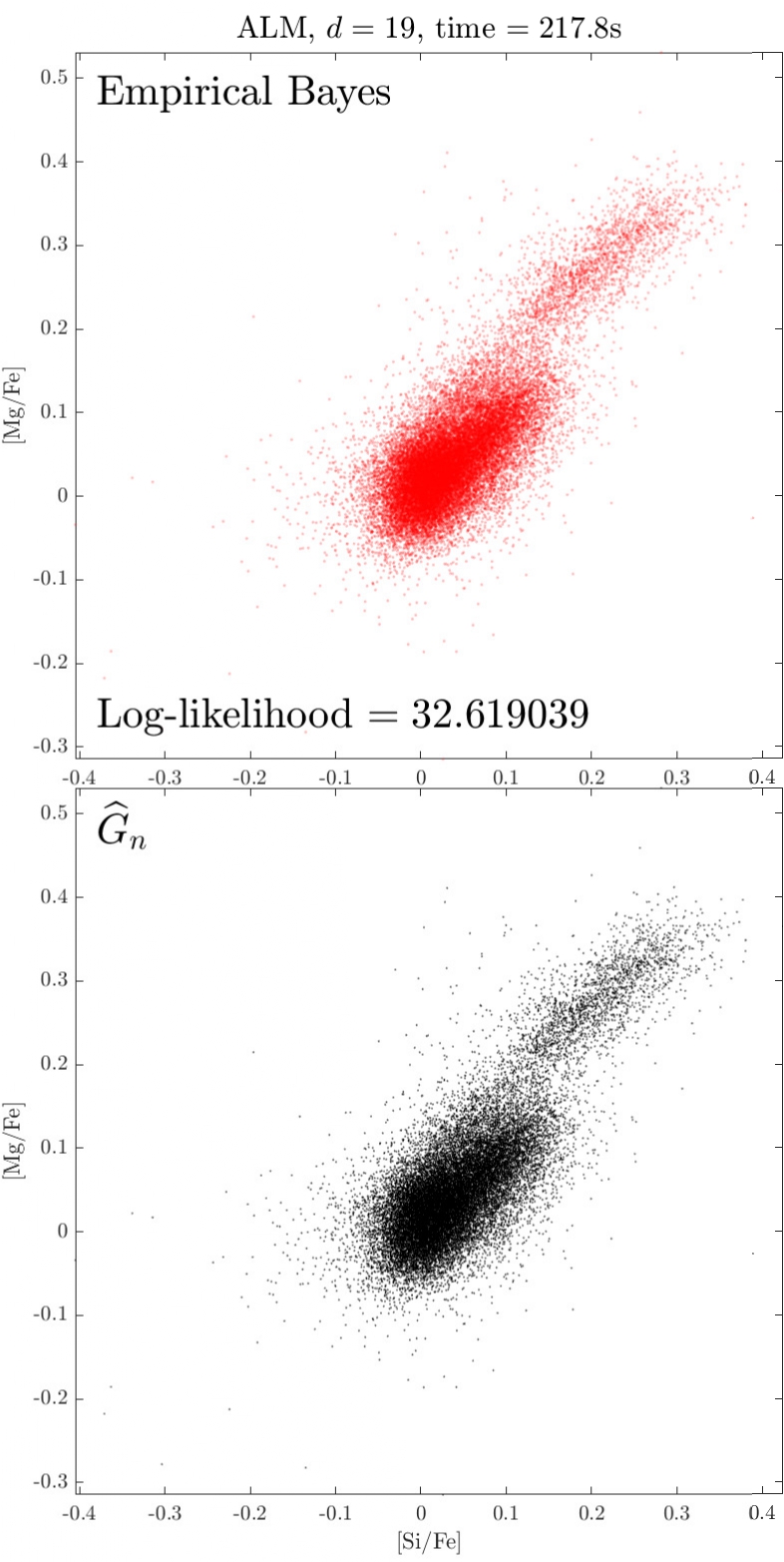}\,
  \includegraphics[width=0.30\textwidth]{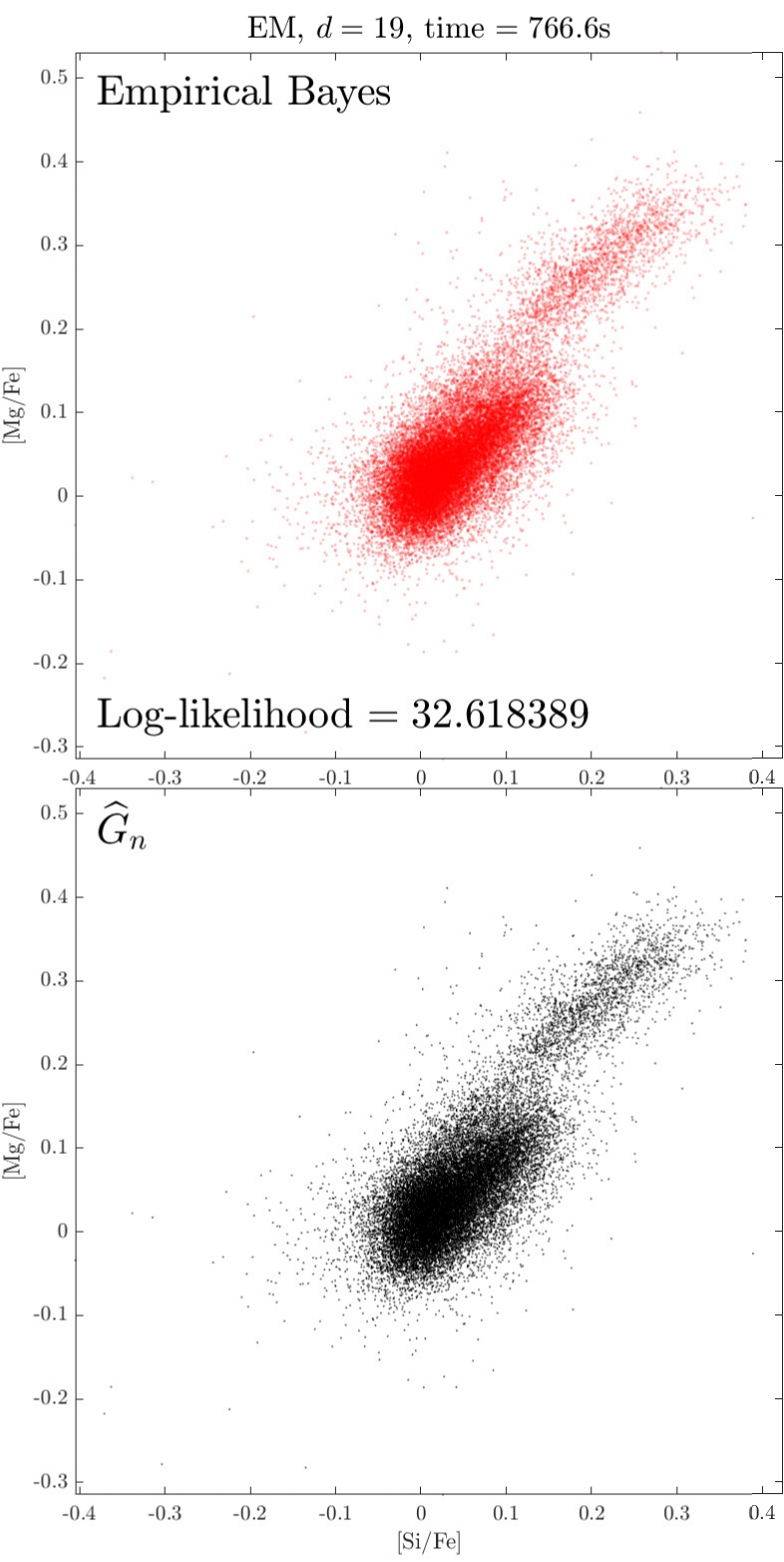}\,
  \includegraphics[width=0.30\textwidth]{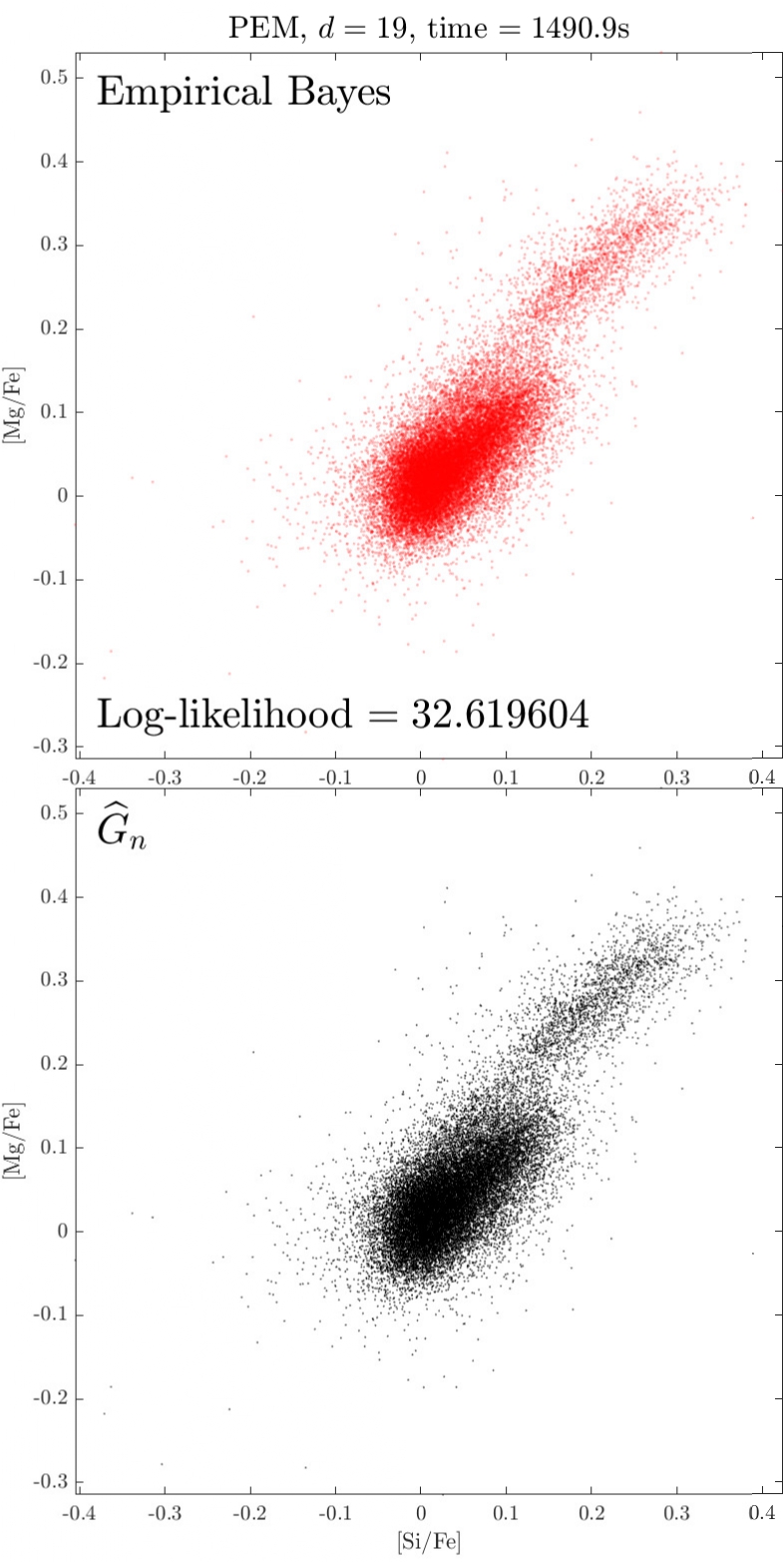}
  \caption{\small Results for the APOGEE data with all  $d=19$ variables projected in the [Mg/Fe]-[Si/Fe] plane. The left/middle/right plots are obtained from the ALM/EM/PEM methods, respectively. {The plots show that when $d = 19$, the NPMLE (as computed via the three methods) cannot denoise the observations effectively; in fact, the denoised points look very similar to the original data.}}\label{fig-APOGEE-PEM-d19}
\end{figure}
\begin{figure}[htbp]
  \centering
  \includegraphics[width=0.32\textwidth]{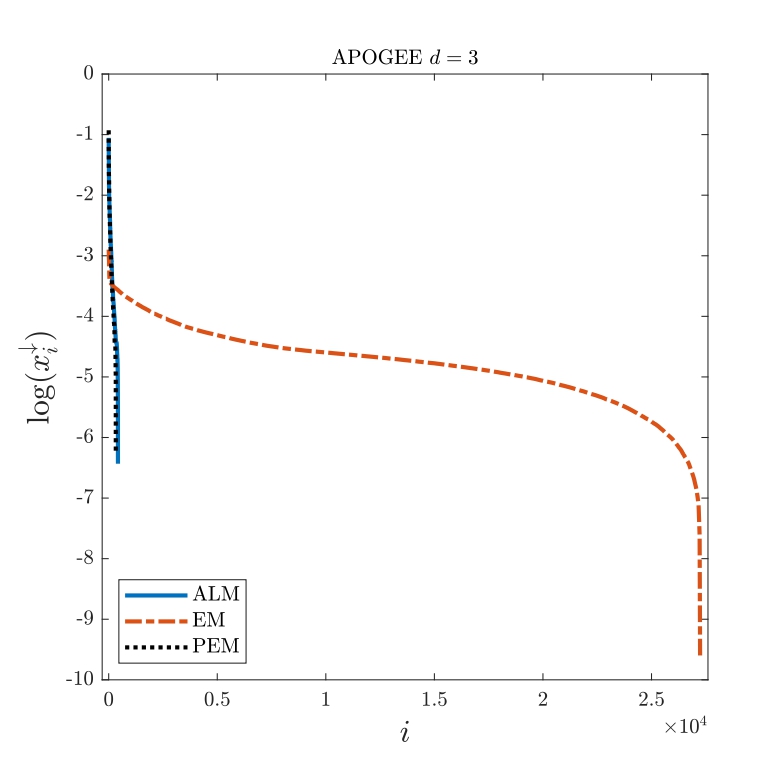}
  \includegraphics[width=0.32\textwidth]{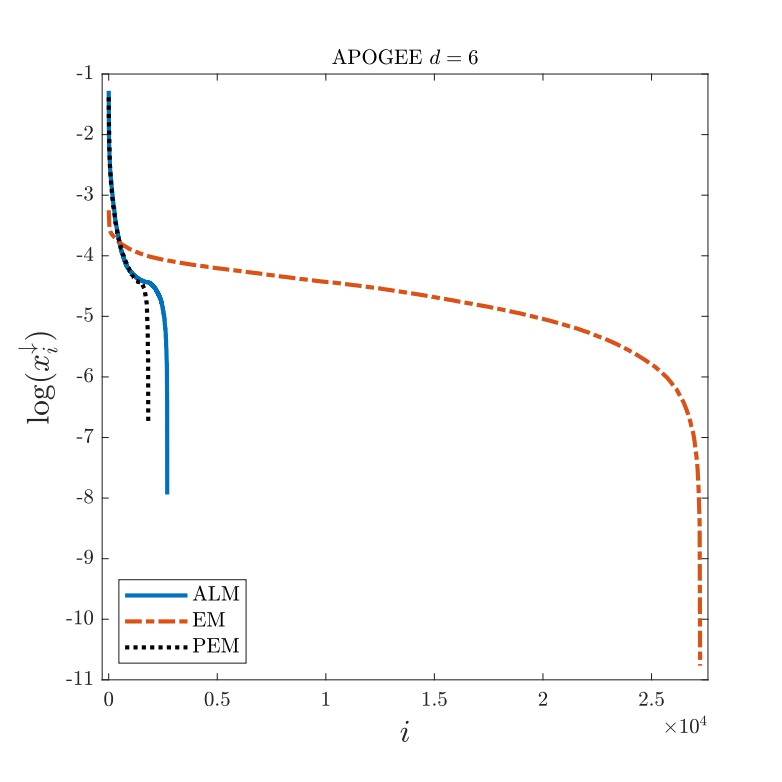}
  \includegraphics[width=0.32\textwidth]{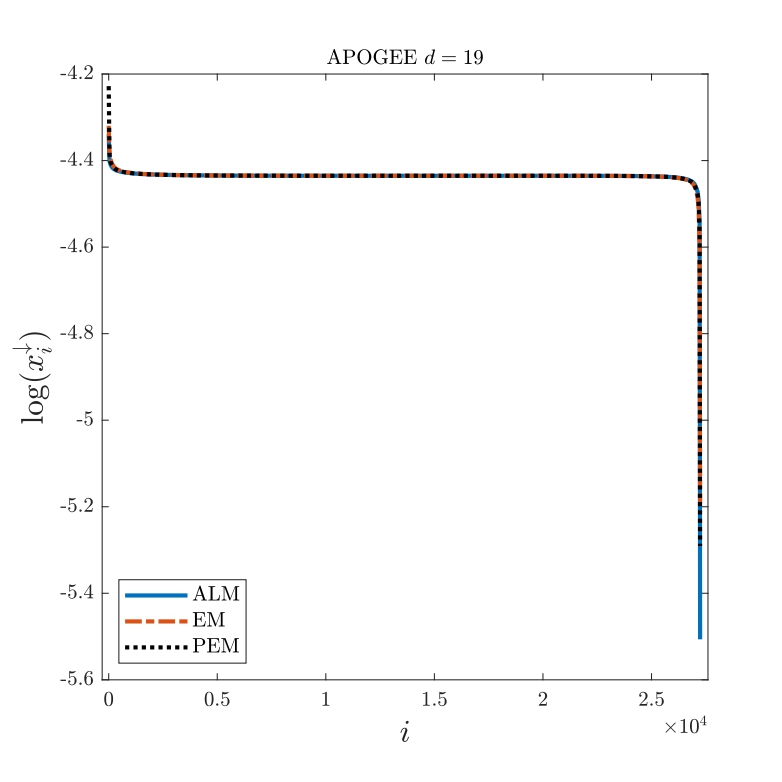}
  \caption{\small Plot of $\log(x^{\downarrow}_{i})$ for different methods --- the ALM, the EM and the PEM --- for the APOGEE survey data solved in $d=3$ (left), $d = 6$ (middle), and $d=19$ (right). Here $x^{\downarrow}$ consists of the sorted elements of the solution vector $x$ in descending order. {The plots reveal that when $d$ is small (e.g., $d=3$ or 6) the ALM and the PEM yield sparse solutions for the mixing proportion $x$; however as $d$ increases, all the components of $x$ essentially converge to $1/m$, for all the 3 methods.} }\label{fig-APOGEE-dd19-atoms}
\end{figure}

\vskip 0.1in
Next we illustrate the performance of these algorithms on the APOGEE survey~\cite{majewski2017apache} as we increase the dimension of the fitted model. We first consider the 3 variables
 [Mn/Fe], [Mg/Fe] and [Si/Fe]. The results, projected in the [Mg/Fe]-[Si/Fe] plane, are given in Figure~\ref{fig-APOGEE-PEM-d3}. It can be immediately seen that although all the three methods yield similar empirical Bayes estimates, the performance of the algorithms in computing $\widehat{G}_n$ vary considerably. The PEM algorithm yields a $\widehat{G}_n$ with the fewest support points, and performs similarly to ALM; in comparison, the EM algorithm cannot provide a sparse solution for $\widehat{G}_n$ as can be expected from theory (see e.g.,~\cite{polyanskiy2020self}). It is also evident that the PEM algorithm achieves the highest objective value for this real data. 

\vskip 0.1in
\noindent{\bf Advantages of our PEM algorithm over the classical EM}: Although the PEM algorithm is a tweak of the classical EM algorithm it has several advantages over the EM: (1) The PEM results in a higher objective (aka log-likelihood) value; (2) the atoms of the PEM are much sparser compared to that of the EM; (3) the PEM has comparable computational runtime as the EM; it can be faster than the EM in some instances, especially when $d$ is not too large (as seen from Table~\ref{table-EM1}).

\vskip 0.1in
\noindent{\bf Curse of dimensionality and the NPMLE}: We next select $d=6$ (here we choose the variables [Mn/Fe], [Mg/Fe], [Si/Fe], [N/Fe], [Ti/Fe] and [K/Fe]) and all the $d=19$ variables from the APOGEE survey~\cite{majewski2017apache} and run the three algorithms --- the ALM, EM and the PEM --- in these two settings; see Figures~\ref{fig-APOGEE-PEM-d6} and \ref{fig-APOGEE-PEM-d19} respectively. It can be seen from the plots that the performance of all the methods deteriorates as the dimension of the problem increases, i.e., the solutions for $\widehat{G}_n$ have more support points and the empirical Bayes estimates reveal more spread. In particular, when $d=19$, the original data, the empirical Bayes estimates and $\widehat{G}_n$ are all essentially the same, illustrating that the NPMLE cannot successfully denoise the observations for such a large $d$ with $n \approx 2.7 \times 10^4$; see Figure~\ref{fig-APOGEE-PEM-d19}. Further, the plots of the weights on the fitted atoms for the three methods in Figure~\ref{fig-APOGEE-dd19-atoms} reveals that the solution for $x$ converges to $\frac{1}{m}{\bf 1}_m$ as $d$ grows, i.e., every data point (which is also chosen as a grid point) has equal mass. Additionally, in our simulation studies, when $d$ is large, we observe that the NPMLE $\widehat{G}_n$ (computed via all the EM and PEM methods) has much higher likelihood value than the truth $G^*$ --- this provides further evidence for the overfitting by the NPMLE when $d$ is large; see Table~\ref{table:ln}.

\begin{table}
  \centering
  \begin{tabular}{lllll}
  \toprule
  Example & ALM & EM & PEM & $G^*$ \\
  \midrule
  3(a) & -30.884461 & -26.885783 & -26.885773 &  -30.638825 \\
  3(c) & -29.952639 & -26.843792 & -26.843515 & -29.527145 \\
  \bottomrule
  \end{tabular}
  \caption{The log-likelihood value at the NPMLE $\widehat{G}_n$ computed by ALM, EM, PEM, and the log-likelihood value at the true $G^*$ on Example 3(a) and 3(b) with $n=m=5,000$ and $d=20$. }\label{table:ln}
\end{table}

\vskip 0.1in
Thus, it seems that in high dimensions the NPMLE cannot effectively extract the `shared information' between the data points and hence the algorithms cannot achieve any shrinkage/denoising. The main reason for this seems to be the fact that in high dimensions most points are far from each other. This possible overfitting by the NPMLE, as defined in~\eqref{NPMLE1}, when $d$ is {\it large} suggests that some form of regularization/penalization is needed to improve the performance of the (marginal) likelihood based approach.

\section{Conclusion}
In this paper we have described two approaches to computing the NPMLE of a (heteroscedastic) Gaussian location mixture model. Our first algorithm solves the dual of the optimization problem~\eqref{primal-0} using a semismooth Newton based ALM. This approach is highly scalable (e.g., we can solve problems with $n \approx 10^6$ and $m \approx 10^4$) and it exploits the second order sparsity in the generalized Hessian matrix arising in the ALM subproblem. We believe that this semismooth Newton based ALM approach is a powerful method for solving large scale optimization problems whose solutions are intrinsically structured sparse (i.e., the solution itself or a linear transformation of the solution is sparse). In fact, this algorithmic framework has already been shown to be effective for the Lasso problem and its variants; see e.g., \cite{li2016highly, li2017efficiently, zhang2020efficient}.


In our second approach, which allows for joint optimization of both the support points $\mu_j$'s and the probability weights $x_j$'s in~\eqref{adaptive supports 0}, we develop the PEM method by combining the advantages of the convex program~\eqref{primal-0} and the classical EM algorithm. Being a grid-free method, the PEM can handle any dimension $d$. Our PEM algorithm is more generally applicable in any mixture model problem, and it can lead to algorithms that produce sparse solutions for the mixing proportions. We plan to investigate this in more detail in future papers.

In this paper we have focussed our attention on fitting the Gaussian location mixture model~\eqref{denoising data}. However, the scope of our approach is much more general. In fact, one could consider the following $d$-dimensional ($d \ge 1$) observation model:
\begin{equation}\label{eq:MixMdl}
Y_i|\theta_i \sim p_i (\cdot|\theta_i), \qquad \mbox{with} \;\; \theta_i \stackrel{iid}{\sim} G^*, \quad \mbox{for} \; i \in \{1,\ldots, n\}
\end{equation}
where $\{p_i(\cdot|\cdot)\}_{i=1}^n$ is a sequence of known probability densities and $\{\theta_i\}^n_{i=1} \subset \R^p$ ($p \ge 1$) is the sequence of i.i.d.~(from $G^*$) underlying latent parameters. The algorithms developed in this paper immediately generalize to this setting as the NPMLE of $G^*$ in~\eqref{eq:MixMdl} can be computed similarly. See {Example 2} in Appendix~\ref{sec:syn} in the supplementary file where we illustrate this for a {\it scale} mixture of centered Gaussian distributions. 

The effectiveness of both our methods --- the ALM and the PEM --- in estimating $\widehat{G}_n$ and the $\widehat{\theta}_i$'s (as illustrated via simulations and theory) show the power and scope of nonparametric empirical Bayes as a methodology in multivariate problems.
However, when $d$ is large (e.g., $d \ge 20$), we also see that the NPMLE in~\eqref{NPMLE1} can overfit the data; see e.g., Figure~\ref{fig-APOGEE-PEM-d19}. This leaves open the study of regularization methods for estimating the unknown $G^*$ when $d$ is large. We expect this to be a fruitful direction for future research.

\section*{Acknowledgements}
The third author would like to thank Alessandro Grande for helpful comments.

\bibliographystyle{abbrv}
\bibliography{Notebib}

\newpage
\appendix

\section{Proofs of the main results}\label{sec:Appendix-A}

\subsection{Proof of Proposition~\ref{prop:rate}}\label{app:rate ALM}
In order to use the general results on the convergence rate of the ALM in \cite[Theorem 5]{rockafellar1976augmented}, we prove that the quadratic growth condition of the dual problem  in \eqref{eq:quadratic growth} holds.
Since each entry of the matrix $L$ is nonnegative and each row of $L$ has at least one nonzero entry, one may obtain from the constraint $\displaystyle\frac{1}{n} L^\top v \leq {\bf 1}_m$  and the nonnegativity of $\bar{v}$ that $\|\bar{u}\|_\infty = \|\bar{v}\|_\infty<+\infty$.
Therefore, we may assume without loss of generality that $c:=\displaystyle\sup_{(u,v)\in {\cal N}} \{\|u\|_\infty, \|v\|_\infty\}<+\infty$ so that $\nabla^2 h(u)=\displaystyle\frac{1}{n}\mbox{Diag}\left(\frac{1}{u_i^2}\right)\geq \frac{1}{nc^2}I_n$ for any $(u,v)\in {\cal N}$, where $I_n$ is the $n\times n$ identity matrix. It can be derived that for any $(u,v)\in {\cal N}$ that is feasible to \eqref{dual},
\[
\begin{array}{rl}
h(u) \geq & h(\bar{u}) + \nabla h(\bar{u})^\top (u-\bar{u}) + \displaystyle\frac{1}{nc^2}\|u-\bar{u}\|_2^2 \\[0.15in]
= & h(\bar{u})  - \displaystyle\sum_{i=1}^n \frac{1}{n\bar{u}_i} (u_i-\bar{u}_i) + \frac{1}{nc^2}\|u-\bar{u}\|_2^2\\[0.15in]
= & h(\bar{u}) - \displaystyle\frac{1}{n}\sum_{i=1}^n (L_{i\bullet}\bar{x}) (v_i-\bar{v}_i) + \displaystyle\frac{1}{nc^2}\|u-\bar{u}\|_2^2 \\[0.1in]
\geq & h(\bar{u}) + \displaystyle\frac{1}{nc^2}\|u-\bar{u}\|_2^2
= h(\bar{u}) + \displaystyle\frac{1}{2nc^2}\left(\|u-\bar{u}\|_2^2 + \|v-\bar{v}\|_2^2\right),
\end{array}
\]
where the last equality and the last inequality follow from the KKT conditions in \eqref{kkt}: $\displaystyle L_{i\bullet}\bar{x} = \bar{y}_i = \frac{1}{\bar{u}_i}, \; \frac{1}{n}\bar{v}^\top L\bar{x} = {\bf 1}_m^\top \bar{x}$, $\bar{x}\geq 0$ and $\displaystyle\frac{1}{n}L^\top v\leq {\bf 1}_m$ (from the feasibility of $v$). Therefore, the quadratic growth condition \eqref{eq:quadratic growth} for the dual problem holds with $\kappa = (2nc^2)^{-1}$. The asymptotically superlinear convergence rate of the sequence $\{(x^k, y^k)\}_{k \ge 1}$ generated by the ALM is now a consequence of \cite[Theorem 5]{rockafellar1976augmented}.
\qed

\subsection{Proof of Proposition~\ref{prop:PEM}}\label{app:PEM}
\begin{proof}
Obviously $\{x^k\}_{k\geq 1}$ is bounded due to the simplex constraint in \eqref{eq:PEM}. For each $j=1, \cdots, m$, to show that $\{\mu_j^k\}$ is bounded, it suffices to prove the boundedness of the subsequence $\{\mu_j^{k}\}_{k\in \kappa}$  where $x_j^{k}\neq 0$ for any $k\in \kappa$  (so that $\widetilde\gamma_{ij}^{k} \neq 0$ for all $i = 1, \cdots, n$).
Fo any $k\in \kappa$, it holds that
\[
\begin{array}{ll}
\|\mu_j^{k} \|_2
&=  \left\| \left(\,\displaystyle\sum_{i=1}^n \widehat{\gamma}_{ij}^{k} \Sigma_i^{-1}\right)^{-1} \left(\displaystyle \sum_{i=1}^n \widehat{\gamma}_{ij}^{k} \Sigma_i^{-1} Y_i\right) \right\|_2 \\[0.25in]
& = \left\| \displaystyle \sum_{i=1}^n   \left(\,\displaystyle\sum_{i^\prime=1}^n \widehat{\gamma}_{i^\prime j}^{k} \Sigma_{i^\prime}^{-1}\right)^{-1} \widehat{\gamma}_{ij}^{k} \Sigma_i^{-1} Y_i \,\right\|_2 \\[0.25in]
& \leq \displaystyle \sum_{i=1}^n   \left\|  \left(\,\displaystyle\sum_{i^\prime=1}^n \widehat{\gamma}_{i^\prime j}^{k} \Sigma_{i^\prime}^{-1}\right)^{-1} \widehat{\gamma}_{ij}^{k} \Sigma_i^{-1} \right\|_F  \|Y_i\|_2 \\[0.25in]
& \leq \displaystyle \sum_{i=1}^n   \left\|  \left(\,\displaystyle\sum_{i^\prime=1}^n \widehat{\gamma}_{i^\prime j}^{k} \Sigma_{i^\prime}^{-1}\right)^{-1} \,\right\|_F \left\| \widehat{\gamma}_{i j}^{k} \Sigma_{i}^{-1} \right\|_F  \|Y_i\|_2 \\[0.25in]
& \leq \displaystyle \sum_{i=1}^n   \left\| \, \left(\widehat{\gamma}_{i j}^{k} \Sigma_{i}^{-1}\right)^{-1} \,\right\|_F \left\| \widehat{\gamma}_{i j}^{k} \Sigma_{i}^{-1} \right\|_F  \|Y_i\|_2
 = \displaystyle \sum_{i=1}^n   \left\| \, \Sigma_{i} \,\right\|_F  \left\| \,  \Sigma_{i}^{-1} \,\right\|_F \|Y_i\|_2,
\end{array}
\]
where the last inequality holds because $\| \cdot \|_F$ is a monotonic norm (see, e.g.,\cite[2.2-10]{ciarlet1989introduction}) and $ \left(\widehat{\gamma}_{i j}^{k} \Sigma_{i}^{-1}\right)^{-1}   - \left(\,\displaystyle\sum_{i^\prime=1}^n \widehat{\gamma}_{i^\prime j}^{k} \Sigma_{i^\prime}^{-1}\right)^{-1} $ is positive definite. Therefore, the sequence $\{(x^k,\mu^k)\}_{k\geq 1}$ generated by the PEM in \eqref{eq:PEM} is bounded.

Let $(x^*, \mu^*)$ be an accumulation point of a convergent subsequence $\{(x^{k_s}, \mu^{k_s})\}_{s\geq 1}$, which must exist due to the boundedness of the sequence  $\{(x^k,\mu^k)\}_{k\geq 1}$.
One may obtain from the update rule of $\{(x^k, \mu^k)\}$ in \eqref{eq:PEM} that for any $\mu=\{\mu_j\}_{j=1}^m$,
\[
\begin{array}{rl}
& \widehat{\ell}_{n;x} (x^{k_{s}}; x^{k_{s}}, \mu^{k_{s}}) + \widehat{\ell}_{n;\mu} (\mu, x^{k_{s}}; x^{k_{s}}, \mu^{k_{s}})\\[0.05in]

\leq & \widehat{\ell}_{n;x} (x^{k_{s}}; x^{k_{s}}, \mu^{k_{s}}) + \widehat{\ell}_{n;\mu} (\mu^{k_{s}+1}, x^{k_{s}}; x^{k_{s}}, \mu^{k_{s}})\qquad (\mbox{due to the update rule of $\mu^{k_s}$})\\[0.05in]

\leq  & \ell_n(x^{k_{s}}, \mu^{k_{s}+1}) \qquad (\mbox{due to \eqref{eq:Jensen inequality}})  \\[0.05in]
\leq  & \ell_n(x^{k_{s}+1}, \mu^{k_{s}+1}) \qquad (\mbox{due to the update rule of $x^{k_s+1}$})  \\[0.05in]

\leq  & \ell_n(x^{k_{s+1}}, \mu^{k_{s+1}}) \qquad\; (\mbox{due to \eqref{PEM:decrease}}) \\[0.05in]
= & \widehat{\ell}_{n;x} (x^{k_{s+1}}; x^{k_{s+1}}, \mu^{k_{s+1}}) + \widehat{\ell}_{n;\mu} (\mu^{k_{s+1}},x^{k_{s+1}}; x^{k_{s+1}}, \mu^{k_{s+1}}).
\end{array}
\]
%
%
%
For any $x$ satisfying ${\bf 1}_m^\top x = 1$ and $x\geq 0$, we have
\[
\ell_n(x, \mu^{k_s}) \leq \ell_n(x^{k_s}, \mu^{k_s}).
\]
Taking the limit with $s\to \infty$, we can obtain that for any $\mu=\{\mu_j\}_{j=1}^m$ 
and any $x$ satisfying ${\bf 1}_m^\top x = 1$ and $x\geq 0$,
\begin{equation}\label{eq:limit problem}
\left\{\begin{array}{ll}
\widehat{\ell}_{n;\mu}(\mu,x^*; x^*,\mu^*) \leq \widehat{\ell}_{n;\mu}(\mu^*, x^*;x^*, \mu^*)\\[0.1in]
\ell_n(x, \mu^*) \leq \ell_n(x^*, \mu^*).
\end{array}\right.
\end{equation}
If $x_j^*=0$ for some $j$'s, then we directly have $\nabla_{\mu_j} \ell(x^*, \mu^*) = 0$ for these $j$'s. For those $j$'s such that $x_j^*\neq 0$, we have $x_j^{k_s}\neq 0$ for all $s$ sufficiently large and
\[
\nabla_{\mu_j} \, \widehat{\ell}_{n;\mu} (\bullet\,,x^*;x^*, \mu^*)(\mu^*) = \nabla_{\mu_j}\, \ell_{n;\mu}(x^*,\mu^*).
\]
One can then derive the stated results directly from the optimality conditions in \eqref{eq:limit problem}.
\end{proof}

\subsection{Proof of Theorem~\ref{thm:Rate-OT-Map}}\label{pf:Denoising}
\begin{proof}
For notational simplicity, let $G_n$ denote the empirical distribution of $\mathfrak{T}^*(Y_1),\ldots, \mathfrak{T}^*(Y_n)$ (note the slight change in notation compared to~\eqref{eq:G_n}); also, we denote by $\|\cdot\|$ the usual Euclidean norm (instead of $\|\cdot\|_2$). Consider an optimal coupling $\hat \pi$ between $\nu_n$ and $\widehat{G}_n$ minimizing~\eqref{eq:Assign-Est}, and let $\hat{\pi}_{ij}$ denote the resulting probability mass that
is assigned to $Y_i$ and $\hat {a}_j$, for $1 \leq i \leq n$, and $1 \leq j \leq \hat k$. Define further $\pi_j(Y_i) = \hat{\pi}_{ij} n$, for $1 \leq i \leq n$, and $1 \leq j \leq \hat{k}$. Accordingly, we have $\hat {\alpha}_j = \frac{1}{n} \sum_{i=1}^n \pi_j(Y_i) = \int \pi_j(y) \; d\nu_n(y)$, for $1 \leq j \leq \hat{k}$.

Let $\psi^{\star}$ denote the Legendre-Fenchel conjugate of $\psi$ (recall that $\mathfrak{T}^* =  \nabla \psi$ with $\psi:\R^d \to \R$ being a convex function). We first bound $\int \psi^{\star}(\theta) \, d\widehat{G}_n(\theta) - \int \psi^{\star}(\theta) \, d G_n(\theta)$ as
\begin{align}
  &\sum_{j = 1}^{\hat k} \hat{\alpha}_j \psi^{\star}(\hat{a}_j)  -  \int \psi^{\star}(\theta) \, d G_n(\theta) \notag \\
  =&\int \sum_{j = 1}^{\hat k} \pi_j(y) \psi^{\star}(\hat {a}_j) \, d\nu_n(y) - \int \psi^{\star}(\mathfrak{T}^*(y)) \, d\nu_n(y) \notag \\
  \geq&\int  \psi^{\star}\Big(\sum_{j = 1}^{\hat k} \pi_j(y) \hat{a}_j \Big) \, d\nu_n(y)  -  \int \psi^{\star}(\mathfrak{T}^*(y)) \, d\nu_n(y) \notag \\
  =& \int \psi^{\star}(\hat T_n(y)) \; d\nu_n(y) - \int \psi^{\star}(\mathfrak{T}^*(y)) \, d\nu_n(y) \notag \\
  \geq& \int \nabla \psi^{\star}(\mathfrak{T}^*(y))^\top (\hat{T}_n(y) - \mathfrak{T}^*(y)) \; d\nu_n(y) \, + \frac{1}{2L} \int \|\hat{T}_n(y) - \mathfrak{T}^*(y)\|^2 \, d\nu_n(x), \notag \\
  =& \int x^\top (\hat{T}_n(y) - \mathfrak{T}^*(y)) \, d\nu_n(y) + \frac{1}{2L} \int \|\hat{T}_n(y) - \mathfrak{T}^*(y)\|^2 \, d\nu_n(y)    \label{eq:main_lower_bound}
\end{align}
where the two inequalities follow from the convexity of $\psi^{\star}$ (by Jensen's inequality) and the $L$-smoothness of $\psi$, which implies $\frac{1}{L}$-strong convexity of its conjugate $\psi^{\star}$ (see e.g.,~\cite{Hiriart-Urruty-93}); the last equality follows from Brenier's theorem (Theorem \ref{theo:Brenier} in Appendix \ref{app:optimaltransport}) in light of which $\nabla \psi^{\star}$ is the inverse map of $\nabla \psi$.  Moreover, $W_2^2(\nu_n, \widehat{G}_n)$  can be expressed as 
\begin{align}
\sum_{i=1}^n \sum_{j = 1}^{\hat k} \|\hat{a}_j - Y_i\|^2 \hat{\pi}_{ij} &= \sum_{j = 1}^{\hat k} \hat{\alpha}_j \|\hat{a}_j\|^2 + \frac{1}{n} \sum_{i=1}^n \|Y_i\|^2 - 2 \sum_{i=1}^n \sum_{j = 1}^{\hat k} \hat{a}_j^\top Y_i \hat{\pi}_{ij}  \notag \\
&= \int \|\theta\|^2 \, d\widehat{G}_n(\theta) + \int \|y\|^2 \, d\nu_n(y) - \frac{2}{n} \sum_{i=1}^n Y_i^\top \Big(\sum_{j = 1}^{\hat k} n \hat{\pi}_{ij} \hat{a}_j\Big) \notag \\
&= \int \|\theta\|^2 \, d\widehat{G}_n(\theta) + \int \|y\|^2 \;d\nu_n(y) - 2 \int y^\top \hat{T}_n(y) \; d\nu_n(y) \label{eq:main_aux1}.
\end{align}
Similarly,
\begin{equation}\label{eq:main_aux2}
W_2^2(\nu_n, G_n) =\int \|\theta\|^2 \, dG_n(\theta) + \int \|y\|^2 \, d\nu_n(y) - 2 \int y^\top \mathfrak{T}^*(y) \, d\nu_n(y)
\end{equation}
where we note that $\mathfrak{T}^*$ is also the optimal transport map from $\nu_n$ to $G_n$ (as $\mathfrak{T}^* \#\nu_n = G_n$ and $\mathfrak{T}^*$ is the gradient of a convex function).
Combining \eqref{eq:main_lower_bound}, \eqref{eq:main_aux1}, \eqref{eq:main_aux2}, we obtain that 
\begin{align}
\int \|\hat{T}_n(y) - \mathfrak{T}^*(y) \|^2 \, d \nu_n(y) &\leq L \Big[ W_2^2(\nu_n, \widehat{G}_n) - W_2^2(\nu_n, G_n) + 2 \int \psi^{\star}(\theta) \,d (\widehat{G}_n - G_n) (\theta) \notag \\
&\qquad \qquad + \int \|\theta\|^2 \, d(G_n- \widehat{G}_n)(\theta) \Big]. \label{eq:main_intermediate}
\end{align}
Let $\wh{\eta}$ be an optimal coupling between $G_n$ and $\widehat{G}_n$, and let $\eta = (\nabla \psi^{\star}, \textsf{id}) \# \wh{\eta}$ be the push-forward (cf.~Definition \ref{def:pushforward}) of the coupling $\wh{\eta}$; note that  $\eta$ has the two marginals to $\nabla \psi^{\star} \# G_n = \nu_n$ and $\textsf{id} \# \widehat{G}_n = \widehat{G}_n$, where we have
used that $\nabla \psi^{\star}(\mathfrak{T}^*(Y_i)) = Y_i$, $1 \leq i \leq n$, by Brenier's theorem, with $\textsf{id}$ denoting
the identity map.

Accordingly, by the definition of the 2-Wasserstein distance in terms of optimal couplings (cf.~Appendix \ref{app:optimaltransport}), we obtain that
\begin{equation*}
W_2^2(\nu_n, \widehat{G}_n) \leq \int \|y - \theta\|^2 \, d\eta(y, \theta) =
\int \|\nabla \psi^{\star}(\zeta) - \theta\|^2 \, d\wh{\eta}(\zeta, \theta). \label{eq:remainder_term_0}
\end{equation*}
Adding and subtracting $\zeta$ inside the norm on the right-hand side and expanding the square, it follows that
\begin{align}
W_2^2(\nu_n, \widehat{G}_n) &\leq \int \|\nabla \psi^{\star}(\zeta) - \zeta\|^2 \, d G_n (\zeta) +
\int \|\theta - \zeta\|^2 \, d\wh{\eta}(\zeta, \theta) + 2 \int (\nabla \psi^{\star}(\zeta) - \zeta)^\top (\zeta - \theta) \, d\wh{\eta}(\zeta, \theta) \notag \\
&= W_2^2(G_n, \nu_n) +  W_2^2(G_n, \widehat{G}_n) + 2 \int (\nabla \psi^{\star}(\zeta) - \zeta)^\top (\zeta - \theta) \, d\wh{\eta}(\zeta, \theta) \label{eq:remainder_term_bound1},
\end{align}
where we have used that $\nabla \psi^{\star}$ is the OT map pushing forward $G_n$ to $\nu_n$, the definition of the 2-Wasserstein distance in terms of optimal transport and optimal couplings, and the definition of $\wh{\eta}$ as optimal coupling between $G_n$ and $\widehat{G}_n$.

In order to bound the rightmost term in the preceding display, we use the fact that the function $\psi^{\star}$ is $(1 / \lambda)$-smooth. This yields
\begin{align}
2 \int \nabla \psi^{\star}(\zeta)^\top (\zeta - \theta) \, d\wh{\eta}(\zeta, \theta) &\leq  2 \int \left\{\psi^{\star}(\zeta) - \psi^{\star}(\theta) + \frac{1}{2 \lambda} \|\zeta - \theta\|^2 \right\} \, d\wh{\eta}(\zeta, \theta) \notag \\
&= 2 \int \psi^{\star}(\zeta) \, dG_n(\zeta) - 2 \int \psi^{\star}(\theta) \, d\widehat{G}_n(\theta) + \frac{1}{ \lambda} W_2^2(G_n, \widehat{G}_n), \label{eq:remainder_term_bound2}
\end{align}
where we have used the fact that $\wh{\eta}$ be an optimal coupling between $G_n$ and $\widehat{G}_n$.  Finally, note that
\begin{align}
2 \int (-\zeta)^\top (\zeta - \theta) \, d\wh{\eta}(\zeta, \theta) &= \int \left\{ \|\theta\|^2  - \|\theta - \zeta\|^2 - \|\zeta\|^2 \right \} \,d \wh{\eta}(\zeta, \theta) \notag \\
&= \int \|\theta\|^2 \, d\widehat{G}_n(\theta) - \int \|\zeta\|^2 \, d G_n(\zeta) - W_2^2(G_n, \widehat{G}_n). \label{eq:remainder_term_bound3}
\end{align}
Combining \eqref{eq:remainder_term_bound1}, \eqref{eq:remainder_term_bound2}, and \eqref{eq:remainder_term_bound3}, we obtain that
\begin{align*}
W_2^2(\nu_n, \widehat{G}_n) &\leq W_2^2(\nu_n, G_n) + \frac{1}{\lambda} W_2^2(G_n, \widehat{G}_n)  +  2 \int \psi^{\star}(\zeta) \, dG_n(\zeta) - 2 \int \psi^{\star}(\theta) \, d\widehat{G}_n(\theta) \\
&\qquad \quad + \int \|\theta\|^2 \, d\widehat{G}_n(\theta) - \int \|\zeta\|^2 \, dG_n(\zeta).
\end{align*}
Substituting this bound back into \eqref{eq:main_intermediate}, we observe that the right-hand side of the display equals $\frac{L}{\lambda} W_2^2(G_n,  \widehat{G}_n)$. The desired result now follows from the above fact in conjunction with the following result: In~\cite[Theorem 10]{soloff2021multivariate}, it is shown that if $G^*([-M,M]^d) = 1$, for some $M > 0$, then there is a function $n(d, M)$ such that, for all sample sizes $n$ with $n \ge n(d, M)$,
\begin{equation}\label{eq:W_2^2-Rate}
W_2^2(G^*, \widehat{G}_n) \le C_{d, \sigma} \frac{1}{\log n},
\end{equation}
for a constant $C_{d, \sigma}$, with probability at least $1- \frac{4 d}{n^8}$.
\end{proof}

\section{Introduction to the theory of optimal transport}\label{app:optimaltransport}
To make this paper self-contained, we here present notions and results from the theory of optimal transport
as far as needed for the purpose of the paper. This material or slight modifications thereof are accessible from
popular monographs and lecture notes on the subject, e.g., \cite{COT2019, Villani2009, Villani2003, Santambrogio2015, McCann2011}.

\begin{definition}[Push-forward]\label{def:pushforward} Let $\mu$ and $\nu$ be two Borel probability measures on measurable spaces
$(\mc{X}, \mc{B}_{\mc{X}})$ and $(\mc{Y}, \mc{B}_{\mc{Y}})$ respectively, and let $T$ be a measurable map from
$\mc{X}$ to $\mc{Y}$. The map $T$ is said to {\bfseries push forward} $\mu$ to $\nu$, in symbols
$T \# \mu = \nu$, if $T \# \mu(B) \equiv \mu(T^{-1}(B)) = \nu(B)$ for all $B \in \mc{B}_{\mc{Y}}$.
\end{definition}

\begin{definition}[Monge's problem]\label{def:optimaltransport}  Let $\mu$ and $\nu$ be as in the previous definition, and let $c: \mc{X} \times \mc{Y} \rightarrow [0, \infty)$ be a measurable function (`cost function'). The optimal transport problem (Monge's problem) with $\mu$, $\nu$, and $c$ is given by
\begin{equation}\label{eq:Monge-OT}
\inf_T \int_{\mc{X}} c(x, T(x)) \; d\mu(x) \qquad \text{subject to} \quad T\#\mu = \nu.
\end{equation}
Any minimizer of the above problem is called an optimal transport map.
\end{definition}
\noindent The following optimization problem is in general a relaxation of the above problem; under certain conditions, both
problems are equivalent.
\begin{definition}[Kantorovich's problem]\label{def:kantorovich} Let $\mu$ and $\nu$ be as in Definition \ref{def:pushforward}, and let $c$ be a cost function as in Definition~\ref{def:optimaltransport}. Let further $\Pi(\mu, \nu)$ denote
the set of all couplings between $\mu$ and $\nu$, i.e., probability measures on $\mc{X} \times \mc{Y}$ whose marginals
equal to $\mu$ and $\nu$. The Kantorovich problem is given by the optimization problem
\begin{equation*}
\inf_{\gamma \in \Pi(\mu, \nu)} \int_{\mc{X}} \int_{\mc{Y}}
 c(x,y) \; d\gamma(x,y).
 \end{equation*}
 Any minimizer of the above problem is called an optimal transport plan.
\end{definition}
\noindent For measures $\mu$ and $\nu$ on $\R^d$ with finite $k$-th moments ($k \geq 1$), i.e., $\int \|x\|^k \, d\mu(x) < \infty$ and
$\int \|x\|^k \, d\nu(x) < \infty$, the {\it $k$-Wasserstein distance} between $\mu$ and $\nu$ is defined via the above Kantorovich
problem with cost function $c(x,y) = \|x - y\|^k_2$, i.e.,
\begin{equation}
W_k(\mu, \nu) := \left( \inf_{\gamma \in \Pi(\mu, \nu)} \int \int \|x - y\|^k_2 \; d\gamma(x,y) \right)^{1/k}.
\end{equation}
\vskip1.5ex
\noindent A celebrated result due to Brenier characterizes optimal transport maps in the sense of Definition
\ref{def:optimaltransport} for $\mc{X} = \mc{Y} = \R^d$ and quadratic cost, i.e., $c(x,y) = \|x - y\|^2_2$ and $\mu$ absolutely continuous with respect to the Lebesgue measure. In the sequel, we let $g^{\star}(x) := \sup_{y \in \R^d} \{y^\top x - g(y) \}$ denote the Legendre-Fenchel conjugate of a convex function $g: \R^d \rightarrow \R  \cup \{ +\infty \}$.

\begin{theorem}[Brenier's theorem]\label{theo:Brenier} Suppose that $\mu$ and $\nu$ are Borel probability measures on $\R^d$ with finite second moments, and suppose further that $\mu$ is absolutely continuous with respect to the Lebesgue measure. Then the optimal transport problem~\eqref{eq:Monge-OT} with the quadratic cost, i.e., $c(x,y) = \|x - y\|^2_2$ has a ($\mu$-a.e.) unique minimizer $T = \nabla \psi$ for a convex function $\psi: \R^d \rightarrow \R \cup \{ +\infty \}$. Furthermore, the optimal transport problem and its Kantorovich relaxation are equivalent in the sense
that the optimal coupling in Definition \ref{def:kantorovich} is of the form $(\text{\emph{id}} \times T) \#\mu$. Moreover, if
in addition $\nu$ is absolutely continuous, then $\nabla \psi^{\star}$ is the ($\nu$-a.e.) minimizer of the Monge problem transporting
$\nu$ to $\mu$, and it holds that $\nabla \psi^{\star} \circ \nabla \psi(x) = x$ ($\mu$-a.e.), and $\nabla \psi \circ \nabla \psi^{\star}(y) = y$ ($\nu$-a.e.).
\end{theorem}


\end{document}